\documentclass[reprint,amsmath,amssymb,aps,superscriptaddress]{revtex4-1}

\usepackage{graphicx}
\usepackage{dcolumn}
\usepackage{bm}
\usepackage{braket}
\usepackage{color}
\usepackage{pstricks,pst-grad,color}
\usepackage{appendix}
\usepackage{here}
\usepackage{url}
\usepackage{hyperref}

\newcommand{\m}[1]{\mathcal{#1}}
\newcommand{\mr}[1]{\mathrm{#1}}
\newcommand{\tr}[1]{\mathrm{Tr}\Bigl[#1\Bigr]}

\newcommand{\bk}{\bm{k}}
\newcommand{\p}{\partial}

\usepackage{tikz}
\usetikzlibrary{shapes,calc}
\usetikzlibrary{shapes.geometric,arrows.meta,decorations.markings}

\begin{document}
\preprint{APS}

\title{Effects of renormalization and non-Hermiticity on nonlinear responses in strongly-correlated electron systems}
\author{Yoshihiro Michishita}
\email{michishita.yoshihiro.56e@st.kyoto-u.ac.jp}
 \affiliation{Department of Physics, Kyoto University, Kyoto 606-8502, Japan}
\author{Robert Peters}%
 \email{peters@scphys.kyoto-u.ac.jp}
 \affiliation{Department of Physics, Kyoto University, Kyoto 606-8502, Japan}

\date{\today}

\begin{abstract}
Nonlinear responses in condensed matter are intensively studied because they provide rich information about the materials and hold the possibility of being applied in diodes or high-frequency optical devices.
While nonlinear responses in noninteracting models have been explored widely, the effect of strong correlations on the nonlinear response is still poorly understood, even though it has been suggested that correlations can enhance the nonlinear response. In this work, we first give an analytical derivation of nonlinear responses using Green's function methods at finite temperature.
Then, we discuss the difficulties of considering dissipation using conventional methods, such as the reduced density matrix method. We reveal that the relaxation time approximation leads to severe limitations when considering optical responses. 
Finally, we demonstrate that correlation effects, such as the renormalization of the band structure and different lifetimes in orbitals or sublattices, can significantly enhance nonlinear responses and even change the sign of the nonlinear conductivity.
\end{abstract}

\maketitle

\section{Introduction}
Nonlinear responses in condensed matter theory have attained great interest because of their rich information about the symmetries of materials and their various functionalities.
For example, the breaking of the inversion symmetry in a material can be detected by measuring the second harmonic generation of the electric susceptibility\cite{Petersen2006,Zhao2016,Harter295}.
Moreover, in non-centrosymmetric materials, the shift current and non-reciprocal(rectification) current can occur in nonlinear responses\cite{Morimotoe1501524,Tokura2018}. It was extensively studied due to its application in solar cells, photodetectors, and high-frequency rectification devices\cite{doi:10.1063/1.107968,PhysRevB.95.041104,deJuan2017,Isobeeaay2497,PhysRevApplied.13.024053}.

Although nonlinear responses in condensed matter systems have many possible applications, the magnitude of the nonlinear response, which is usually small, poses a significant obstacle for most applications. 
Thus, much effort has been put into enhancing the amplitude of the nonlinear response. It has been proposed that the shift current can be magnified in Dirac systems\cite{Wu2017,PhysRevB.95.041104,Morimotoe1501524,Ma2019} and that superconducting fluctuations can enhance the nonreciprocity\cite{Wakatsukie1602390,Itahashieaay9120,Ando2020}.
Another possibility to enhance nonlinear responses might be correlation effects. 
A strong high-harmonic generation was revealed in strongly-correlated electron systems both in  experiments\cite{Kishida2000,Liu2017} and numerical calculations\cite{PhysRevB.95.035416,Silva2018,PhysRevA.100.043839,PhysRevLett.121.057405,PhysRevLett.121.097402}. A nonlinear Hall effect, which is almost $10^3$ times as large as the $ab$ $initio$ calculation result, has been measured in the Weyl-Kondo semimetal candidate Ce$_3$Bi$_4$Pd$_3$\cite{Dzsabere2013386118}.
Moreover, it has been suggested from a Hartree-analysis that the strong Coulomb interaction may enhance nonreciprocity\cite{Morimoto2018}.
Although these works show that correlation effects give large nonlinear responses, a systematic analysis of strong correlation effects on nonlinear responses is still missing.

In this paper, we first derive a formalism based on Green's functions for calculating the nonlinear response at finite temperature and formulate a diagrammatic method to use them. We note that Parker {\it et al.}\cite{PhysRevB.99.045121} derived a similar diagrammatic method for nonlinear responses focusing on the zero dissipation limit and Jo$\tilde{\mr{a}}$o {\it et al.}\cite{Jo_o_2019} introduced a diagrammatic method based on Keldysh Green's functions.
Neglecting vertex corrections, we can derive equations based on the single-particle Green's function, including correlation effects via the self-energy. Because there are many methods available to calculate the self-energy of correlated materials, the here derived formalism makes it easy to analyze correlation effects on nonlinear responses.
Next, we discuss difficulties of including the dissipation effect in conventional methods, such as the reduced density matrix(RDM) method\cite{PhysRevB.48.11705,PhysRevB.61.5337,PhysRevB.96.035431,PhysRevB.97.235446,PhysRevResearch.2.043081}.
 In these methods, dissipation is often introduced phenomenologically by using the relaxation time approximation(RTA). We reveal that the RTA breaks the gauge invariance and is only justified in the DC limit, the high-frequency limit, and at high-temperatures, while dissipation is appropriately included in the Green's function method.
 
 Furthermore, while the RDM method for nonlinear responses mainly focuses on noninteracting systems, we demonstrate that it is possible to include correlation effects into the RDM using Green's functions. By including correlation effects into the RDM, we are able to retrieve the equations of the Green's function method in the DC limit.
Finally, we use our Green's function formalism to analyze correlation effects on nonlinear responses. Notably, we look at the impact of the renormalization of the band structure and the effect of different lifetimes on the nonlinear response functions.
We show that renormalization effects can enormously enhance the nonlinear response. Considering a renormalization uniform in all orbitals, the renormalization factor $z (<1)$ enhances the $n$-th order response by a factor of $z^{-(n-1)}$. 
Furthermore, we study the effect of different lifetimes in different orbitals using the non-Hermitian band-index of the effective non-Hermitian Hamiltonian describing the single-particle Green's function. 
We show that the occurrence of different lifetimes can not only enhance terms already existing in the Hermitian case, but also creates novel non-Hermitian terms in the nonlinear response function originating in the coalescence of several bands. Our framework can be applied to most correlated electron systems, such as heavy fermions, magnetic systems, Mott insulators, and so on. However, we note that it cannot be directly used for systems with strong spatial fluctuations because we ignore vertex corrections and the momentum dependence of the self-energy. On the other hand, by using the Nambu formalism, we can also expand our framework to superconducting systems.

The rest of the paper is organized as follows: In Sec.~\ref{Derivation}, we derive the Green's function formalism for the nonlinear response at finite temperature. Next, we discuss the difficulties of including the dissipation in the RDM method in Sec.~\ref{RDM_section}.
We reveal that the RTA under an AC electric field is a severe approximation, although it is often used in previous works. In Sec.~\ref{Extension}, we extend the RDM method to interacting systems by using Green's functions. Finally, we analyze correlation effects, such as the renormalization of the band structure and the occurrence of different lifetimes in different orbitals, on the nonlinear response in Sec.~\ref{Correlation}.

\section{nonlinear response using the Green's function method\label{Derivation}}
 In this section, we introduce the Matsubara formalism to express nonlinear response functions by Green's functions, which are common and easy to handle in the context of correlated systems at finite temperature. Throughout this paper, we set the Planck constant and the lattice constant to unity,  $\bar{h}= a =1$. We also set the electron charge $e=1$ in the numerical calculations.

We here use the velocity gauge, in which the effect of electric fields is described in the Hamiltonian as
 \begin{eqnarray}
    \m{H}(\bk)&\rightarrow& \m{H}(\bk\!-\!q\bm{A}(t))\nonumber\\
    &&= \m{H}(\bk)\!+\!\sum_{n=1}^{\infty}\frac{1}{n!}\prod_{i=1}^{n} \Bigl(-qA^{\alpha_i}(t)\p_{\alpha_i}\Bigr) \m{H}(\bm{k}),
 \end{eqnarray}
where $q$ is the charge of the electron and $\alpha_i$ is a direction in the momentum space. In this paper, we suppose that there is no magnetic field and we use the Coulomb gauge $\bm{A}(\bm{x},t) = \bm{A}(t)$. We note that there is another choice of gauge, namely the length gauge. 
Under the length gauge, electric fields can be described by the dipole Hamiltonian, and it is often used in the semi-classical Boltzmann equation and the RDM. It is known that both gauges give the same results for noninteracting systems when calculating exactly\cite{PhysRevB.96.035431}. 

The action of the system in the imaginary time is given as
\begin{eqnarray}
&&S[\bm{A}]\nonumber\\
&&=\!\int_{0}^{\beta} d\tau \Bigl[ \sum_{\bk,a}\Bigl\{\bar{\psi}_{a,\bk}\partial_{\tau}\psi_{a,\bk} + \m{H}(\bk\!-\!q\bm{A}(-i\tau))\Bigr\}\!+\!H_{\mr{int}}\Bigr]\nonumber\\
\label{action_A0}\\
&&=\!\int_{0}^{\beta} d\tau \Bigl[\sum_{\bk,a}\Bigl\{\bar{\psi}_{a,\bk}\partial_{\tau}\psi_{a,\bk} + \m{H}(\bk)\nonumber\\
&& \ \ \ \ \ + \sum_{n=1}^{\infty}\frac{(-1)^n}{n!}\prod_{i=1}^{n} \Bigl(A^{\alpha_i}(-i\tau)\Bigr)\hat{\m{J}}_{\alpha_1\dots\alpha_n}(\bk)\Bigr\} +H_{\mr{int}} \Bigr]\label{action_A1}\\
&& \ \ \ \hat{\m{J}}_{\alpha_1\dots\alpha_n}(\bk)=q^n\p_{\alpha_1}\dots\p_{\alpha_n}\m{H}(\bk) \label{CO}
\end{eqnarray}
where $\bar{\psi}_{a},\psi_{a}$ are fermionic creation and annihilation  operators which construct the Hamiltonian $\m{H}$, $a$ is the orbital index, $\bm{A}(t)$ is the vector potential, $\hat{\m{J}}_{\alpha_1\dots\alpha_n}(\bk)=q^n\p_{\alpha_1}\dots\p_{\alpha_n}\m{H}(\bk)$ and $H_{\mr{int}}$ is the interaction part of the Hamiltonian. In this paper, we suppose that there is only a local interaction. We note that for general nonlocal interactions, the interaction part of the Hamiltonian also depends on the vector potential.

The partition function with applied electric field is written in the path integral formalism as
\begin{eqnarray}
    Z[\bm{A}] = \int \m{D}\bar{\psi}\m{D}\psi\exp\Bigl[-S[\bm{A}]\Bigr].
\end{eqnarray}
The expectation value of the current is 
\begin{eqnarray}
    \langle\m{J}_{\alpha}(\tau)\rangle &=& \frac{\delta}{Z[\bm{A}]\delta A^{\alpha}(-i\tau)}Z[\bm{A}],
\end{eqnarray}
which can be written using response functions as
\begin{eqnarray}
    \langle\m{J}_{\alpha}(\tau)\rangle\nonumber&=&
    \int d\tau' \m{K}^{1}_{\alpha\beta}(\tau,\tau')A^{\beta}(-i\tau')\nonumber\\
    &+&  \int d\tau'\int d\tau''\m{K}^{2}_{\alpha\beta\gamma}(\tau\,\tau',\tau'')A^{\beta}(-i\tau')A^{\gamma}(-i\tau'')\nonumber\\
    &+& \dots,
    \end{eqnarray} where
    \begin{eqnarray}
    &&\m{K}^n_{\alpha\alpha_1\dots\alpha_n}(\tau_1,\dots,\tau_n)\nonumber\\
    &&= \frac{1}{Z[A]}\Bigl(\prod_{i=1}^{n} \frac{\delta}{\delta A^{\alpha_i}(-i\tau_i)}\Bigr)\frac{\delta}{\delta A^{\alpha}(-i\tau)}Z[A]\Bigr|_{A=0}.\label{Res_Matsu}
\end{eqnarray}
The results for the response functions in imaginary time are explicitly written in the Appendix \ref{app:Matsubara}.

After Fourier transformation to Matsubara frequencies, the current is given as
\begin{align}
&\langle\m{J}_{\alpha}(i\omega_n)\rangle\nonumber\\
&= \mr{K}^{(1)}_{\alpha\beta}(i\omega_n;i\omega_n) A^{\beta}(i\omega_n)\nonumber\\
& \ \  +\!\sum_{\omega_m,\omega_l}\mr{K}^{(2)}_{\alpha\beta\gamma}(i\omega_n;i\omega_m,i\omega_l)A^{\beta}(\omega_m)A^{\gamma}(\omega_l)\delta(\omega_n\!-\!\omega_m\!-\!\omega_l)\nonumber\\
& \ \  +\dots
\end{align}
The frequency before the semicolon in the response function $\mr{K}^{(n)}_{\alpha\beta}(i\omega_n;i\omega_n,\ldots)$ represents the frequency of the output response, and the frequencies after the semicolon represent the frequencies of the input forces, i.e. of the vector potentials.

Analytical continuation and using $\bm{E}(\omega_i)=i\omega_i\bm{A}(\omega_i)$ finally yields 
\begin{align}
&\langle\m{J}_{\alpha}(\omega)\rangle\nonumber\\
&= K^{(1)}_{\alpha\beta}(\omega;\omega)A^{\beta}(\omega)\nonumber\\
& \ +\!\int d\omega_1\int d\omega_2 K^{(2)}_{\alpha\beta\gamma}(\omega;\omega_1,\omega_2)A^{\beta}(\omega_1)A^{\gamma}(\omega_2)\delta(\omega\!-\!\omega_{12})\nonumber\\
& \ + \dots\\
&=\sigma^{(1)}_{\alpha\beta}(\omega)E^{\beta}(\omega)\nonumber\\
& \ + \int d\omega_1 \sigma^{(2)}_{\alpha\beta\gamma}(\omega;\omega_1,\omega_2)E^{\beta}(\omega_1)E^{\gamma}(\omega_2)\delta(\omega\!-\!\omega_{12})\nonumber\\
& \ + \dots
\end{align}
\begin{eqnarray}
\sigma^{(n)}_{\alpha\beta\dots}(\omega;\{\omega_s\}) &=& K^{(n)}_{\alpha\beta\dots}(\omega;\{\omega_i\})/\bigl(\prod_{s=1}^ni\omega_i\bigr),\label{Cond_RF}
\end{eqnarray}
where  $\omega_{12}=\omega_1+\omega_2$.
The first- and second-order conductivities can be expressed via single-particle Green's functions as
\begin{widetext}
\begin{align}
&\sigma^{(1)}_{\alpha\beta}(\omega_1;\omega_1)\nonumber\\
&=-\frac{1}{\omega_1}\int_{-\infty}^{\infty} \frac{d\omega}{2\pi}f(\omega)\sum_{\bk}\Biggl\{\tr{ \m{J}_{\alpha\beta}(\bk) \bigl(G^R(\omega,\bk)-G^A(\omega,\bk)\bigr)}\nonumber\\
& \ \ \ + \tr{\m{J}_{\alpha}(\bk)G^R(\omega\!+\!\omega_1,\bk)\m{J}_{\beta}(\bk) \Bigl(G^R(\omega,\bk)\!-\!G^A(\omega,\bk)\Bigr)\!+\!\m{J}_{\alpha}(\bk)\Bigl(G^R(\omega,\bk)\!-\!G^A(\omega,\bk)\Bigr)\m{J}_{\beta}(\bk) G^A(\omega\!-\!\omega_1,\bk)}\Biggr\}\label{cond_first}
\end{align}
\begin{align}
&\sigma^{(2)}_{\alpha\beta\gamma}(\omega_1+\omega_2;\omega_1,\omega_2)\nonumber\\
&= \frac{1}{\omega_1\omega_2}\int_{-\infty}^{\infty} \frac{d\omega}{2\pi i}f(\omega)\sum_{\bk}\Biggl\{\frac{1}{2}\tr{\m{J}_{\alpha\beta\gamma}\bigl(G^R(\omega)-G^A(\omega)\bigr)}\nonumber\\
& \ \ \ \ \ \ \ \ + \tr{\m{J}_{\alpha\beta}G^R(\omega\!+\!\omega_2)\m{J}_{\gamma}\bigl(G^R(\omega)\!-\!G^A(\omega)\bigr) +\m{J}_{\alpha\beta}\bigl(G^R(\omega)\!-\!G^A(\omega)\bigr)\m{J}_{\gamma}G^A(\omega\!-\!\omega_2)}\nonumber\\
& \ \ \ \ \ \ \ \ + \frac{1}{2}\tr{\m{J}_{\alpha}G^R(\omega\!+\!\omega_{12})\m{J}_{\beta\gamma}\bigl(G^R(\omega)\!-\!G^A(\omega)\bigr) +\m{J}_{\alpha}\bigl(G^R(\omega)\!-\!G^A(\omega)\bigr)\m{J}_{\beta\gamma}G^A(\omega\!-\!\omega_{12})}\nonumber\\
& \ \ \ \ \ \ \ \ +\mr{Tr}\Bigl[\m{J}_{\alpha}G^R(\omega\!+\!\omega_{12})\m{J}_{\beta}G^R(\omega\!+\!\omega_2)\m{J}_{\gamma} \bigl(G^R(\omega)\!-\!G^A(\omega)\bigr)\nonumber\\
& \ \ \ \ \ \ \ \ \ \ \ \ \ \ + \m{J}_{\alpha}G^R(\omega\!+\!\omega_{1})\m{J}_{\beta}\bigl(G^R(\omega)\!-\!G^A(\omega)\bigr)\m{J}_{\gamma}G^A(\omega\!-\!\omega_2) \nonumber\\
& \ \ \ \ \ \ \ \ \ \ \ \ \ \ + \m{J}_{\alpha}\bigl(G^R(\omega)\!-\!G^A(\omega)\bigr)\m{J}_{\beta}G^A(\omega\!-\!\omega_1)\m{J}_{\gamma}G^A(\omega\!-\!\omega_{12})\Bigr]\nonumber\\
& \ \ \ \ \ +\Bigl[(\beta,\omega_1) \leftrightarrow (\gamma,\omega_2)\Bigr]\Biggr\},\label{cond_second}
\end{align}
\end{widetext}
where $\m{J}_{\alpha\beta\dots}$ is the matrix representation of $\hat{\m{J}}_{\alpha\beta\dots}$, $G^{R/A}(\omega,k)$ is the retarded/advanced Green's function, and $f(\omega)$ is the Fermi distribution function. $[(\beta,\omega_1)\leftrightarrow (\gamma,\omega_2)]$ means a term in which the index and the variable have been replaced by the other set.
Further details of the derivation are given in the Appendix \ref{app:Matsubara} and \ref{app:AnalyticContinuation}. Throughout this paper, we omit the $\bk$-index of the Green's function and the velocity operator, $\m{J}_{\alpha\beta\dots}$. Furthermore, we ignore vertex corrections in the many-particle Green's functions, which allows us to express the conductivity as a product of single-particle Green's functions. This approximation is also commonly used in the semi-classical Boltzmann equation and the RDM formalism. The results above are consistent with the results in \cite{PhysRevB.99.045121}, and \cite{Jo_o_2019}. Specifically, in the dissipationless limit, the results are consistent with\footnote{We note that the results in \cite{PhysRevB.99.045121} seem to include a typo in Eq.~(B18) where $1/(\omega-\epsilon_{ab})$ should be changed to $1/(\omega-\epsilon_{ba})$ in the third term in Eq.~(43).} Eqs.(26) and (43) in \cite{PhysRevB.99.045121}. The detail is written in Appendix \ref{app:free_limit}.
The here presented procedure to derive the nonlinear optical conductivity can be summarized into a diagrammatic method, which is given in Appendix \ref{app:DiagramMethods}. We note that this diagrammatic method is a generalization of the diagrammatic method at zero temperature in Parker {\it et al.} \cite{PhysRevB.99.045121} to nonlinear response functions using real-frequencies at finite temperature.

If we take the DC limit $\omega_1,\omega_2\rightarrow 0$, the first- and second-order conductivities become
\begin{widetext}
\begin{eqnarray}
\sigma^{(1)}_{DC;\alpha\beta}&=& \int_{-\infty}^{\infty} \frac{d\omega}{2\pi}  \Biggl\{\Bigl(-\frac{\partial f(\omega)}{\partial \omega}\Bigr)\mr{Re}\tr{\m{J}_{\alpha} G^R(\omega)\m{J}_{\beta}G^A(\omega)} - 2f(\omega)\mr{Re}\tr{\m{J}_{\alpha} \frac{\p G^R(\omega)}{\p\omega}\m{J}_{\beta}G^R(\omega)}\Biggr\}\label{DC_linear}\\
\sigma^{(2)}_{DC;\alpha\beta\gamma}&=& -2\int_{-\infty}^{\infty} \frac{d\omega}{2\pi}  \Biggl\{\Bigl(-\frac{\partial f(\omega)}{\partial\omega}\Bigr)\mathrm{Im}\Bigl(\tr{\m{J}_{\alpha}\frac{\partial G^R(\omega)}{\partial\omega}\m{J}_{\beta}G^R(\omega)\m{J}_{\gamma}G^A(\omega)} +\frac{1}{2} \tr{\m{J}_{\alpha}\frac{\p G^R(\omega)}{\p\omega}\m{J}_{\beta\gamma}G^A(\omega)}\Bigr)\nonumber\\
&& \ \ \ \ \ \ \ \ \ \ \ \ \ \ -f(\omega)\mathrm{Im}\Bigl(\tr{\m{J}_{\alpha}\frac{\p}{\p\omega}\Bigr(\frac{\partial G^R(\omega)}{\partial\omega}\m{J}_{\beta}G^R(\omega)\Bigr)\m{J}_{\gamma}G^R(\omega)} + \frac{1}{2}\tr{\m{J}_{\alpha}\frac{\p^2G^R(\omega)}{\p\omega^2}\m{J}_{\beta\gamma}G^R(\omega)}\Bigr)\Biggr\}\nonumber\\
&& \ \ \ \ \ \ \ \ \ \ + (\beta \leftrightarrow \gamma)\label{DC_second1}
\end{eqnarray}
\end{widetext}
Interaction effects can be taken into account by including the retarded/advanced self-energy $\Sigma^{R/A}(\omega)$ into the Green's function, $G^R(\omega)=1/[\omega-\m{H}-\Sigma^{R/A}(\omega)]$. Throughout this paper, we ignore the momentum dependence of the self-energy. Including the momentum dependence of the self-energy, we should also consider vertex corrections to satisfy the Ward-Takahashi identities. We note that the momentum dependence of the self-energy can become significant for certain phenomena in strongly correlated materials and, in these cases, must be included in the considerations about nonlinear responses. We also note that we can recover the physical unit by substituting $\omega \rightarrow \bar{h}\omega$ and multiply $a^{n}$ for $n$-th order nonlinear conductivity.

Finally, setting $\Sigma(\omega)=i\gamma/2$ and taking the limit $\gamma\rightarrow 0$, we can perform the frequency integrals and further simplify the results which are summarized in Appendix~\ref{app:free_limit} .

\section{Difficulties describing dissipation effects in the reduced density matrix formalism\label{RDM_section}}
Having introduced the Green's function technique based on a path integral derivation to calculate nonlinear transport, we can compare with different approaches and approximations made to calculate the nonlinear response.
The semi-classical Boltzmann equation and the RDM method are often used to calculate nonlinear responses. In these methods, the dissipation is usually introduced by the relaxation time approximation(RTA). In this section, we briefly introduce the RDM method. Being able to compare it with the Green's function method, we can pinpoint the problems accompanying the RTA and explain in what situation RTA is justified. We note that the results by the semi-classical Boltzmann equation can also be obtained by the RDM results\cite{PhysRevB.99.045121} so that we here consider only the RDM method. We briefly introduce the Boltzmann equation approach to nonlinear transport in Appendix~\ref{app:OtherMethods}.

\subsection{Reduced Density Matrix Formalism\label{RDM_formalism}}
When ignoring two-body correlations, we can write the total density matrix of the lattice system as the tensor product of the reduced density matrices $\rho_{\mr{tot}}(t) = \prod_{\bk} \otimes \rho_{\bk}(t)$. We can now describe the dynamics of the density matrix for each momentum $\bm{k}$ under the electric field by using the von Neumann equation, which reads
\begin{eqnarray}
&&\frac{d}{dt}\rho_{\bk}(t) = -i\bigl[\m{H},\rho_{\bk}(t)\bigr]-(\rho(t)-\rho^{(0)})/\tau\label{VN}\\
&&\m{H}=\m{H}_0+\m{H}_{E}\\
&&\m{H}_{E} = -q \bm{E}\cdot\bm{r}, \ \ \ \bm{r} = -i\bm{\nabla_k},\label{dipole}
\end{eqnarray}
where we introduce the effect of dissipation by using the RTA, $-(\rho(t)-\rho^{(0)})/\tau$, and $\rho^{(0)}$ describes the equilibrium state without the electric field.
In the RDM formalism, we use the length gauge and describe the dynamics with the dipole Hamiltonian in Eq.~(\ref{dipole}). The density matrix under the velocity gauge can be obtained by using the transformation $\rho_E = U\rho_A U^{\dagger}, \ \ U= \exp[-iq\bm{A}(t)\cdot\bm{r}]$, where $\rho_{E/A}$ is the density matrix under the length/velocity gauge. We note again that results obtained by the length gauge are equivalent to those obtained under the velocity gauge without dissipation\cite{PhysRevB.96.035431}. The recurrence equation of the $n$-th order density matrix $\rho^{(n)}(t)$ about the electric field can be written as
\begin{eqnarray}
&&\frac{d\rho^{(n)}(t)}{dt} = -i[\m{H}_0,\rho^{(n)}(t)] -i [\m{H}_E,\rho^{(n-1)}(t)]-\gamma\rho^{(n)}(t)\nonumber\\
&& \ \ \ \ \ \ \ = -i \m{L}\rho^{(n)}(t) + q \bm{E}(t)\cdot\bm{\nabla} \rho^{(n-1)}(t) -\gamma\rho^{(n)}(t)\\
&&\mathrm{F.T.} \Leftrightarrow \rho^{(n)}_{\bm{k}}(\omega) = \frac{iq E^{\mu}(\omega_n)}{\omega-\m{L} + i\gamma}\nabla_{\mu}\rho^{(n-1)}_{\bm{k}}(\omega-\omega_n),\label{RDM1}
\end{eqnarray}
where $-i\m{L}\rho = -i[\m{H}_0,\rho]$, $\gamma=1/\tau$, $\omega_n$ describes the frequency of the electric field which leads to the $n$-th order density matrix $\rho^{(n)}$, $\bm{E}(\omega_n)$ is the Fourier component of $\bm{E}(t)$ and $F.T.$ means Fourier transformation. In the length gauge, the current operator $\m{J}$ can be written as,
\begin{eqnarray}
 \m{J} = q \dot{\bm{r}} = -iq[\bm{r},\m{H}] = q \bm{\nabla_{\bk}}\m{H},
\end{eqnarray}
and, therefore, the $n$-th order conductivity can be calculated by
\begin{eqnarray}
    \sigma^{(n)}_{\alpha;\{\alpha_i\}}(\omega;\{\omega_{i}\}) = \sum_{\bk}\tr{\m{J}_{\alpha}\rho^{(n)}_{\bk}(\omega)/(\prod_i E^{\alpha_i}(\omega_i))}.\nonumber\\
\end{eqnarray}
Detailed expressions can be found in \cite{PhysRevB.61.5337,PhysRevX.11.011001}. 
We note that the equations of the RDM method using RTA can be derived from the Green's function technique in the DC-limit and in the dissipation-free limit for $\omega_i\gg\epsilon_{nm},\gamma$. Details about this correspondence are given in Appendix~\ref{app:free_limit}. The RDM method introduced here is exact except for the RTA, and therefore, the necessary conditions we listed above are caused by the RTA.

\subsection{Velocity gauge vs Length gauge under the relaxation time approximation\label{VvsL}}
In an isolated system without dissipation, physical quantities calculated by the velocity and length gauge are the same, which was shown in \cite{PhysRevB.96.035431}. In this subsection, we show that this correspondence between both gauges breaks down when using the RTA. The density matrix in each gauge can be written as \cite{PhysRevB.96.035431}
\begin{eqnarray}
    \rho_E(t) &=& U \rho_A(t) U^{\dagger}, \ \ \ \ U = \exp [-i\bm{A}(t)\cdot \bm{r}],\label{LvsV}\\
    \rho^{(n)}_E(t)&=& \rho^{(n)}_A(t)\nonumber\\
    && +\sum_{l=1}^{n}(\prod_{m=1}^{l}-iA^{\alpha_m}(t))[r_{\alpha_l},[r_{\alpha_{l-1}},\dots [r_{\alpha_1},\rho^{(n-l)}]]] \nonumber\\
    &=& \rho^{(n)}_A(t) - i\bm{A}(t)\cdot [\bm{r},\rho^{(n-1)}_A(t)] - \dots,\label{equality}
\end{eqnarray}
where $\rho_E(t)$ is the density matrix under the length gauge, $\rho_A(t)$ is the density matrix under the velocity gauge, and $\rho^{(n)}(t)$ represent the density matrix with the $n$-th order perturbation by the electric fields. By applying the RTA, the density matrices under both gauges change as $\rho^{(n)}_{E/A}(t)\rightarrow \rho^{(n)}_{E/A}(t)e^{-\gamma t}$ when $n\geq1$. The equality in Eq.~(\ref{equality}) for the $n=1$-order density matrix using the RTA becomes
\begin{eqnarray}
    \rho^{1}_E(t)e^{-\gamma t}&\overset{?}{=}& \rho^{(1)}_A(t)e^{-\gamma t} - i\bm{A}(t)\cdot [\bm{r},\rho^{(0)}_A]\label{eq?}
\end{eqnarray}
However, because $\rho^{(0)}_A$ does not include dissipation, the equality in Eq.~(\ref{eq?}) has to break down. 

One possible strategy to avoid this breakdown is to ignore the dissipation in the system and instead include photon dissipation or adiabatic switching as $\bm{A}(t)\rightarrow \bm{A}(t)e^{-\gamma t}$. In this case, the equality in Eq.~(\ref{equality}) holds true. However, it gives different results from the RTA, especially in the regime $\omega_i \ll \gamma$ \cite{PhysRevB.97.235446}. When substituting $\bm{A}(t)\rightarrow \bm{A}(t)e^{-\gamma t}$, we do not consider the dissipation and scattering of electrons in the system. Thus, a  current must not occur because there is no mechanism to change the momentum of electrons, $\bm{k}\rightarrow \bm{k'}$, and to induce a non-equilibrium steady-state state. 
Therefore, when including dissipation of electrons by applying the RTA, a breakdown of the equality between the velocity gauge and the length gauge is inevitable. 
We note that, in the Green's function method, this breakdown does not occur when we use $G^R(\omega)=1/(\omega-\m{H}+i\gamma/2)$ and $G^A(\omega)=1/(\omega-\m{H}-i\gamma/2)$ because it just supposes that the dissipation is constant in the absence of an electric field.

\subsection{Problems of the relaxation time approximation in an AC electric field\label{RTA_problem}}

In this part, we introduce the dissipation into the RDM method without using the RTA and show under which conditions the RTA is a good approximation. This analysis reveals the problems of using the RTA in an AC electric field. Finally, we compare the RDM using the RTA with the Green's function formalism numerically.

The easiest way to introduce the dissipation microscopically is to couple the system with a dissipative bath. For the sake of simplicity, we consider the single-band case and the coupling Hamiltonian $\m{H}_c=\lambda(\psi^{\dagger}_{\bk}\m{B} + \mr{H.c.})$, where $\m{B}^{(\dagger)}$ is the annihilation(creation) operator in the dissipative bath. In that case, the dynamics of the system can be described by the quantum master equation, which reads
\begin{eqnarray}
    \frac{d}{dt}\rho^I_{\bk}(t) &=& \!-\!\int_{t_0}^{t}ds \mr{Tr}_B\Bigl(\bigl[\m{H}^I_c(t),[\m{H}^I_c(s),\rho^I_{\bk}(s)\otimes\rho_B]\bigr]\Bigr),\nonumber\\
    \label{QME}
\end{eqnarray}
where $\rho_B$ is the density matrix of the bath and $\mr{Tr}_B$ corresponds to the trace over the bath degrees of freedom.
The operators are in the interaction representation,
$\m{O}^I(t) = T_{\leftarrow}\exp [i\int_{t_0}^{t} dt' \m{H}(t')]\m{O}T_{\rightarrow}\exp [-i\int_{t_0}^{t} dt' \m{H}(t')]$, where $\m{H}(t)=\m{H}_S(t)\otimes\m{H}_B$, $\m{H}_S(t)=\m{H}_0-q\bm{E}(t)\cdot\bm{r}$ is the system Hamiltonian, $\m{H}_B$ is the bath Hamiltonian, $T_{\rightarrow(\leftarrow)}$ represents the (anti-)time ordering operator. Although we take here the length gauge, the correspondence between the length and the velocity gauge holds exactly in this formulation. The proof is written in Appendix~\ref{app:Gauge}.
Equation (\ref{QME}) includes the dissipation term, the energy shift term, and the gain and loss terms which describe the dynamics of a particle leaving or entering the system.
Here, we suppose that a particle that leaves the system loses the information about the acceleration due to the electric fields, and the electric fields do not accelerate the particles in the bath. Under this assumption, the gain and loss terms do not affect the dynamics of $\rho^{(n)} \ (n\neq0)$, and therefore, they do not affect the conductivity. Now, we focus on the dissipation term and ignore the energy shift term. Then, Eq.~(\ref{QME}) can be rewritten as
\begin{widetext}
\begin{align}
&\frac{d}{dt}\Big(\rho^I_{\bk}(t)\Bigr)^{(n)}=-\lambda^2\int_{t_0}^{t}ds\mr{Re}\Bigl[\bigl\{iG^{l}_B(t-s)\psi^{\dagger I}_{\bk}(t)\psi^{ I}_{\bk}(s),\rho^I_{\bk}(s)\bigr\}\Bigr]^{(n)},\label{QME2}
\end{align}
\end{widetext}
where $G^{l}_B(t-s) = -i\mr{Tr}_B\bigl[\m{B}^{I}(t)\m{B}^{\dagger I}(s)\rho_B\bigr]$ and $\{\m{O},\rho\}=\m{O}\rho+\rho\m{O}^{\dagger}$.

Now, we use the Markov approximation to simplify Eq.~(\ref{QME2}), in which we take the limit 
$t_0\rightarrow -\infty$ and approximate $\rho^I_{\bk}(s)\simeq \rho^I_{\bk}(t)$. The Markov approximation is justified when $\lambda \tau_B \ll 1$, where $\tau_B$ is the relaxation time of the bath and $G_B(t-s)\propto \exp[-(t-s)/\tau_B]$. Under the Markov approximation, Eq.~(\ref{QME2}) can be rewritten as
\begin{widetext}
\begin{eqnarray}
    &&\frac{d}{dt}\rho_{\bk}(t) = -i[\m{H},\rho_{\bk}(t)] -\lambda^2\Bigl(\int_{0}^{\infty} d(t-s) \mr{Re}\Bigl[ \bigl\{iG^{l}_B(t-s)\psi^{\dagger}_{\bk}U'(t,s)\psi_{\bk}U'^{\dagger}(t,s),\rho_{\bk}(t)\bigr\}\Bigr] \Bigr)\label{QME3}\\
    &&U'(t,s) = T_{\rightarrow}\exp\Bigl[-i\int_{s}^{t} dt'(\m{H}_0 -q\bm{E}e^{i\omega_0t'}\cdot\bm{r} )\Bigr].
\end{eqnarray}
\end{widetext}
Finally, we consider in what situation we can derive the RTA  from Eq.~(\ref{QME3}). RTA should be a good approximation to describe transport when the integral in Eq.~(\ref{QME3}) becomes time-independent, thus, when $U'(t,s)$ becomes a function of $(t-s)$ or is constant. We see that in the DC limit $\omega_0\rightarrow 0$ or when the temperature of the bath is infinite and $G_B(t-s)\propto \delta(t-s)$, or when $\omega_0$ is large enough so that $q\bm{E}\cdot\bm{r}/\omega_0$ can be ignored, the integral $\bigl(\int d(t-s) \sim \bigr)$ becomes a constant and Eq.~(\ref{VN})  can be derived from Eq.~(\ref{QME3}).

After having analyzed the validness of the RTA, we will now directly compare the linear and nonlinear(photogalvanic) optical conductivity calculated by the Green's function method with the RDM using the RTA for a simple model. For this purpose, we use a model describing two-dimensional transition metal dichalcogenides(TMD) in which nonlinear optical response was discussed in the literature \cite{Zhang_2018,PhysRevB.99.201410,PhysRevApplied.13.024053}.  Details about the model are given in Appendix \ref{app:Model} and the details about how to perform the numerical calculations is given in  Appendix \ref{app:Numerical}.

\begin{figure}[t]
   \includegraphics[width=0.49\linewidth]{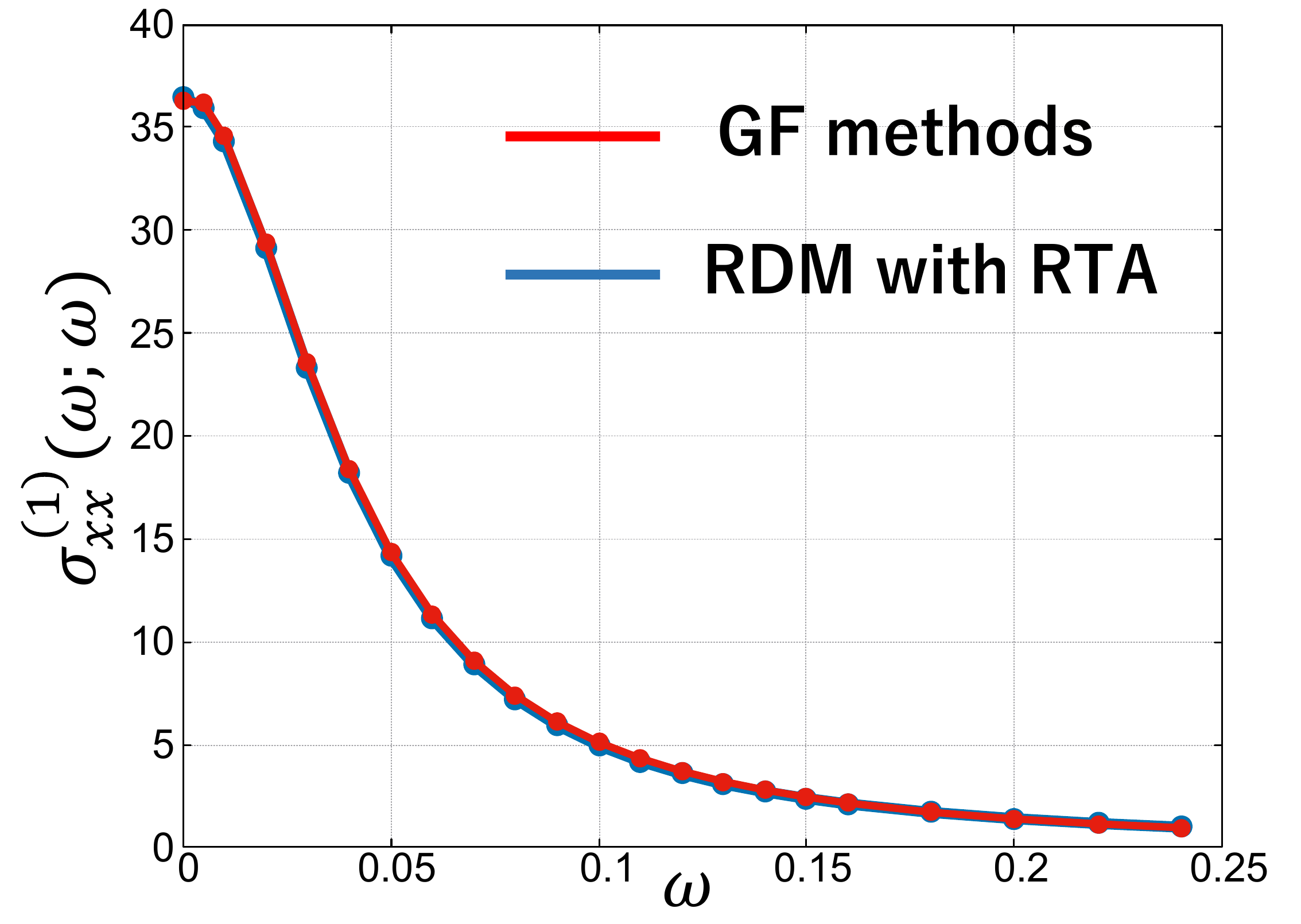}
   \includegraphics[width=0.49\linewidth]{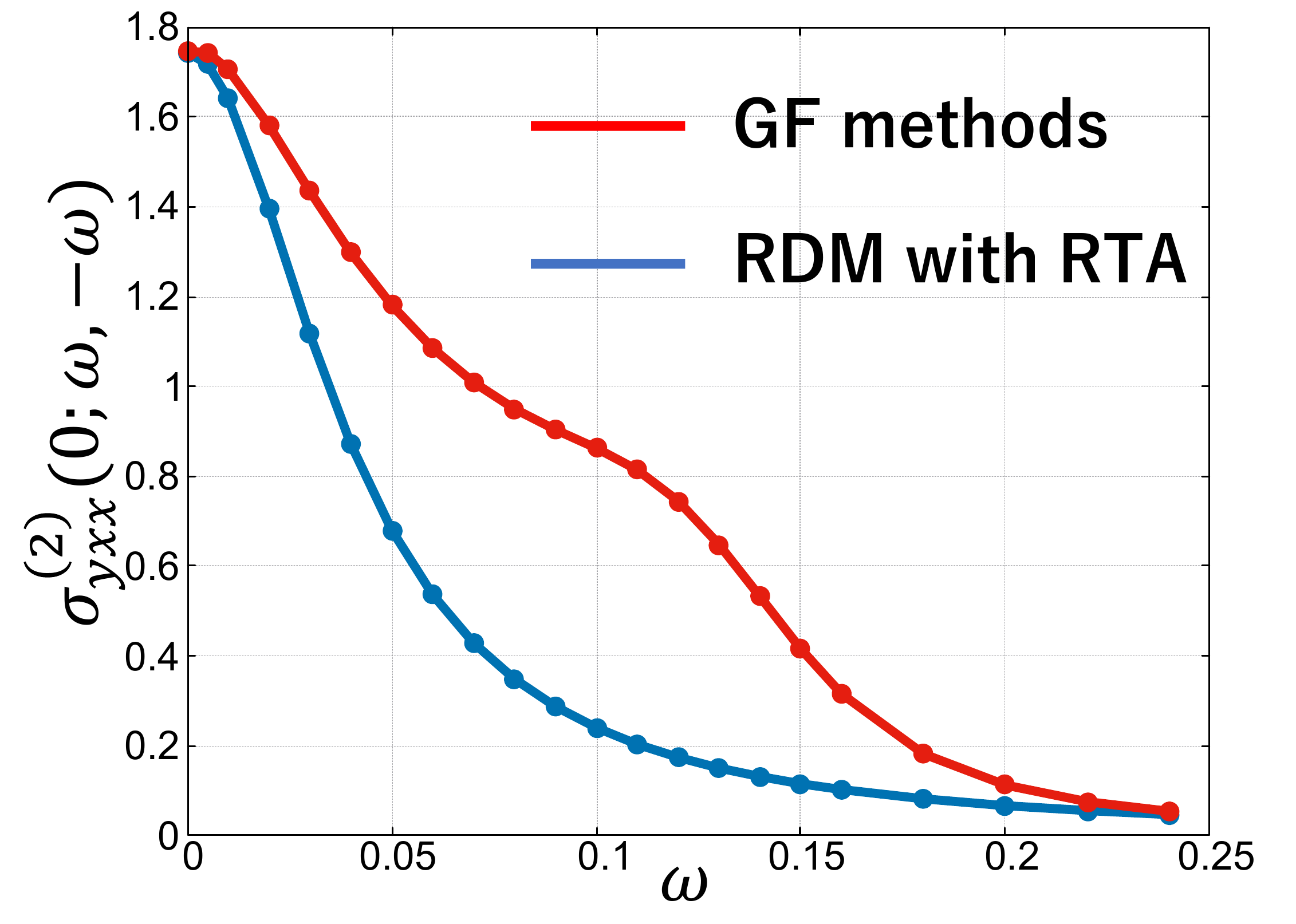}\\
  \caption{Comparison between the RDM method using RTA (blue) and the Green's function method (red) for the linear (left panel) and nonlinear (right panel) optical conductivities, which is the photogalvanic effect.
\label{RTAvsGreen}
The parameters for the monolayer TMD model are
 $t=0.5$, $\mu=0.7$, $p=0.7$, $\alpha_1=0.08$, $\alpha_2=0.06$, $\delta=0.7$, $\gamma=0.05$, $T=0.02$.\label{fig:NH_cond}}
  \label{fig:Fig1}
\end{figure}

The numerical results of the optical conductivity by the RDM method using the RTA and by the Green's function method are shown in Fig.~\ref{RTAvsGreen}. For the linear optical conductivity, the results of both methods agree with each other over the full frequency range.
On the other hand, for the nonlinear optical conductivity, the results only match in the DC limit, and for large frequencies, $\omega_i\gg\gamma$, as has been discussed above.
We thus find that while RTA is a good approximation for the linear optical conductivity, it leads to severe problems for the nonlinear optical conductivity except  in the DC limit and for $\omega_i\gg\gamma$. Again, we note that the RTA supposes that all non-equilibrium states decay equally by $\gamma$. On the other hand, the Green's function method only assumes that the dissipation is constant in the absence of an electric field. The RTA is a more severe approximation, which affects nonlinear responses. We note that the relaxation time in most materials is usually about $1\sim100 [\mr{ps}]$\cite{NatureComm.10.3047}. Thus, when analyzing a Terahertz laser as input force, $\omega_i \tau \sim1$, and the error of the RTA becomes large.

\section{Extension of the reduced density matrix formalism to interacting system\label{Extension}}
Having derived the Green's function method for nonlinear responses, we are able to extend the RDM method to interacting systems, mainly in the DC limit, and reproduce the results of the Green's function method. For free electron systems, we use $\rho^{(0)}_{\bk} = \sum_{n} f(\epsilon_n(\bk)) \ket{n}\bra{n}$. However, when we consider interacting systems, the pole of the Green's function includes the information of the quasi-particle's energy level, and therefore, the density matrix can be written as
\begin{eqnarray}
    \rho^{(0)}_{\bk}&=& \int\frac{d\omega}{2\pi i}\sum_n \ket{n}\bra{n}\Bigl(G^A_n(\omega)\!-\!G^R_n(\omega)\Bigr)f(\omega)\\
    &=&\int\frac{d\omega}{2\pi i}\sum_{\alpha\beta} \ket{\alpha}\bra{\beta}\Bigl(G^A(\omega)\!-\!G^R(\omega)\Bigr)_{\alpha\beta}f(\omega),\nonumber\\
\end{eqnarray}
where $\ket{\alpha},\bra{\beta}$ are states of an arbitrary basis, and $()_{\alpha\beta}$ represent the elements of the Green's function in this basis. We note that we again omit  the momentum-dependence of the Green's function. Here, we can choose a momentum-independent basis, $\p_{\alpha}\ket{\alpha}=0$. 
In this case, the correction of the density matrix by the electric fields only affects the Green's function matrices because $f(\omega)$ does not depend on $\bk$.
Therefore, the density matrix corrected by $n$-th order electric fields can be written as
\begin{align}
    &\Bigl(\rho^{(n)}_{\bk}\Bigr)_{\alpha\beta}\nonumber\\
    &= \int\frac{d\omega}{2\pi i}\sum_{l=0}^n \Bigl(G^{R(l)}(\omega)\Bigl(\bigl(G^{R(0)}(\omega)\bigr)^{-1}\!-\!\bigl(G^{A(0)}(\omega)\bigr)^{-1}\Bigr)\nonumber\\
    & \ \ \ \ \ \ \ \ \ \ \ \ \ \ \ \ \ \times f(\omega)G^{A(n-l)}(\omega)\Bigr)_{\alpha\beta}.\label{DC_RDM}
\end{align}
Although we need the Green's function corrected by the $n$-th order of the electric field, we can easily derive an equation for this using the RDM method. Here, we note that in our previous work\cite{PhysRevLett.124.196401}, we showed that the dynamics of the matrix elements $\rho^G_{\bk,\alpha}(0)=\ket{\bm{k},\alpha}\bra{0}$ corresponds to the retarded Green's function $G^R(\bm{k})$, which reads 
\begin{eqnarray}
G^{R}_{\alpha\beta}(t) &=& -i\theta(t)\tr{\Bigl(\psi_{\alpha}(t)\psi^{\dagger}_{\beta}+\psi^{\dagger}_{\beta}\psi_{\alpha}(t)\Bigr)\rho^{(0)}_{\bk}}\\
&=& -i \tr{\psi_{\alpha}\rho^G_{\beta}(t)},
\end{eqnarray}
where $\rho^G_{\bk\beta}(0) = \psi^{\dagger}_{\beta}\rho^{(0)}_{\bk}+\rho^{(0)}_{\bk}\psi^{\dagger}_{\beta} = \ket{\bk,\beta}\bra{0}$ and the dynamics of $\rho^G$ can be described\cite{PhysRevLett.124.196401} as
\begin{eqnarray}
\frac{d}{dt}\rho^{GI}_{\bk}(t) &=& - \int_{t_0}^{t}dsi\Sigma^{RI}(t-s)\rho^{GI}_{\bk}(s)\\
\frac{d}{dt}\rho^{G}_{\bk}(t) &=& -i[\m{H}_0+\m{H}_E,\rho^{G}_{\bk}(t)]\nonumber\\
&&-\!\int_{t_0}^{t}\!ds i\Sigma^{R}(t\!-\!s)\Bigl(\sum_{n=0}^{\infty}\frac{\bigl(iq(t\!-\!s)\bm{E}\cdot\bm{r}\bigr)^n}{n!}\Bigr)\rho^{G}_{\bk}(s)\nonumber\\
&&\label{DC_RDM2}\\
\rho^{G(n)}_{\bk}(\omega)&=&\frac{iq\bm{E}}{\omega-(\m{H}_0+\Sigma^R(\omega))}\Bigl\{\bigl(1-\frac{\p\Sigma^R}{\p\omega}\bigr)\bm{\nabla}\rho^{G(n-1)}_{\bk}(\omega)\nonumber\\
&&\!+\!\sum_{m=2}^{\infty}\frac{1}{m!}\frac{\p^m\Sigma^R}{\p\omega^m}(-\!iq\bm{E}\cdot\bm{\nabla})^{m}\rho^{G(n\!-\!m)}_{\bk}(\omega)\Bigr\}.\label{DC_RDM1}
\end{eqnarray}
To derive Eq.~(\ref{DC_RDM2}), we approximate $U'(t,s)\simeq\exp[-i\m{H}_0(t-s)]\exp[-iq\bm{E}\cdot\bm{r}(t-s)]$ in the dissipation term, which should correspond to ignoring the vertex correction. By using this equation, $G^{R(1)}(\omega)$ (the first-order correction of an electric field to the single-particle Green's function) can be derived as
\begin{align}
&\Bigl(G^{R(1)}(\omega)\Bigr)_{\alpha\beta}\nonumber\\
&= iE^{\mu}\Bigl(G^{R(0)}(\omega)\bigl(1-\frac{\p\Sigma^R}{\p\omega}\bigr)G^{R(0)}(\omega)\m{J}_{\mu}G^{R(0)}(\omega)\Bigr)_{\alpha\beta}\nonumber\\
&=-iE^{\mu}\Bigl(\frac{\p G^{R(0)}}{\p\omega}\m{J}_{\mu}G^{R(0)}(\omega)\Bigr)_{\alpha\beta},\label{RDM_first}
\end{align}
By inserting Eq.~(\ref{RDM_first}) into Eq.~(\ref{DC_RDM}), we can derive the equation for the linear conductivity as given  by the path integral method in Eq.~(\ref{DC_linear}). We can also calculate the higher-order DC conductivity in the same way. We note that using the RDM methods might be easier than the path integral methods for higher-order DC conductivities. However, in the AC case, it is hard to derive an equation equivalent to Eq.~(\ref{DC_RDM1}) so that the path integral method should be used.

\section{Correlation effects on the nonlinear response\label{Correlation}}
Finally, we use the Green's function formalism and analyze the effect of renormalization and different lifetimes in different orbitals, which were not considered in previous studies. We reveal that both effects can enhance the nonlinear conductivity. 
\subsection{Renormalization effect}
Intuitively, the renormalization effect seems to be a disadvantage for obtaining a large conductivity because it decreases the Fermi velocity. However, as the density of states might be enhanced by the renormalization at the Fermi surface, one should properly analyze how the renormalization affects the linear and the nonlinear conductivities.

First, we analyze the simple case where $\Sigma^R(\omega)\simeq \Sigma^R_0 + \alpha\omega\bm{1}$. Under this approximation, the Green's function can be written as
\begin{eqnarray}
    {G^{R}}^{-1}(\omega)&=& \omega - \m{H}_0-\Sigma^R(\omega) \simeq (1-\alpha)\omega - \m{H}'_0\nonumber\\
    &\equiv& {G'^{R}}^{-1}(Z^{-1}\omega),
\end{eqnarray}
where $Z^{-1} = 1-\alpha$, $\m{H}'_0 = \m{H}_0 + \Sigma^R_0$, and ${G'^{R}}^{-1}(\omega)=\omega-\m{H}'_0$. 
We can now analyze the effect of the renormalization on the conductivities calculated by the Green's function method.
By the variable transformation $Z^{-1}\omega\rightarrow\omega'$ and $Z^{-1}\omega_i\rightarrow\omega'_i$, the functions which appear in the linear and nonlinear conductivities change as follows:
\begin{eqnarray}
G^{R/A}(\omega)&=&G'^{R/A}(\omega'),\\
f(\omega)\simeq \theta(-\omega)&=&\theta(-\omega')\simeq f(\omega'),\label{sim1}\\
\frac{\p f(\omega)}{\p\omega}\simeq\delta(\omega)&=&Z^{-1}\delta(\omega') = \frac{\p f(\omega')}{\p\omega'},\label{sim2}\\
 \frac{\p G^R(\omega)}{\p\omega} &=& Z^{-1}\frac{\p G^R(\omega')}{\p\omega'},\\
 d\omega=Zd\omega', \ \ && \ \ \frac{1}{\omega_i} = \frac{Z^{-1}}{\omega'_i},
\end{eqnarray}
where the equality in Eqs.~(\ref{sim1}) and (\ref{sim2}) are justified at zero temperature. By inserting the above equations into Eqs.~(\ref{cond_second}) and (\ref{DC_second1}), we can derive $\sigma^{(2)} = Z^{-1}\sigma'^{(2)}$ in both the AC and the DC case, where $\sigma^\prime$ is the conductivity described by $\omega',\omega'_i,G'^{R/A}(\omega')$, which includes the energy shift by $\Sigma^R_0$. We note that we should compare the renormalized conductivity $\sigma^{(n)}(\omega;\{\omega_i\})$ with $\sigma'^{(n)}(\omega';\{\omega_i'\})=\sigma'^{(n)}(Z^{-1}\omega;\{Z^{-1}\omega_i\})$ in the AC case.
In the optical conductivity, the interband contribution becomes large when $\omega_i\simeq \epsilon_{nm}$. To focus on the same interband transition, we set the frequency $Z^{-1}\omega$ for the renormalized band. We can generalize this analysis for higher order conductivities and find
\begin{eqnarray}
    \sigma^{(n)}(\omega;\{\omega_i\}) \simeq Z^{-(n-1)}\sigma'^{(n)}(\omega';\{\omega'_i\}).\label{renorm}
\end{eqnarray}
By remembering that $Z^{-1}>1$ holds for correlated systems around the Fermi energy, we conclude that the renormalization effect enhances  the higher-order nonlinear conductivity more strongly, while it does not affect the linear conductivity. 

\begin{figure}[t]
   \includegraphics[width=0.49\linewidth]{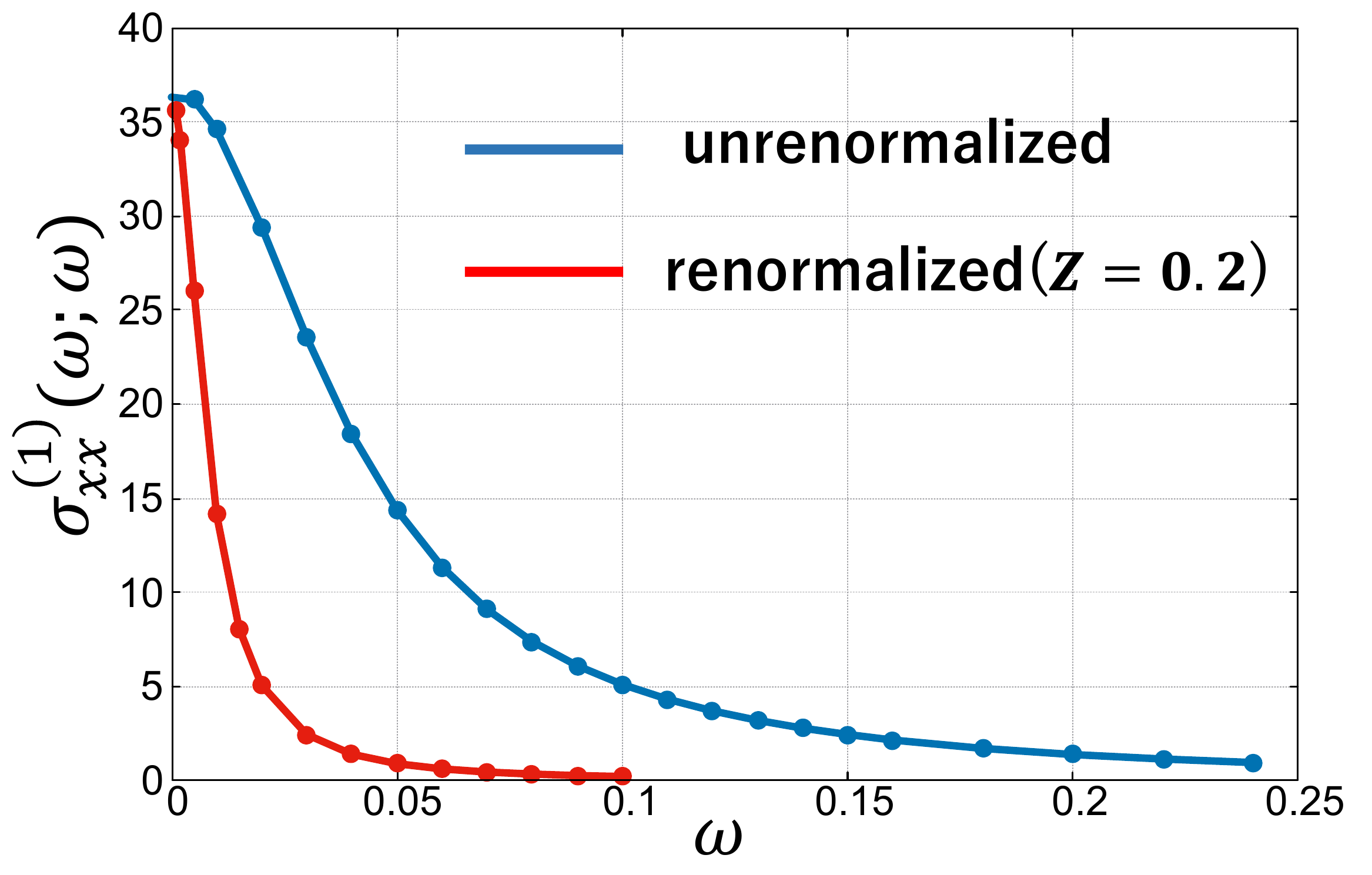}
   \includegraphics[width=0.49\linewidth]{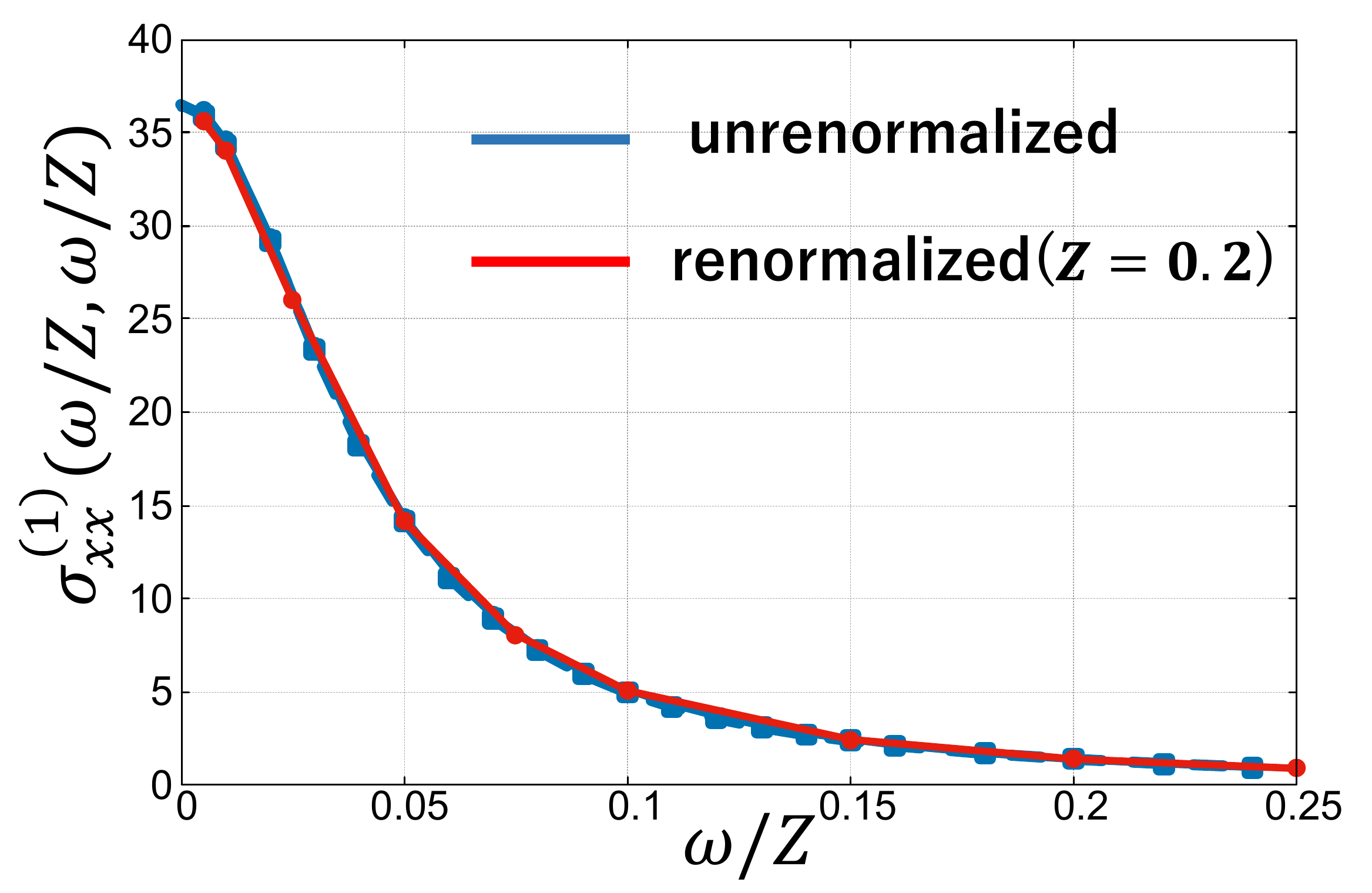}\\
   \includegraphics[width=0.49\linewidth]{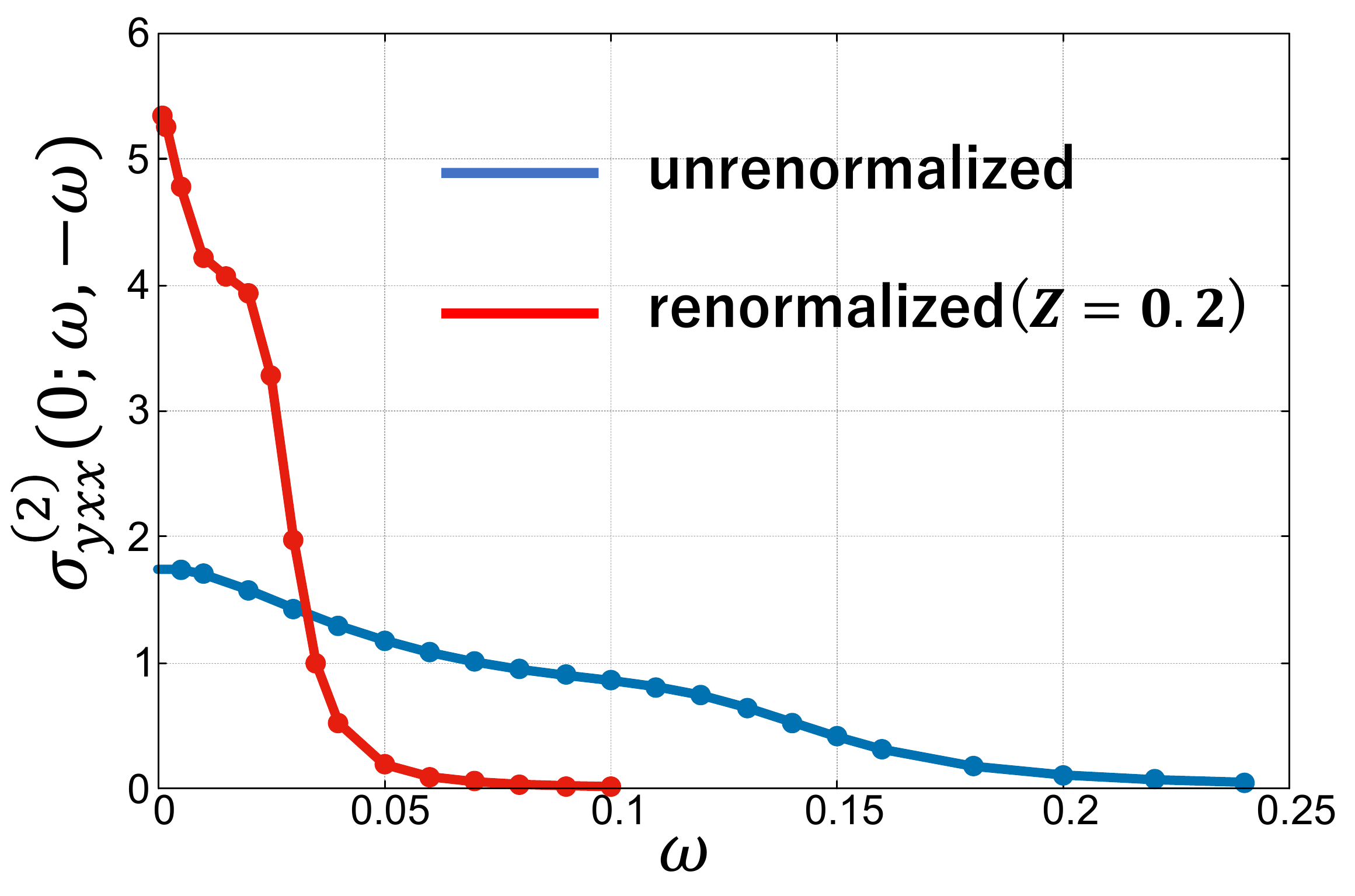}
   \includegraphics[width=0.49\linewidth]{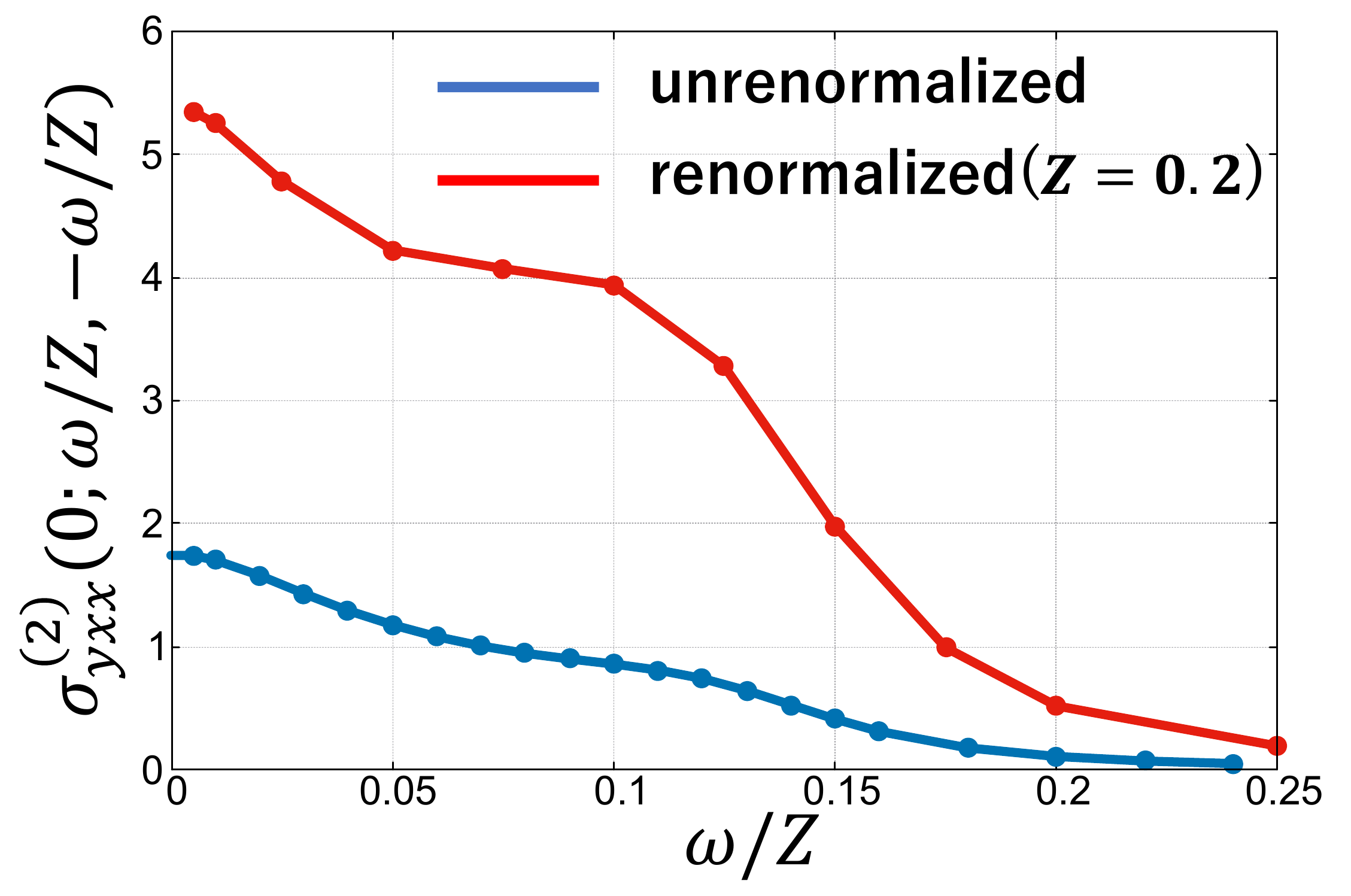}\\
  \caption{Renormalization effect on the linear and nonlinear optical conductivity for the monolayer TMD model.\\
The upper figures show the linear optical conductivity, and the lower figures show the second-order nonlinear optical conductivity(photogalvanic effect) using an  unrenormalized (blue lines) and a renormalized (red lines) band. In the right figures, we use input frequencies normalized by the renormalizion factor. The parameters are $t=0.5$, $\mu=0.7$, $p=0.7$, $\alpha_1=0.08$, $\alpha_2=0.06$, $\delta=0.7$,  $\gamma=0.05$, $T=0.02$, $Z=1$ or $0.2$. The details about how to perform the numerical calculations are written in Appendix~\ref{app:Numerical}.}\label{fig:TMD_renorm}
\end{figure}

Using the Green's function technique, we can easily confirm our general discussion above by calculating the linear and the nonlinear optical conductivity for the monolayer TMD model. 
The results for these calculations using an unrenormalized band ($Z=1$) and a renormalized band ($Z=0.2$) are shown in Fig.~\ref{fig:TMD_renorm}.
 As we derived analytically, the numerical results confirm that the nonlinear optical response is strongly enhanced by the renormalization effect, while the linear optical response is not enhanced. We note that the renormalized nonlinear optical conductivity is not as strongly enhanced as predicted ($Z^{-1}\sigma'^{(2)}(0;-\omega,\omega)$) in Fig.~\ref{fig:TMD_renorm}, which can be attributed to a finite temperature, $T=0.02$, where the Fermi-function does not correspond to the step-function.

\subsection{Different lifetimes in different orbitals\label{NHeffect}}
In this section, we analyze the effect of different lifetimes in different orbitals, which is not considered within the RTA. We note that there is the study by Kaplan $et \ al.$ \cite{PhysRevResearch.2.033100,PhysRevLett.125.227401}, where the authors analyzed the effect of different lifetimes on the nonlinear response, assuming that the conventional band-index representation is justified.

When using the RTA, the non-Hermitian (dissipation) term is described by the identity matrix. Therefore, the eigenvectors are the same as that of the Hermitian Hamiltonian. However, when different lifetimes are present in different orbitals, as in a material consisting of strongly-correlated electrons coupled to weakly-correlated electrons, the eigenvectors are distinct from the Hermitian case.
The eigenvectors are then determined by the effective non-Hermitian Hamiltonian, which describes the single-particle Green's function. In that case, the conventional band index representation breaks down, and one should use a non-Hermitian band index.
In this section, we first derive the non-Hermitian band index and then analyze its effect.

\subsubsection{Band index representation using an effective non-Hermitian Hamiltonian}
In this paper, we suppose that the effective non-Hermitian Hamiltonian can be diagonalized. We note that, in general, there are situations when a non-Hermitian Hamiltonian cannot be diagonalized, which generates novel and interesting phenomena\cite{PhysRevLett.112.203901,xu2016topological,chen2017exceptional,PhysRevLett.120.146402,PhysRevB.101.085122,PhysRevResearch.2.033018}. 
For a non-Hermitian Hamiltonian, its left eigenstates are different from its right eigenstates, while in the Hermitian case they correspond to each other by Hermitian conjugation. By describing the left and right eigenstates as $\bra{n_L}\m{H}=\epsilon_n\bra{n_L},\ \m{H}\ket{n_R}=\epsilon_n\ket{n_R}$, the following equations are satisfied:
\begin{eqnarray}
&& \braket{n_L|m_R} = \braket{n_R|m_L} = \delta_{nm},\label{Ide0}\\
&&\bm{1} = \sum_n\ket{n_R}\bra{n_L} =\sum_n\ket{n_L}\bra{n_R},\label{Ide}
\end{eqnarray}
where $\bra{n_R} = (\ket{n_R})^{\dagger}$ and $\ket{m_L} = (\bra{m_L})^{\dagger}$. We note that $\bra{n_R}\neq\bra{n_L}$ and $\braket{n_R|m_R}\neq \delta_{nm}$. 
In Eq.~(\ref{Ide0}) and (\ref{Ide}), we can construct the orthonormal basis by the left and right eigenstate, and we represent the Green's functions by the band index as
\begin{eqnarray}
&&\m{H}_{nm}(\bm{k}) \equiv \bra{n_L}\m{H}(\bm{k})\ket{m_R} = \delta_{nm} \epsilon_m,\\
 && G^R_{nm}(\omega,\bm{k}) \equiv \bra{n_L}G^R(\omega,\bm{k})\ket{m_R} = \frac{\delta_{nm}}{(\omega-\epsilon_m)},\label{GR_diag}\\
 && G^A_{nm}(\omega,\bm{k}) \equiv \bra{n_R}G^R(\omega,\bm{k})\ket{m_L} = \frac{\delta_{nm}}{(\omega-\epsilon^*_m)},\label{GA_diag}
\end{eqnarray}
where $\m{H}=\m{H}_0 + \Sigma^R$ includes the lifetime of the particles and is thus a non-Hermitian operator.
In the following, we consider the effect of non-Hermiticity on the conductivity through the non-Hermitian band-index representation.

\subsubsection{Non-Hermitian effect on the conductivity}

First, we consider the linear conductivity using the non-Hermitian band-index representation, which reads,
\begin{eqnarray}
   \m{K}^{(1)}_{\alpha\beta}(\omega_1)\!&=&\!\sum_{n,m}\Bigl\{\int_{-\infty}^{\infty} \frac{d\omega}{2\pi} \mr{Im}\Bigl[\m{J}^{nn}_{LR;\alpha\beta}G^R_n(\omega)\Bigr]f(\omega)\nonumber\\
    && \ \ + \int_{-\infty}^{\infty} \frac{d\omega}{2\pi i} \ \Bigl[\m{J}^{nm}_{LR;\alpha}G^R_m(\omega+\omega_1) \m{J}^{mn}_{LR;\beta}G^R_n(\omega)\nonumber\\
    && \ \ \ \ \ \ \ -\m{J}^{nm}_{RR;\alpha}G^R_m(\omega+\omega_1) \m{J}^{mn}_{LL;\beta}G^A_n(\omega)\nonumber\\
    && \ \ \ \ \ \ \ +\m{J}^{nm}_{RR;\alpha}G^R_m(\omega)\m{J}^{mn}_{LL;\beta} G^A_n(\omega-\omega_1)\nonumber\\
    && \ \ \ \ \ \ \ - \m{J}^{nm}_{RL;\alpha}G^A_m(\omega)\m{J}^{mn}_{RL;\beta} G^A_n(\omega-\omega_1)\Bigr]f(\omega)\Bigr\},\nonumber\\
\end{eqnarray}
where $\m{J}^{mn}_{AB;i}= \bra{m_A}\m{J}_{i}\ket{n_B}$. In the DC limit, this becomes
\begin{align}
&\sigma^{(1)}_{DC;\alpha\beta}\nonumber\\
&=2\int_{-\infty}^{\infty} \frac{d\omega}{2\pi}\Bigl(-\frac{\p f(\omega)}{\p\omega}\Bigr) \mr{Re}\Bigl[\m{J}^{nm}_{RR;\alpha}G^R_m(\omega) \m{J}^{mn}_{LL;\beta}G^A_n(\omega)\Bigr]\nonumber\\
& \ \ \ - f(\omega)\mr{Re}\Bigl[\m{J}^{nm}_{LR;\alpha}\frac{\p G^R_m(\omega)}{\p\omega} \m{J}^{mn}_{LR;\beta}G^R_n(\omega)\Bigr].\label{DC_NH_linear}
\end{align}
In the non-Hermitian band-index representation, four different types of velocity operators appear, which are $\m{J}_{LL},\m{J}_{LR},\m{J}_{RL},\m{J}_{RR}$. We note that the conventional velocity operator in the Hermitian case corresponds to $\m{J}_{LR}$ and $\m{J}_{RL}$. $\m{J}_{LL}$ and $\m{J}_{RR}$ are unique in the Fermi surface contribution to transport in a non-Hermitian system.
To compare to the conventional results, we can write $\m{J}_{LL/RR}$ by $\m{J}_{LR}$ as
\begin{eqnarray}
\m{J}^{nm}_{LL} &=& \sum_l \m{J}^{nl}_{LR} \braket{l_L|m_L}\\
\m{J}^{nm}_{RR} &=& \sum_l \m{J}^{lm}_{LR} \braket{n_R|l_R}
\end{eqnarray}
By using this relation, the Fermi surface term in Eq.~(\ref{DC_NH_linear}) can be rewritten as
\begin{eqnarray}
&&\mr{Re}\Bigl[\m{J}^{nm}_{RR;\alpha}G^R_m(\omega) \m{J}^{mn}_{LL;\beta}G^A_n(\omega)\Bigr]\nonumber\\
&&= \mr{Re}\Bigl[\braket{n_L|n_L}\braket{n_R|n_R}\m{J}^{nm}_{LR;\alpha}G^R_m(\omega) \m{J}^{mn}_{LR;\beta}G^A_n(\omega)\Bigr]\nonumber\\
&& \ \ + \mr{Re}\Bigl[\braket{l_L|n_L}\braket{n_R|l'_R}\m{J}^{lm}_{LR;\alpha}G^R_m(\omega) \m{J}^{ml'}_{LR;\beta}G^A_n(\omega)\Bigr].
\end{eqnarray}
We note that the term includes $\p f(\omega)/\p\omega$ is said as “the Fermi surface term." 
The first term is the conventional term multiplied by the factor $\gamma_{NH;n} \equiv \braket{n_L|n_L}\braket{n_R|n_R}$. We can easily show that $\gamma_{NH;n}\geq1$ is always satisfied. (See Appendix \ref{app:NH_proof}.) Therefore, we reveal that, when the system is described by a non-Hermitian Hamiltonian, with different lifetimes in different orbitals, the conventional Fermi surface term can be enhanced by the factor $\gamma_{NH;n}$.
The second term is unique in the non-Hermitian band-index representation, which describes the mixture of eigenstates in the decay dynamics. We call this term the “band-coalescent term" in this paper.
For the second-order conductivity, we perform the same analysis and find
\begin{eqnarray}
     &&\sum_{l,m,n}\gamma_{NH;n}\m{J}^{nl}_{LR;\alpha}\frac{\partial G^R_l(\omega)}{\partial\omega}\m{J}^{lm}_{LR;\beta}G^R_m(\omega)\m{J}^{mn}_{LR;\gamma}G^A_n(\omega)\nonumber\\
     && \ +\sum_{l,m,n}\sum_{k,k'(\neq n)}\braket{k'_L|n_L}\braket{n_R|k_R}\nonumber\\
     && \ \ \ \times \m{J}^{kl}_{LR;\alpha}\frac{\partial G^R_l(\omega)}{\partial\omega}\m{J}^{lm}_{LR;\beta}G^R_m(\omega)\m{J}^{mk'}_{LR;\gamma}G^A_n(\omega),\label{NH4}
\end{eqnarray}
where the first term is the conventional term with non-Hermitian factor and the second term describes the band-coalescent term for nonlinear conductivity.
Finally, we numerically check these results and how the non-Hermiticity changes the conventional terms and the band-coalescent terms by explicitly calculating the linear and nonlinear conductivity for two different models, including orbital(sublattice) dependent lifetimes. 
First, we show the results for the one-dimensional non-Hermitian Rice-Mele model, in which the dissipation depends on the sub-lattice.
A detailed explanation of the model is given in Appendix~\ref{app:Model}. Here, we note that $\Gamma$ denotes the difference of the dissipation strength at each sublattice.
The upper panels in Fig.~\ref{fig:NH_cond} show the  $\Gamma$-dependence of the linear and nonlinear DC-conductivity in the non-Hermitian Rice-Mele model.
We see that the conventional conductivity with the non-Hermitian factor is dominant for the linear conductivity, while the band-coalescent term is dominant for the nonlinear conductivity. We note that the band-coalescent term can be determined by subtracting the conventional term from the total conductivity. 

Next, we analyze the monolayer TMD model with uniaxial strain and spin-dependent scattering rates, where $\Gamma_{\uparrow/\downarrow}=\pm\Gamma$.
The lower panels in Fig.~\ref{fig:NH_cond} show that the conventional conductivity with the non-Hermitian factor is dominant for the linear conductivity, while the band-coalescent term prevails for the nonlinear conductivity. Notably,  the sign of the nonlinear conductivity changes due to the non-Hermitian effect, and the absolute value is strongly enhanced.
We note that the small spike in the conventional term of the nonlinear Hall conductivity originates from numerical errors due to exceptional points. The non-Hermitian band-index is very sensitive in parameter regions, including exceptional points, where the non-Hermitian Hamiltonian cannot be diagonalized.

Although we have analyzed the effect of different lifetimes in orbitals(sublattices) in two specific models, it seems to be clear that the non-Hermitian effect on nonlinear responses is highly model-dependent.
Our results, however, suggest that non-Hermiticity due to a difference of lifetimes in orbitals(sublattices) can strongly enhance nonlinear transport. This enhancement of nonlinear responses should also become important for correlated materials, where the self-energy depends on the orbital and atom. 

\begin{figure}[t]
   \includegraphics[width=0.49\linewidth]{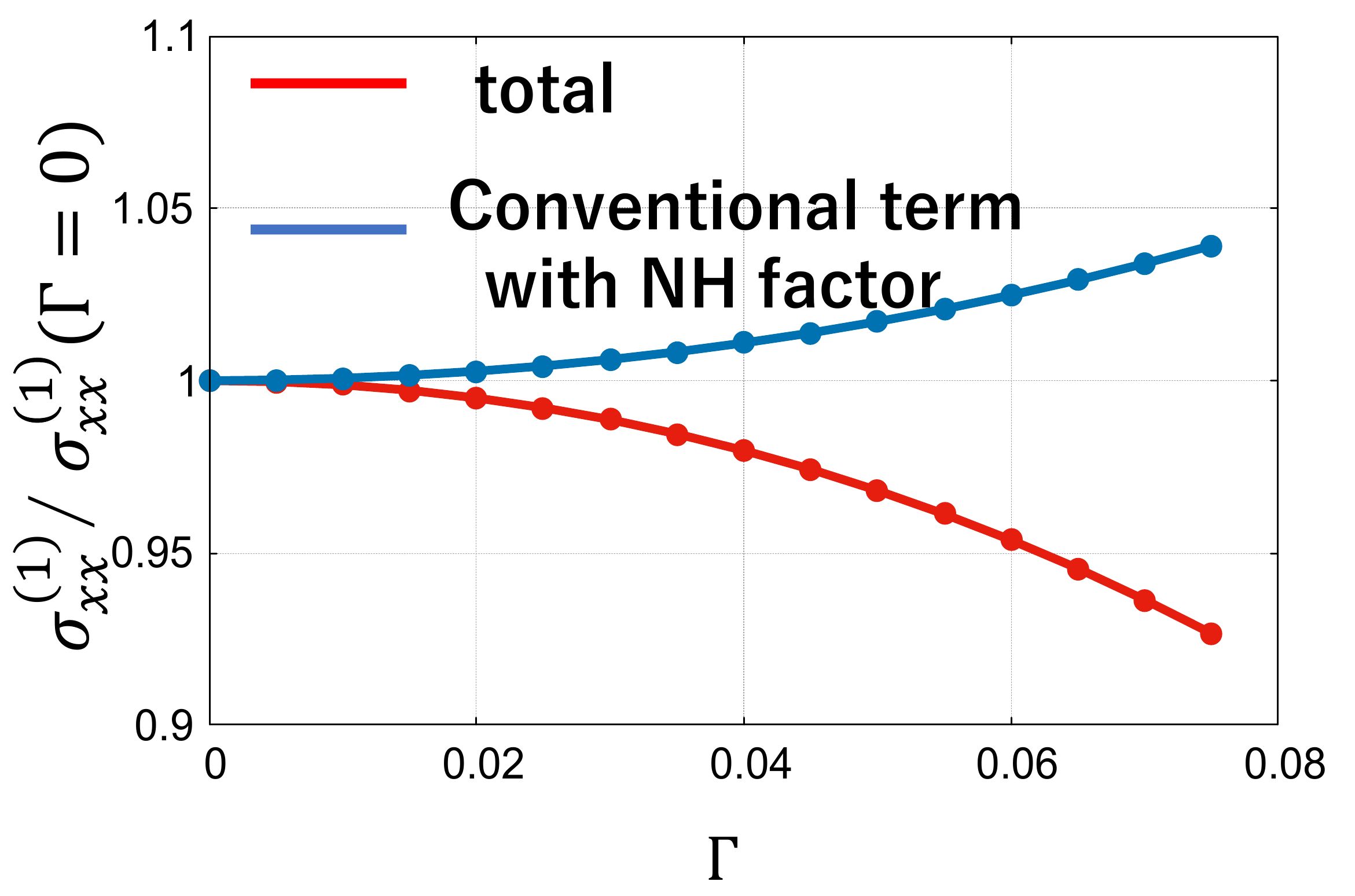}
   \includegraphics[width=0.49\linewidth]{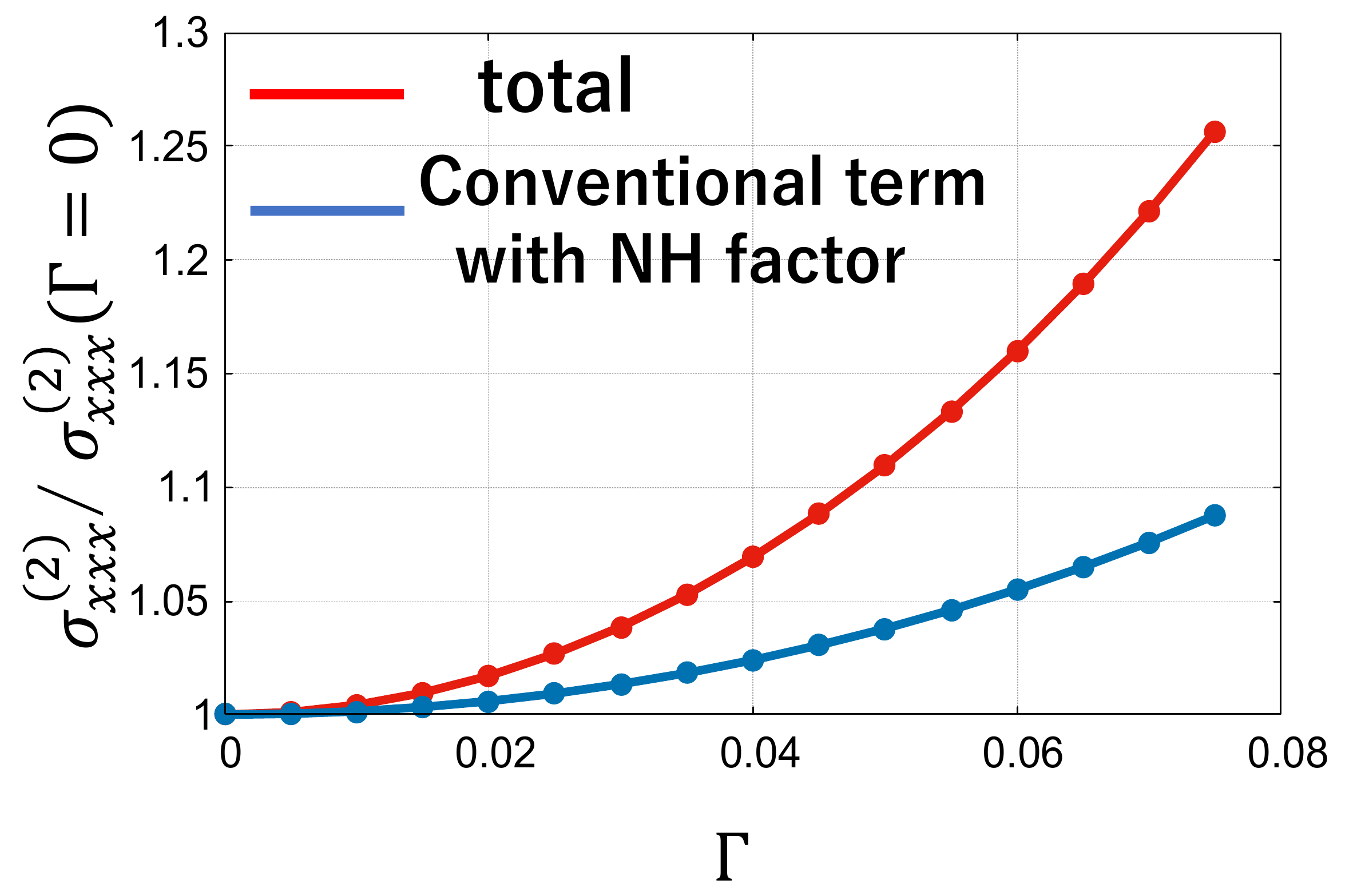}\\
   \includegraphics[width=0.49\linewidth]{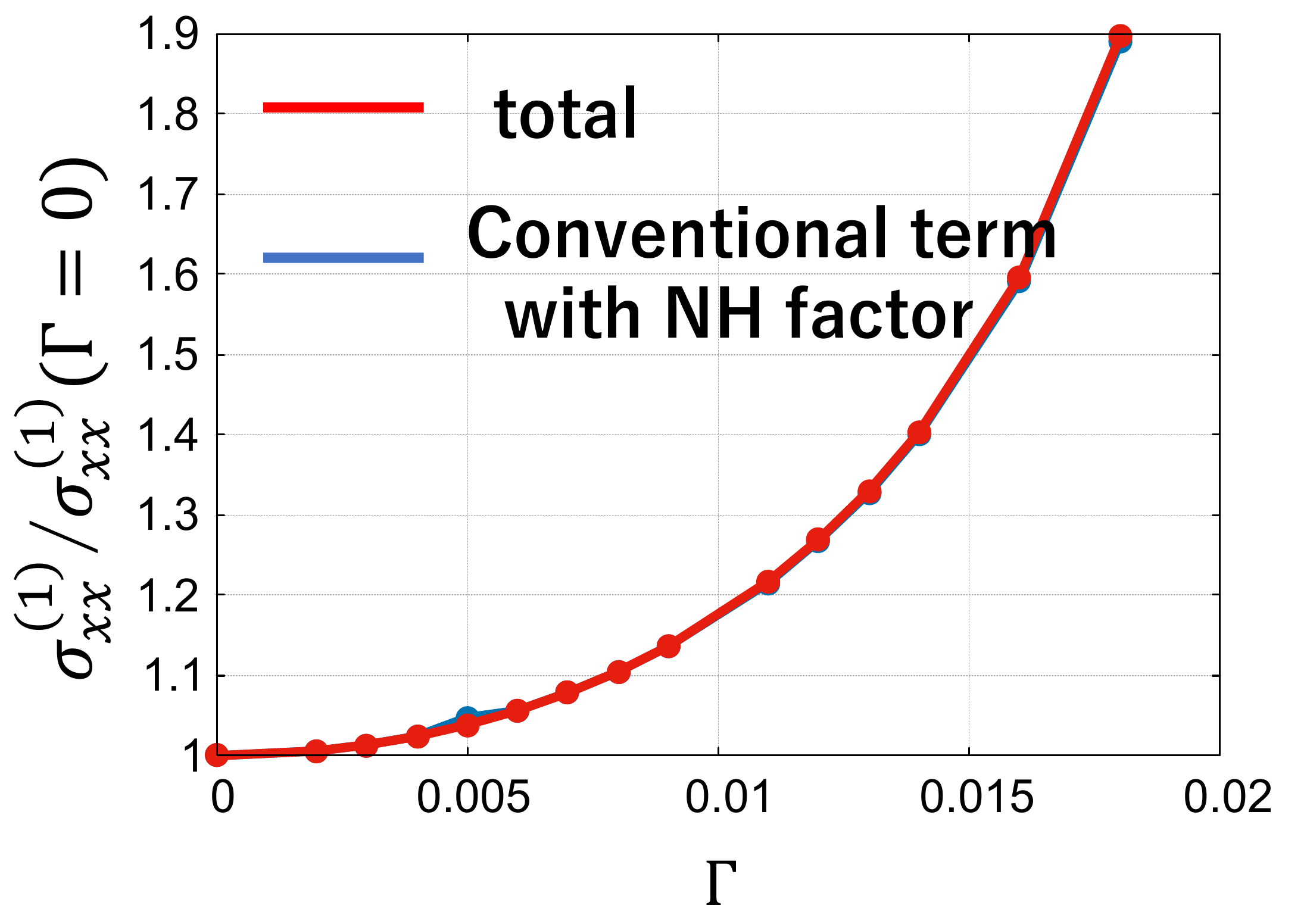}
   \includegraphics[width=0.49\linewidth]{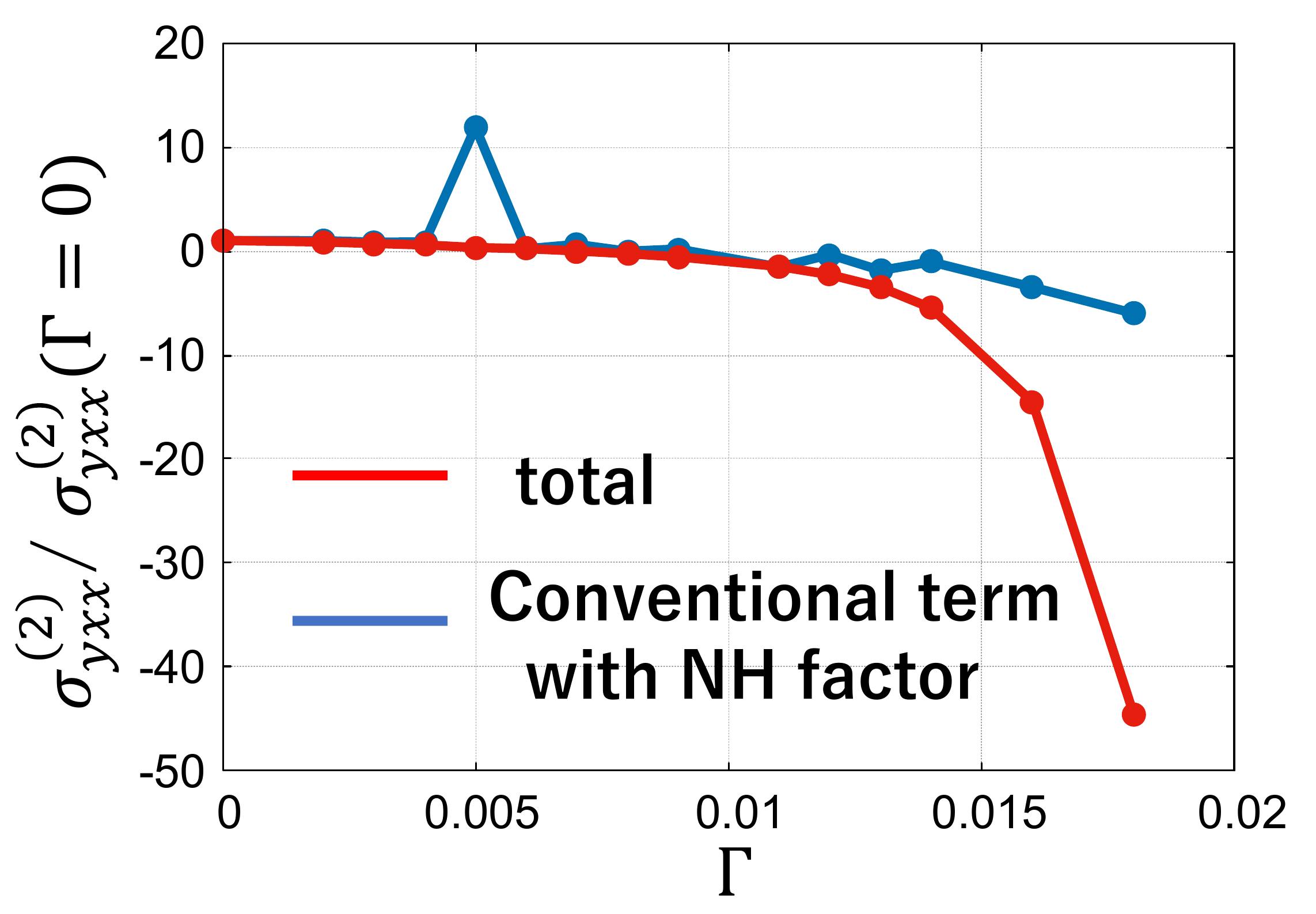}\\
  \caption{ $\Gamma$-dependence of the linear and non-linear conductivities in the non-Hermitian Rice-Mele model and the the monolayer TMD materials under uniaxial strain\label{fig:NH_cond}.\\
  The upper figures show the linear conductivity and the non-reciprocal conductivity in the 1D Rice-Mele model, and the lower figures depict the linear conductivity and the non-linear Hall conductivity in the monolayer TMD model under uniaxial strain. The blue lines represent the original terms (also appearing in the Hermitian model) now modified by the non-Hermitian factor as in Eq.~(\ref{NH4}). The red lines describe the total conductivitywhich is the sum of the conventional term with non-Hermitian factor and the band-coalescent term.. The parameters are $t=1.0$, $\delta t = 0.3$, $\Delta=0.3$, $\eta=0.05$, $T=0.02$ for the 1D Rice-Mele model, and $t=0.5$, $\mu=0.7$, $p=0.7$, $\alpha_1=0.08$, $\alpha_2=0.06$, $\delta=0.7$,  $\eta=0.05$, $T=0.02$ for the monolayer TMD model. The normalization coefficients are $\sigma^{(1)}_{xx}(\Gamma=0)=0.0801$, $\sigma^{(2)}_{xxx}(\Gamma=0)=-0.0160$ in the Rice-Mele model and $\sigma^{(1)}_{xx}(\Gamma=0)=36.21$, $\sigma^{(2)}_{yxx}(\Gamma=0)=1.417$ in the monolayer TMD model.}\label{fig:NH_cond}
\end{figure}

\section{summary and discussion\label{summary}}
In this paper, we constructed a formalism based on Green's functions to calculate the nonlinear response at finite temperature and generally analyze the impact of correlations on nonlinear response. 
By using a formalism based on Green's functions, correlations and electron scattering can be easily included via the self-energy.
 Previous studies on nonlinear response mainly focused on noninteracting systems using the semi-classical  Boltzmann equation and the reduced density matrix formalism.  In these methods, dissipation, which is necessary for the generation of a current, is introduced phenomenologically by the RTA. We reveal that the RTA is justified for nonlinear optical response only in the DC limit and in the free limit $\omega_i\gg \gamma,\epsilon_{nm}$, while the RTA seems to be a good approximation for the linear optical conductivity. We note that although Parker $et \ al.$\cite{PhysRevB.99.045121} also derived a Green's function formalism for noninteracting systems, they considered mostly photon decay $\omega_i\rightarrow\omega_i+i\gamma$  and  neglect correlations and electron scattering.

After having established the Green's function formalism, we analyze the renormalization effect and the impact of different lifetimes in a multi-orbital system as common correlation effects, which are not considered in  previous studies. We demonstrate that the enhancement generated by the renormalization effect increases with the order of the nonlinear response. When considering a single-band model, the renormalization coefficient $z(<1)$ enhances the $n$-th order response by a factor of $z^{-(n-1)}$.  Thus, the nonlinear response is more strongly increased than linear transport. 
Finally, we analyzed systems with different lifetimes, which commonly occur in materials where strongly correlated electrons couple to weakly interacting. 
The effect of different lifetimes can be analyzed by the band index of the non-Hermitian Hamiltonian. It causes the enhancement of terms that can also be derived in the Hermitian case and the emergence of a new term in which several bands coalesce. We analyzed these non-Hermitian effects on the conductivity in two specific models. In both models, the conventional term with the non-Hermitian factor is dominant for the linear conductivity, while the band-coalescent term is dominant for the nonlinear conductivity. The non-Hermitian effect can enhance the (non)linear conductivity and can even change the sign, although it depends on the model.
Although the non-Hermitian band index is not well-defined at exceptional points, where the non-Hermitian factor $\gamma_{NH}$ diverges, different lifetimes might give rise to novel transport. For example, in photonic crystals, the emergence of exceptional points induces non-reciprocal transport\cite{Regensburger2012,PhysRevLett.106.213901,Doppler2016,Choi2017}. It should be possible to observe related phenomena in correlated materials. However, these questions are left for future works.
\section*{acknowledge}
YM deeply appreciates Hikaru Watanabe, Yoichi Yanase, Shun Okumura, and Yukitoshi Motome for fruitful discussions. This work is supported by the WISE program, MEXT. Y. M. is supported by a JSPS research fellowship and by JSPS KAKENHI (Grant No. 20J12265).  R.P. is supported by JSPS, KAKENHI Grant No. JP18K03511. Computer simulations were done on the supercomputer of Tokyo University at the ISSP. 
\newpage

\appendix

\section{Derivation of the Matsubara formalism\label{app:Matsubara}}
In this section, we derive the conductivities using Green's function in Eqs.~(\ref{cond_first}) and (\ref{cond_second})
starting from Eqs.~(\ref{Res_Matsu}) and (\ref{Cond_RF}).  The first- and second-order response functions in the imaginary time are written as
\begin{eqnarray}
 &&\m{K}^{(1)}_{\alpha\beta}(\tau,\tau_1)\nonumber\\
 &&= \frac{1}{Z[A]}\frac{\delta}{\delta \m{A}^{\beta}(-i\tau_1)}\frac{\delta}{\delta \m{A}^{\alpha}(-i\tau)}Z[A]|_{A=0}\\
 &&= -< \bar{\psi}_{\mu}(\tau)\m{J}^{\mu\nu}_{\alpha}\psi_{\nu}(\tau)\bar{\psi}_{\lambda}(\tau_1)\m{J}^{\lambda\eta}_{\beta}\psi_{\eta}(\tau_1) >\nonumber\\
 && \ \ + <\bar{\psi}_{\mu}(\tau)\m{J}^{\mu\nu}_{\alpha\beta}\psi_{\nu}(\tau)>\delta(\tau\!-\!\tau_1)\label{app:C9}\\
 &&= - \delta(\tau\!-\!\tau_1)\tr{\m{J}_{\alpha\beta}\m{G}(0)}-\tr{\m{J}_{\alpha} \m{G}(\tau\!-\!\tau_1)\m{J}_{\beta} \m{G}(\tau_1\!-\!\tau)},\nonumber\\
 &&\label{app:C10}\\
&&\m{K}^{(2)}_{\alpha\beta\gamma}(\tau,\tau_1,\tau_2)\nonumber\\
&&=\frac{1}{Z[A]}\frac{\delta}{\delta \m{A}_{\gamma}(\tau_2)}\frac{\delta}{\delta \m{A}_{\beta}(\tau_1)}\frac{\delta}{\delta \m{A}_{\alpha}(\tau)}Z[\m{A}]|_{\m{A}=0}\label{app:2ndC}\\
&& = \delta(\tau\!-\!\tau_1)\delta(\tau\!-\!\tau_2)\tr{\m{J}_{\alpha\beta\gamma}\m{G}(0)}\nonumber\\
&& \ \ \ \ \ \  +\delta(\tau\!-\!\tau_1)\tr{\m{J}_{\alpha\beta}\m{G}(\tau-\tau_2)\m{J}_{\gamma}\m{G}(\tau_2-\tau)}\nonumber\\
&& \ \ \ \ \ \  +\delta(\tau\!-\!\tau_2)\tr{\m{J}_{\alpha\gamma}\m{G}(\tau-\tau_1)\m{J}_{\beta}\m{G}(\tau_1-\tau)}\nonumber\\
&& \ \ \ \ \ \  +\delta(\tau_1\!-\!\tau_2)\tr{\m{J}_{\alpha}\m{G}(\tau-\tau_1)\m{J}_{\beta\gamma}\m{G}(\tau_1-\tau)}\nonumber\\
&& \ \ \ \ \ \ +\tr{\m{J}_{\alpha} \m{G}(\tau\!-\!\tau_2)\m{J}_{\gamma} \m{G}(\tau_2\!-\!\tau_1)\m{J}_{\beta} \m{G}(\tau_1\!-\!\tau)}\nonumber\\
&& \ \ \ \ \ \ +\tr{\m{J}_{\alpha} \m{G}(\tau\!-\!\tau_1)\m{J}_{\beta} \m{G}(\tau_1\!-\!\tau_2)\m{J}_{\gamma} \m{G}(\tau_2\!-\!\tau)},\label{app:2ndCW}
\end{eqnarray}
where we used Wick's theorem to derive Eqs.~(\ref{app:C10}) and (\ref{app:2ndCW}) from Eqs.~(\ref{app:C9}) and (\ref{app:2ndC}). When calculating conductivities for correlated systems, Eqs.~(\ref{app:C10}) and (\ref{app:2ndCW}) are exact except for vertex corrections. Correlations are included via the self-energy in the single-particle Green's function in imaginary time, $\m{G}$. We note that physical quantities obtained within the length gauge correspond to those from the velocity gauge when calculating exactly\cite{PhysRevB.96.035431}. Therefore, taking the length gauge, we can derive the same results. 

After Fourier transformation, we can derive the linear and second-order nonlinear response function in the Matsubara frequency as
\begin{eqnarray}
&&\m{K}^{(1)}_{\alpha\beta}(i\omega_m;i\omega_m)\nonumber\\ 
&&=-\frac{1}{\beta}\sum_{\omega_l}\tr{\m{J}_{\alpha\beta}\m{G}(i\omega_l)+ \m{J}_{\alpha} \m{G}(i\omega_l\!+\!i\omega_m)\m{J}_{\beta} \m{G}(i\omega_l)},\nonumber\\
&&\label{app:K1}\\
&&\m{K}^{(2)}_{\alpha\beta\gamma}(i\omega_s(=i\omega_n+i\omega_m);i\omega_n,i\omega_m)\nonumber\\
&&= \frac{1}{\beta}\sum_{\omega_l}\Biggl\{\frac{1}{2}\tr{\m{J}_{\alpha\beta\gamma}\m{G}(i\omega_l)}\nonumber\\
&& \ \ \ \ + \Bigl(\tr{\m{J}_{\alpha\beta} \m{G}(i\omega_m\!+\!i\omega_l)\m{J}_{\gamma} \m{G}(i\omega_l)}\nonumber\\
&& \ \ \ \ +\frac{1}{2}\tr{ \m{J}_{\alpha} \m{G}(i\omega_n\!+\!i\omega_m\!+\!i\omega_l)\m{J}_{\beta\gamma} \m{G}(i\omega_l)}\nonumber\\
&& \ \ \ \ + \tr{\m{J}_{\alpha} \m{G}(i\omega_n\!+\!i\omega_m\!+\!i\omega_l)\m{J}_{\beta} \m{G}(i\omega_m\!+\!i\omega_l) \m{J}_{\gamma} \m{G}(i\omega_l)}\Bigr)\nonumber\\
&& \ \ \ \ \ \ \ \ \ \ \ \ \   + \Bigl((\beta,i\omega_n) \leftrightarrow )\gamma,i\omega_m)\Bigr) \Biggr\},\label{app:K2}
\end{eqnarray}
where $\omega_l = (2l+1)\pi/\beta$ are Fermionic Matsubara frequencies  and $\omega_{m} = 2m\pi/\beta,\omega_{n} = 2n\pi/\beta$ are Bosonic Matsubara frequencies, which originate from the photons.

\section{Analytic continuation of the nonlinear response function\label{app:AnalyticContinuation}}
\begin{figure*}[t]
   \includegraphics[width=0.45\linewidth]{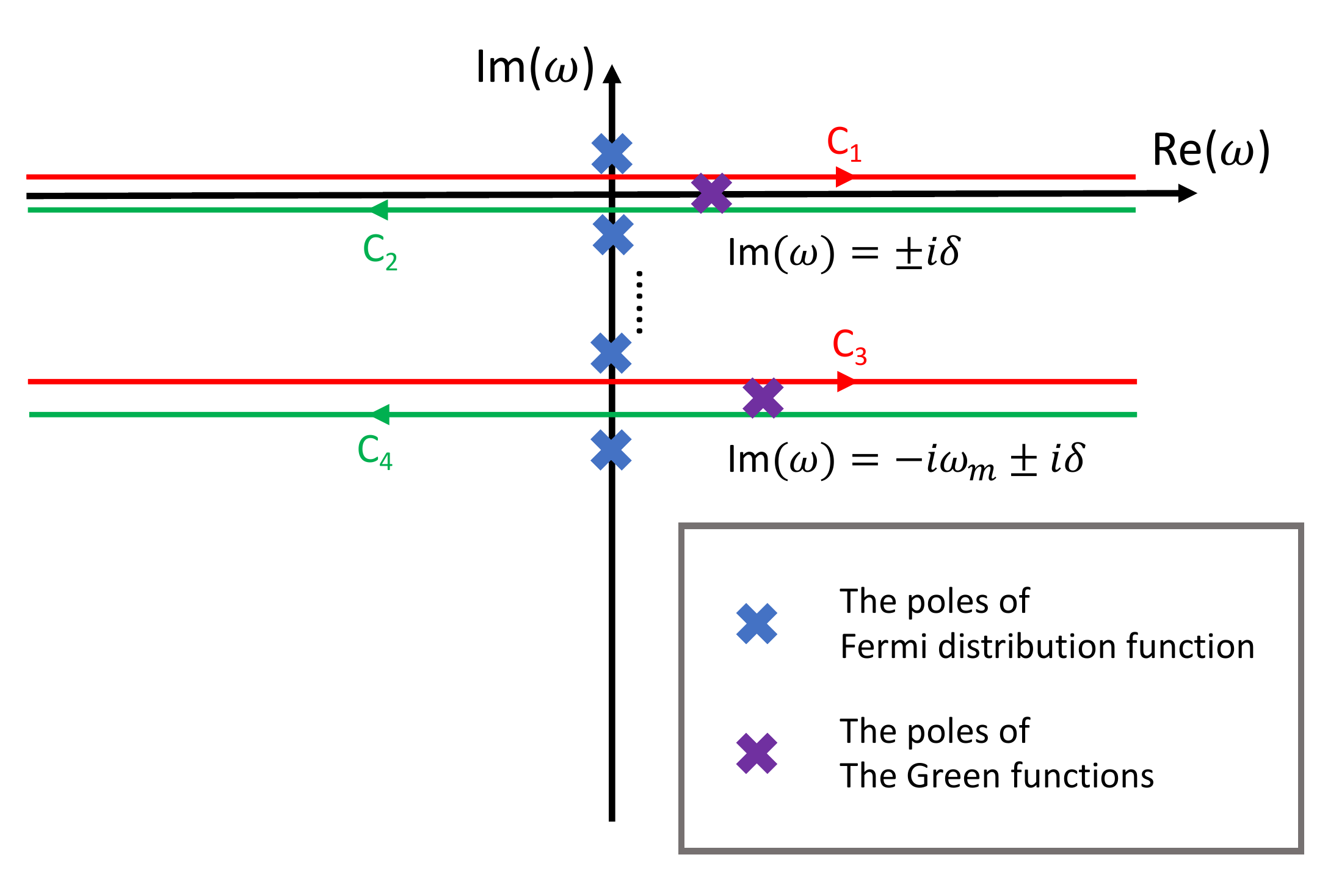}
   \includegraphics[width=0.45\linewidth]{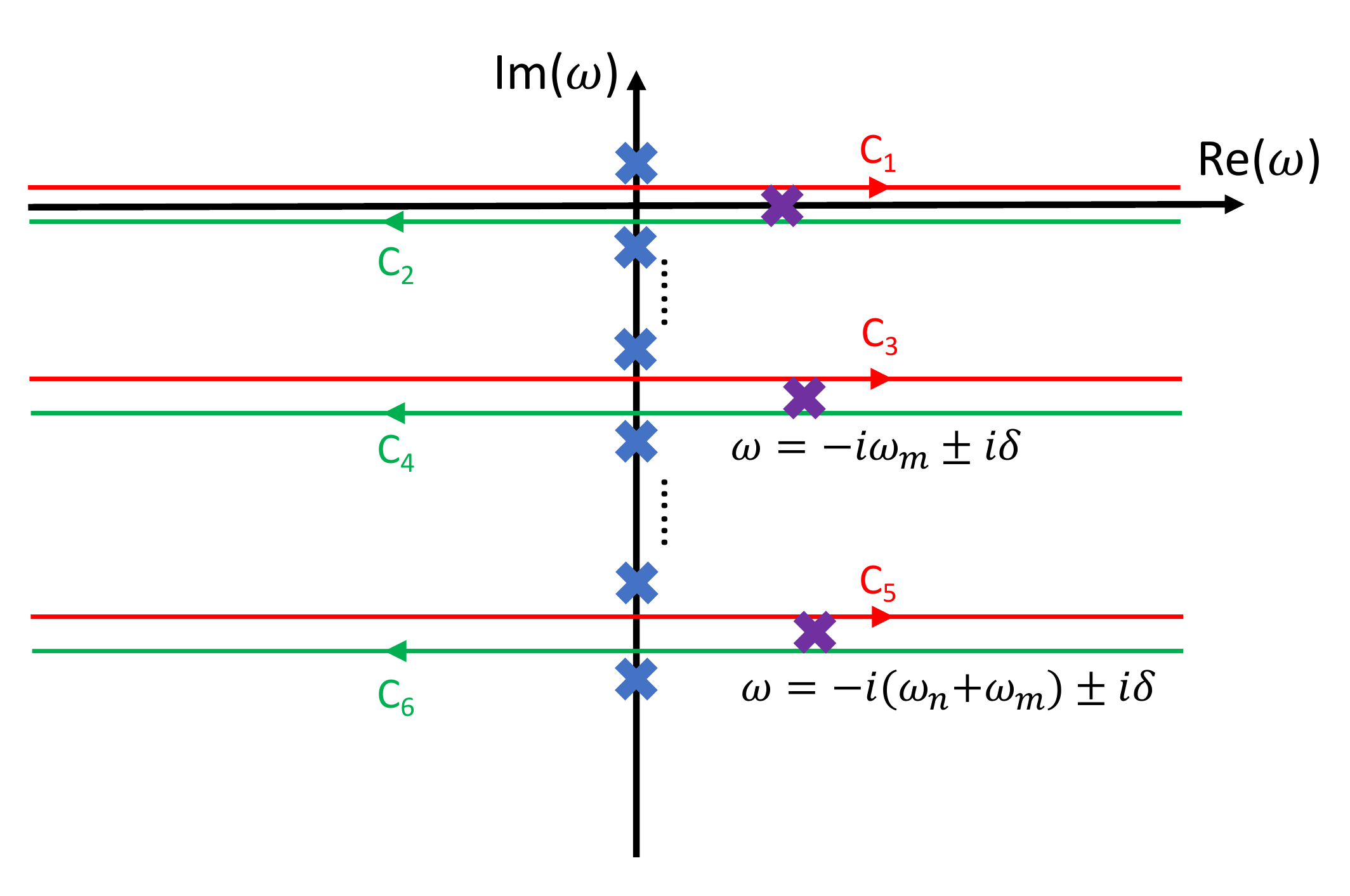}\\
  \caption{Paths in the complex $\omega$-plane for the analytic continuation of the linear and second-order nonlinear response functions.\\
 By constructing the paths, which surround the poles of the Fermi distribution function and avoid the poles of the Green's functions, we can derive Eq.~(\ref{app:K1_r}) and Eq.~(\ref{app:K2_r}) from Eq.~(\ref{app:K1}) and Eq.~(\ref{app:K2}). }
  \label{fig:FigS3}
\end{figure*}

We can calculate the (non-)linear response in real frequency by using analytic continuation.
By considering the paths in the complex frequency plane shown in Fig.~\ref{fig:FigS3}, the (non-)linear response functions can be written as
\begin{widetext}
\begin{eqnarray}
    \m{K}^{(1)}_{\alpha\beta}(i\omega_m;i\omega_m) &=& (\oint_{up+C_1}+\oint_{C_2+C_3}+\oint_{C_4+low})\frac{d\omega}{2\pi i} f(\omega)\tr{ \m{J}_{\alpha\beta}\m{G}(\omega) + \m{J}_{\alpha} \m{G}(\omega+i\omega_m)\m{J}_{\beta} \m{G}(\omega)}\label{app:K1_r}\\
    &=& \int_{-\infty}^{\infty}\frac{d\omega}{2\pi i} f(\omega)\Biggl\{\tr{\m{J}_{\alpha\beta}\Bigl(G^R(\omega)-G^A(\omega)\Bigr)}+\tr{ \m{J}_{\alpha} G^R(\omega+i\omega_m)\m{J}_{\beta} \Bigl(G^R(\omega)-G^A(\omega)\Bigr)}\nonumber\\ 
    && \ \ \ \  + \tr{ \m{J}_{\alpha} \Bigl(G^R(\omega)-G^A(\omega)\Bigr)\m{J}_{\beta} G^A(\omega-i\omega_m)}\Biggr\}\\
    \Leftrightarrow K^{(1)}_{\alpha\beta}(\omega_1;\omega_1) &=&\int_{-\infty}^{\infty}\frac{d\omega}{2\pi i} f(\omega)\Biggl\{\tr{\m{J}_{\alpha\beta}\Bigl(G^R(\omega)\!-\!G^A(\omega)\Bigr)}+ \tr{ \m{J}_{\alpha} G^R(\omega\!+\!\omega_1)\m{J}_{\beta} \bigl(G^R(\omega)\!-\!G^A(\omega)\bigr)}\\
    && \ \ \ \ \ \ \ \ \ +\tr{ \m{J}_{\alpha} \bigl(G^R(\omega)\!-\!G^A(\omega)\bigr)\m{J}_{\beta} G^A(\omega\!-\!\omega_1)}\Biggr\},\label{app:K1_r}\\
    \nonumber\\
    \m{K}^{(2)}_{\alpha\beta\gamma}(i\omega_n+i\omega_m;i\omega_n,i\omega_m) &=& -(\oint_{up+C_1}+\oint_{C_2+C_3}+\oint_{C_4+c_5}+\oint_{C_6+low})\frac{d\omega}{2\pi i} f(\omega)\Bigl\{\frac{1}{2}\tr{\m{J}_{\alpha\beta\gamma}\m{G}(i\omega_n)}\nonumber\\
    && \ \ \ \ +\frac{1}{2}\tr{\m{J}_{\alpha} \m{G}(\omega+i\omega_n+i\omega_m)\m{J}_{\beta\gamma} \m{G}(\omega)}+\tr{\m{J}_{\alpha\beta} \m{G}(\omega+i\omega_m)\m{J}_{\gamma} \m{G}(\omega)}\nonumber\\
    && \ \ \ \ +\tr{ \m{J}_{\alpha} \m{G}(\omega+i\omega_n)\m{J}_{\beta} \m{G}(\omega+i\omega_n+i\omega_m)\m{J}_{\gamma} \m{G}(\omega) + \Bigl((i\omega_n,\beta)\leftrightarrow(i\omega_m,\gamma)\Bigr)}\Bigr\}\label{app:K2_r}\\
    &=& - \int_{-\infty}^{\infty} \frac{d\omega}{2\pi i}f(\omega)\sum_{\bk}\Biggl\{\frac{1}{2}\tr{\m{J}_{\alpha\beta\gamma}\bigl(G^R(\omega)-G^A(\omega)\bigr)}\nonumber\\
&& \ \ \ \ \ \ + \tr{\m{J}_{\alpha\beta}G^R(\omega\!+\!i\omega_m)\m{J}_{\gamma}\bigl(G^R(\omega)\!-\!G^A(\omega)\bigr) +\m{J}_{\alpha\beta}\bigl(G^R(\omega)\!-\!G^A(\omega)\bigr)\m{J}_{\gamma}G^A(\omega\!-\!i\omega_m)}\nonumber\\
&& \ \ \ \ \ \ + \frac{1}{2}\tr{\m{J}_{\alpha}G^R(\omega\!+\!i\omega_{nm})\m{J}_{\beta\gamma}\bigl(G^R(\omega)\!-\!G^A(\omega)\bigr) +\m{J}_{\alpha}\bigl(G^R(\omega)\!-\!G^A(\omega)\bigr)\m{J}_{\beta\gamma}G^A(\omega\!-\!i\omega_{nm})}\nonumber\\
&& \ \ \ \ \ \ +\mr{Tr}\Bigl[\m{J}_{\alpha}G^R(\omega\!+\!i\omega_{nm})\m{J}_{\beta}G^R(\omega\!+\!i\omega_m)\m{J}_{\gamma} \bigl(G^R(\omega)\!-\!G^A(\omega)\bigr)\nonumber\\
&& \ \ \ \ \ \ \ \ \ \ \ \ \ \ + \m{J}_{\alpha}G^R(\omega\!+\!i\omega_n)\m{J}_{\beta}\bigl(G^R(\omega)\!-\!G^A(\omega)\bigr)\m{J}_{\gamma}G^A(\omega\!-\!i\omega_m) \nonumber\\
&& \ \ \ \ \ \ \ \ \ \ \ \ \ \ + \m{J}_{\alpha}\bigl(G^R(\omega)\!-\!G^A(\omega)\bigr)\m{J}_{\beta}G^A(\omega\!-\!i\omega_n)\m{J}_{\gamma}G^A(\omega\!-\!i\omega_{nm})\Bigr]\nonumber\\
&& \ \ \ \ \ +\Bigl[(\beta,\omega_1) \leftrightarrow (\gamma,\omega_2)\Bigr]\Biggr\}
\end{eqnarray}
\begin{eqnarray}
    \Leftrightarrow K^{(2)}_{\alpha\beta\gamma}(\omega_1+\omega_2;\omega_1,\omega_2) &=&-\int_{-\infty}^{\infty} \frac{d\omega}{2\pi i}f(\omega)\sum_{\bk}\Biggl\{\frac{1}{2}\tr{\m{J}_{\alpha\beta\gamma}\bigl(G^R(\omega)-G^A(\omega)\bigr)}\nonumber\\
&& \ \ \ \ \ \ \ \ + \tr{\m{J}_{\alpha\beta}G^R(\omega\!+\!\omega_2)\m{J}_{\gamma}\bigl(G^R(\omega)\!-\!G^A(\omega)\bigr) +\m{J}_{\alpha\beta}\bigl(G^R(\omega)\!-\!G^A(\omega)\bigr)\m{J}_{\gamma}G^A(\omega\!-\!\omega_2)}\nonumber\\
&& \ \ \ \ \ \ \ \ + \frac{1}{2}\tr{\m{J}_{\alpha}G^R(\omega\!+\!\omega_{12})\m{J}_{\beta\gamma}\bigl(G^R(\omega)\!-\!G^A(\omega)\bigr) +\m{J}_{\alpha}\bigl(G^R(\omega)\!-\!G^A(\omega)\bigr)\m{J}_{\beta\gamma}G^A(\omega\!-\!\omega_{12})}\nonumber\\
&& \ \ \ \ \ \ \ \ +\mr{Tr}\Bigl[\m{J}_{\alpha}G^R(\omega\!+\!\omega_{12})\m{J}_{\beta}G^R(\omega\!+\!\omega_2)\m{J}_{\gamma} \bigl(G^R(\omega)\!-\!G^A(\omega)\bigr)\nonumber\\
&& \ \ \ \ \ \ \ \ \ \ \ \ \ \ + \m{J}_{\alpha}G^R(\omega\!+\!\omega_{1})\m{J}_{\beta}\bigl(G^R(\omega)\!-\!G^A(\omega)\bigr)\m{J}_{\gamma}G^A(\omega\!-\!\omega_2) \nonumber\\
&& \ \ \ \ \ \ \ \ \ \ \ \ \ \ + \m{J}_{\alpha}\bigl(G^R(\omega)\!-\!G^A(\omega)\bigr)\m{J}_{\beta}G^A(\omega\!-\!\omega_1)\m{J}_{\gamma}G^A(\omega\!-\!\omega_{12})\Bigr]\nonumber\\
&& \ \ \ \ \ +\Bigl[(\beta,\omega_1) \leftrightarrow (\gamma,\omega_2)\Bigr]\Biggr\},
\end{eqnarray}
\end{widetext}
where $up(low)$ means the path in the complex plane surrounding the upper(lower) plane, and $f(\omega)$ is the Fermi distribution. We use the relation $\oint_C \frac{d\omega}{2\pi i}f(\omega)A(\omega)=-\frac{1}{\beta}\sum_n A(i\omega_n)$, where $\oint_C$ corresponds to the path integral only around the poles of the Fermi distribution function, while avoiding the poles of $A(\omega)$.
Using the definitions of the response functions for real frequencies 
\begin{eqnarray}
\sigma^{(1)}_{\alpha\beta}(\omega_1;\omega_1) &=& K^{(1)}_{\alpha\beta}(\omega_1;\omega_1)/i\omega_1\\
\sigma^{(2)}_{\alpha\beta\gamma}(\omega_1+\omega_2;\omega_1,\omega_2) &=& -K^{(2)}_{\alpha\beta\gamma}(\omega_1+\omega_2;\omega_1,\omega_2)/\omega_1\omega_2\nonumber,\\
\end{eqnarray}
 we can derive Eq.~(\ref{cond_first}) and (\ref{cond_second}) in the main text. 

\section{DC-limit\label{app:DClimit}}
In this section, we explicitly perform the DC-limit ($\omega_i\rightarrow 0$) and derive Eqs.~(\ref{DC_linear}) and (\ref{DC_second1}) starting from Eqs.~(\ref{cond_first}) and (\ref{cond_second}). We thereby show that performing the DC-limit under the velocity gauge does not yield any artificial divergence.

When $\omega_i$ is small enough, in the sense that $\beta\omega_i\ll1$ and $\tau\omega_i\ll1$ [$\tau$ is the inverse of the imaginary part of ${G^{R}}^{-1}(\omega)$], we can expand the single-particle Green's function as follows:
\begin{eqnarray}
    G^{a}(\omega\!+\!\omega_1) &\simeq& G^{a}(\omega) + \frac{\p G^{a}}{\p\omega}\omega_1,\label{app:Ex1}\\
    G^{a}(\omega\!+\!\omega_1\!+\!\omega_2) &\simeq& G^{a}(\omega) + \frac{\p G^{a}}{\p\omega}(\omega_1\!+\!\omega_2) + \frac{\p^2 G^{a}}{\p\omega^2}\omega_1\omega_2,\nonumber\\
    &&\label{app:Ex2}\\
    f(\omega+\omega_1)-f(\omega) &\simeq& \frac{\p f(\omega)}{\p\omega}(\omega_1),\label{app:Ex3}
\end{eqnarray}
where $ a = R ,A$ (retarded and advanced Green's function). By using this expansion, Eq.~(\ref{cond_first}) becomes
\begin{eqnarray}
    &&\sigma^{(1)}_{\alpha\beta}(\omega_1) = \frac{1}{\omega_1}\int\frac{d\omega}{2\pi}\Bigl(A_0(\omega)\!+\!A_1(\omega)\omega_1\Bigr) + \m{O}(\omega_1^2),\\
    && A_0(\omega) = \int \frac{d\bk}{(2\pi)^d}f(\omega)\tr{\m{J}_{\alpha\beta}\Bigl(G^R(\omega)-G^A(\omega)\Bigr)\nonumber\\
    && \ \ \ \ \ \ \ \ \ \ \ \ \ + \m{J}_{\alpha}G^R(\omega)\m{J}_{\beta}G^R(\omega)-\m{J}_{\alpha}G^A(\omega)\m{J}_{\beta}G^A(\omega)},\nonumber\\
    &&\label{app:D5}\\
    && A_1(\omega) = \int \frac{d\bk}{(2\pi)^d}\Bigl\{\frac{\p f(\omega)}{\p\omega}\tr{\m{J}_{\alpha}G^R(\omega)\m{J}_{\beta}G^A(\omega)}\nonumber\\
    && \ \ \ \ \ \ \ \ \ \ \ \ \ \ \ \  + f(\omega)\Bigl(\tr{\m{J}_{\alpha}\frac{\p G^R(\omega)}{\p\omega}\m{J}_{\beta}G^R(\omega)\nonumber\\
    && \ \ \ \ \ \ \ \ \ \ \ \ \ \ \ \ \ \ \ \ \ \ \ \ \ \ \ \ +\m{J}_{\alpha}G^A(\omega)\m{J}_{\beta}\frac{\p G^A(\omega)}{\p\omega} }\Bigr\}.\label{app:D6}
\end{eqnarray}
We here used
\begin{align}
&-\int\frac{d\omega}{2\pi}f(\omega)\Bigl(\m{J}_{\alpha}G^R(\omega\!+\!\omega_1)\m{J}_{\beta}G^A(\omega)\nonumber\\
& \ \ \ \ \ \ \ \ \ \ \ \ \ \ \ \ \ \ \ \ \ \ -\m{J}_{\alpha}G^R(\omega)\m{J}_{\beta}G^A(\omega\!-\!\omega_1)\Bigr)\nonumber\\
& = \int\frac{d\omega}{2\pi} \Bigl(f(\omega\!+\!\omega_1)\!-\!f(\omega)\Bigr)\m{J}_{\alpha}G^R(\omega\!+\!\omega_1)\m{J}_{\beta}G^A(\omega)
\end{align}
to derive Eq.~(\ref{app:D6}). If $A_0(\omega)$ would be finite after the integration, the conductivity diverges at $\omega_1\rightarrow0$ even when $1/\tau>0$.  However, by using the identity, $\p_{\alpha}G^{R/A}(\omega) = G^{R/A}(\omega)\m{J}_{\alpha}G^{R/A}(\omega)$, Eq.~(\ref{app:D5}) can be rewritten as
\begin{eqnarray}
    A_0(\omega) &=& f(\omega)\int \frac{d\bk}{(2\pi)^d}\p_{\beta}\Bigl\{\m{J}_{\alpha}\Bigl(G^R(\omega)-G^A(\omega)\Bigr)\Bigr\}\nonumber\\
    &=&0.
\end{eqnarray}
Therefore, $A_0(\omega)$ becomes zero at $\omega_1\rightarrow0$, an artificial divergence does not occur, and we can derive Eq.~(\ref{DC_linear}) using $G^A=(G^R)^*$.

We perform the same procedure for the second-order conductivity. By using the $\omega_i$ expansion in Eqs.~(\ref{app:Ex1}) to (\ref{app:Ex3}), Eq.~(\ref{cond_second}) becomes
\begin{eqnarray}
    &&\sigma^{(2)}_{\alpha\beta\gamma}(\omega_1+\omega_2;\omega_1,\omega_2) \nonumber\\ &&=\frac{1}{\omega_1\omega_2}\int\frac{d\omega}{2\pi}\Bigl(A_0(\omega) + (A_1(\omega)\omega_1\!+\!A'_1(\omega)\omega_2) + A_2(\omega)\omega_1\omega_2\Bigr)\nonumber\\
    && \ \ \ \ \ \ \ \ \ \ \ \ + \m{O}(\omega_1^2,\omega_2^2,\omega_i^3) 
\end{eqnarray}
\begin{eqnarray}
    &&A_0(\omega) = f(\omega)\int\frac{d\bk}{(2\pi)^d}\Bigl\{\frac{1}{2}\tr{\m{J}_{\alpha\beta\gamma}\Bigl(G^R(\omega)-G^A(\omega)\Bigr)}\nonumber\\
    && \ \ \ \ \ \ \ \ + \frac{1}{2}\tr{\m{J}_{\alpha} G^R(\omega)\m{J}_{\beta\gamma}G^R(\omega)-\m{J}_{\alpha} G^A(\omega)\m{J}_{\beta\gamma}G^A(\omega)}\nonumber\\
    && \ \ \ \ \ \ \ \ +\Bigl(\tr{\m{J}_{\alpha\beta}G^R(\omega)\m{J}_{\gamma}G^R(\omega)-\m{J}_{\alpha\beta}G^A(\omega)\m{J}_{\gamma}G^A(\omega)}\nonumber\\
    && \ \ \ \ \ \ \ \ \ \ \ + \tr{\m{J}_{\alpha}G^R(\omega)\m{J}_{\beta}G^R(\omega)\m{J}_{\gamma}G^R(\omega)}\nonumber\\
    && \ \ \ \ \ \ \ \ \ \ \  - \tr{\m{J}_{\alpha}G^A(\omega)\m{J}_{\beta}G^A(\omega)\m{J}_{\gamma}G^A(\omega)}\Bigr) + (\beta\leftrightarrow\gamma)\Bigr\}\nonumber\\
    && \ \ \ \ \ \ \ \ \ = f(\omega)\int\frac{d\bk}{(2\pi)^d}\p_{\gamma}\p_{\beta}\Bigl\{\m{J}_{\alpha}\bigl(G^R(\omega)\!-\!G^A(\omega)\bigr)\Bigr\}\nonumber\\
    && \ \ \ \ \ \ \ \ \ =0\label{app:A00}\\
    &&A_1(\omega) = f(\omega)\int\frac{d\bk}{(2\pi)^d}\p_{\beta}\tr{\m{J}_{\alpha}\frac{\p G^R(\omega)}{\p\omega}\m{J}_{\gamma}G^R(\omega)}\nonumber\\
    && \ \ \ \ \ \ \ \ \ \ + \frac{\p f(\omega)}{\p\omega}\int\frac{d\bk}{(2\pi)^d}\p_{\beta}\tr{\m{J}_{\alpha}G^R(\omega)\m{J}_{\gamma}G^A(\omega)}+c.c.\nonumber\\
    && \ \ \ \ \ \ \ \ \ =0\label{app:A11}\\
    &&A'_1(\omega) = A_1(\omega;\beta\leftrightarrow \gamma) = 0\label{app:A11p}\\
    &&A_2(\omega)\nonumber\\
    &&=\int\frac{d\bk}{(2\pi)^d}\Biggl\{\Bigl(\frac{\p f(\omega)}{\p\omega}\Bigr)\Bigl(\tr{\m{J}_{\alpha}\frac{\partial G^R(\omega)}{\partial\omega}\m{J}_{\beta}G^R(\omega)\m{J}_{\gamma}G^A(\omega)}\nonumber\\
    && \ \ \ \ \ \ \ \ \ \ \ \ + \frac{1}{2}\tr{\m{J}_{\alpha}\frac{\p G^R(\omega)}{\p\omega}\m{J}_{\beta\gamma}G^A(\omega)}\Bigr)\nonumber\\
    && \ \ \ \ \ \ \ \ \ -f(\omega)\mathrm{Im}\Bigl(\tr{\m{J}_{\alpha}\frac{\p}{\p\omega}\Bigr(\frac{\partial G^R(\omega)}{\partial\omega}\m{J}_{\beta}G^R(\omega)\Bigr)\m{J}_{\gamma}G^R(\omega)}\nonumber\\
    && \ \ \ \ \ \ \ \ \ \ \ \ + \frac{1}{2}\tr{\m{J}_{\alpha}\frac{\p^2G^R(\omega)}{\p\omega^2}\m{J}_{\beta\gamma}G^R(\omega)}\Bigr)\nonumber\\
    && \ \ \ \ \ \ \ \ + (\beta \leftrightarrow \gamma)\Biggr\}
\end{eqnarray}
In the same way as for the linear conductivity, $A_0(\omega),A_1(\omega),A'_1(\omega)$ can be written in the form of an integration over a total derivative and thus become zero. Therefore, we can determine $A_2(\omega)$ as the second-order DC-conductivity.  

\section{Diagrammatic formalism for nonlinear response at finite temperature\label{app:DiagramMethods}}

\begin{table}[t]
\begin{ruledtabular}
\begin{tabular}{ccc}
	\toprule
	Component & Diagram & Value\\[2em]
	\hline\\
\begin{minipage}[h]{2.2cm}
	(Classical) \\
	Photon\\
	Propagator
\end{minipage}&
\begin{tikzpicture}[baseline=(a.center)]
	\node (a) {\includegraphics[width=0.3\linewidth]{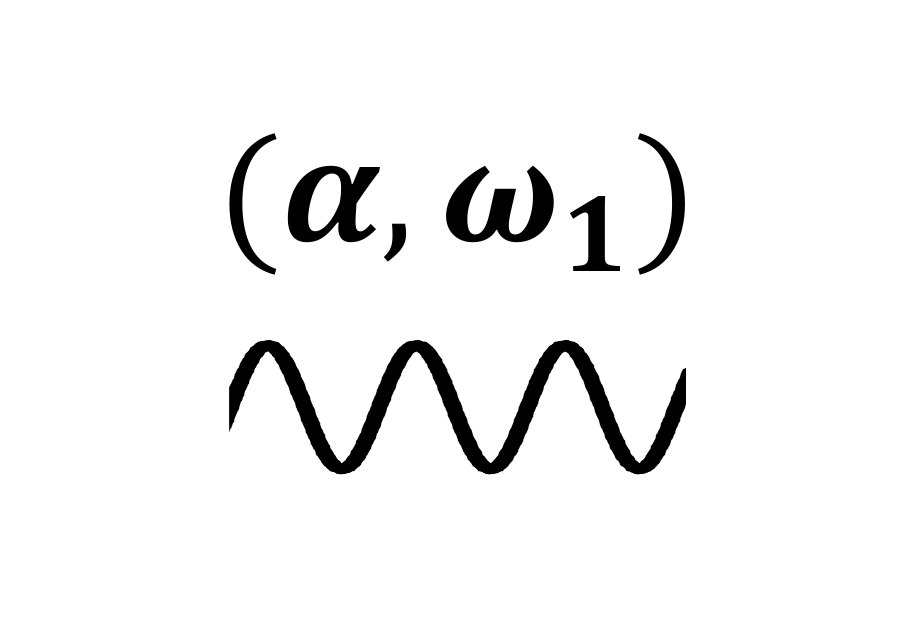}};
\end{tikzpicture}
&
1\\[2em]
\begin{minipage}[h]{2.2cm}
	Electron\\
	Propagator\\
	(Retarded)
\end{minipage}&
\begin{tikzpicture}[baseline=(a.center)]
	\node (a) {\includegraphics[width=0.3\linewidth]{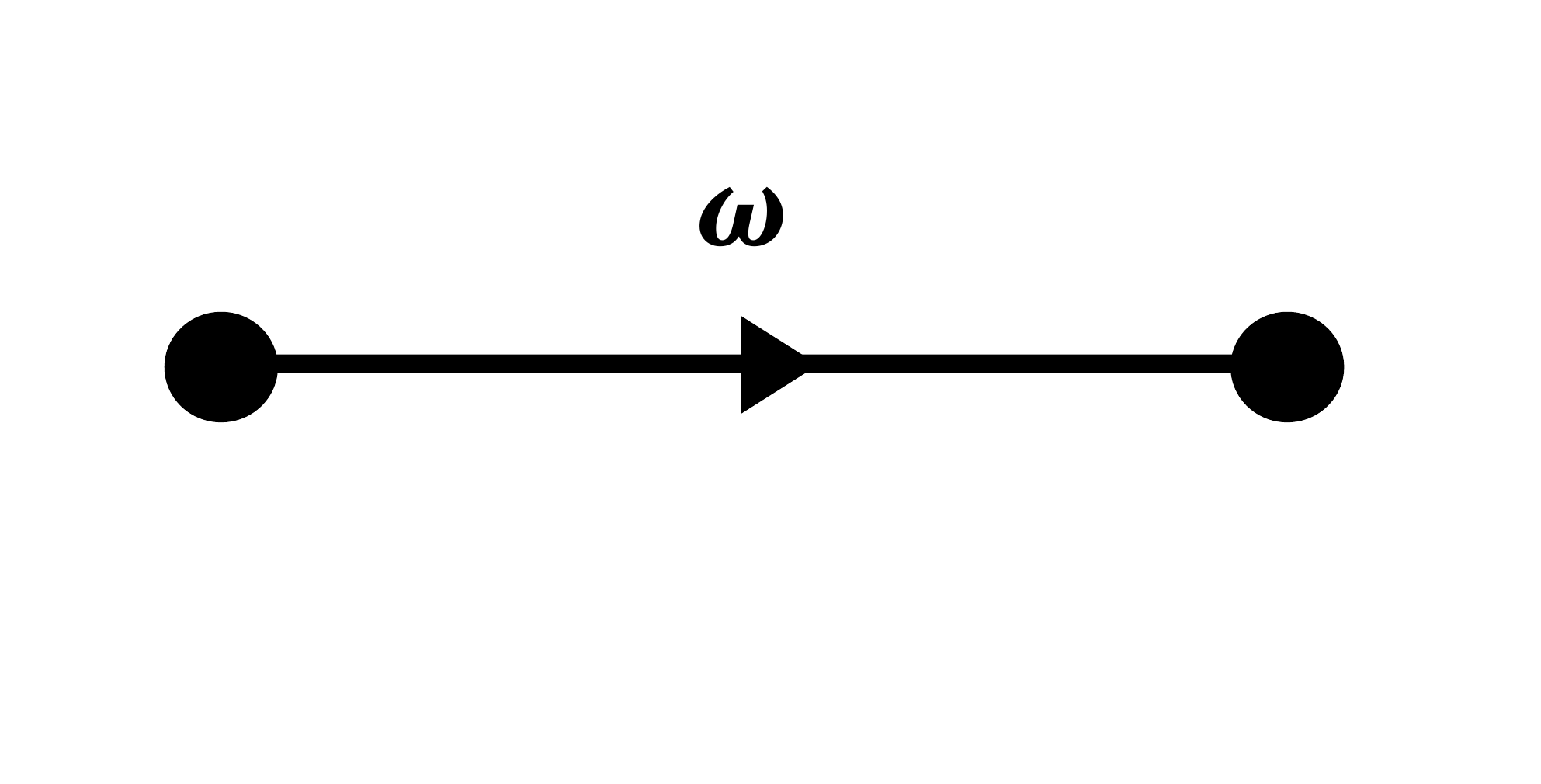}};
\end{tikzpicture}
&
$\displaystyle G^R(\omega)$\\[2em]
\begin{minipage}[h]{2.2cm}
	Electron\\
	Propagator\\
	(Advanced)
\end{minipage}&
\begin{tikzpicture}[baseline=(a.center)]
	\node (a) {\includegraphics[width=0.3\linewidth]{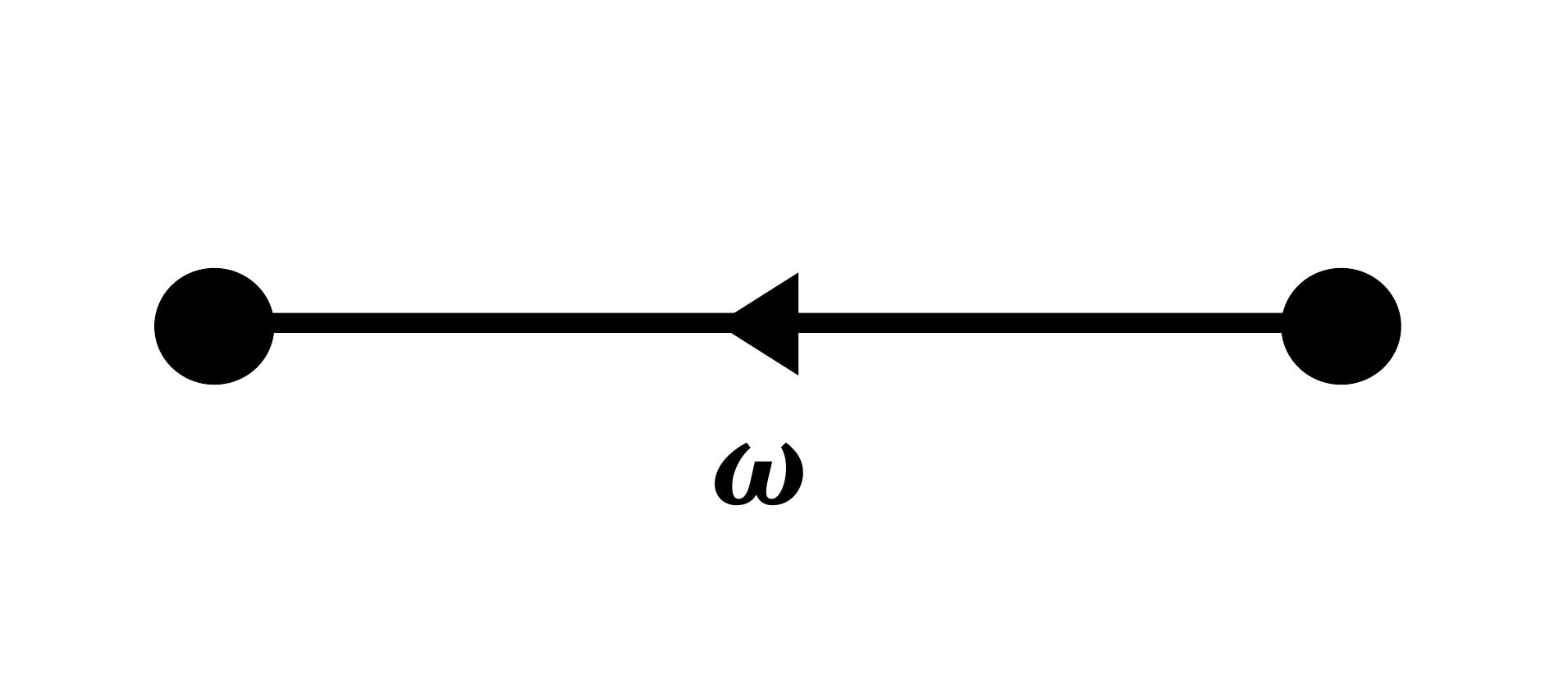}};
\end{tikzpicture}
&
$\displaystyle G^A(\omega)$\\[2em]
\begin{minipage}[h]{2.2cm}
	Distribution\\
	Function
\end{minipage}&
\begin{tikzpicture}[baseline=(a.center)]
	\node (a) {\includegraphics[width=0.3\linewidth]{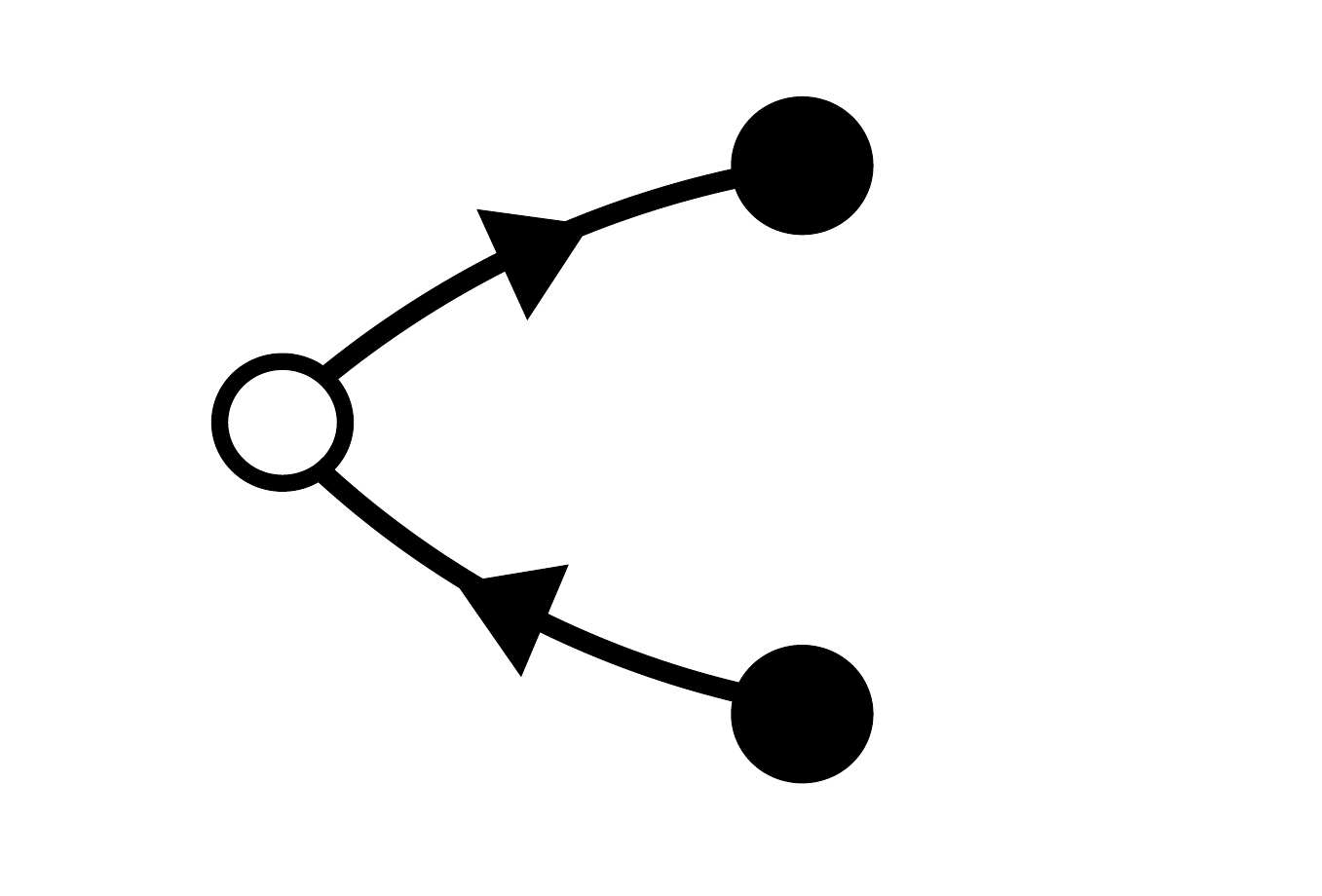}};
\end{tikzpicture}
&
$\Bigl(G^R(\omega)-G^A(\omega)\Bigr)f(\omega)$\\[2em]
\begin{minipage}[h]{2.2cm}
	One-Photon\\
	Input Vertex
\end{minipage}&
\begin{tikzpicture}[baseline=(a.center)]
	\node (a) {\includegraphics[width=0.3\linewidth]{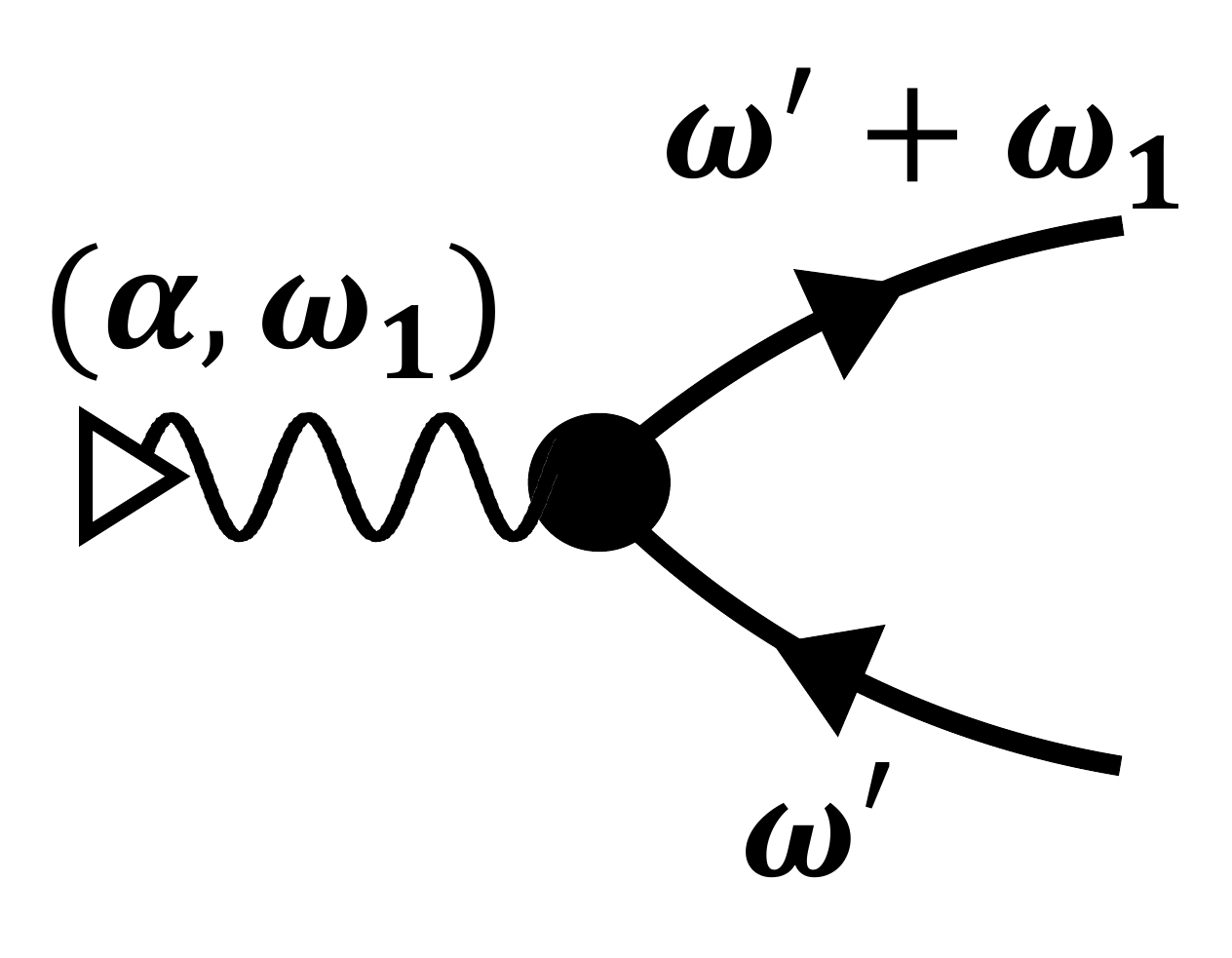}};
\end{tikzpicture}
&
${\displaystyle \frac{1}{i\omega_1}\m{J}_{\alpha}}$\\[2em]
\begin{minipage}[h]{2.2cm}
	Two-Photon\\
	Input Vertex
\end{minipage}&
\begin{tikzpicture}[baseline=(a.center)]
	\node (a) {\includegraphics[width=0.3\linewidth]{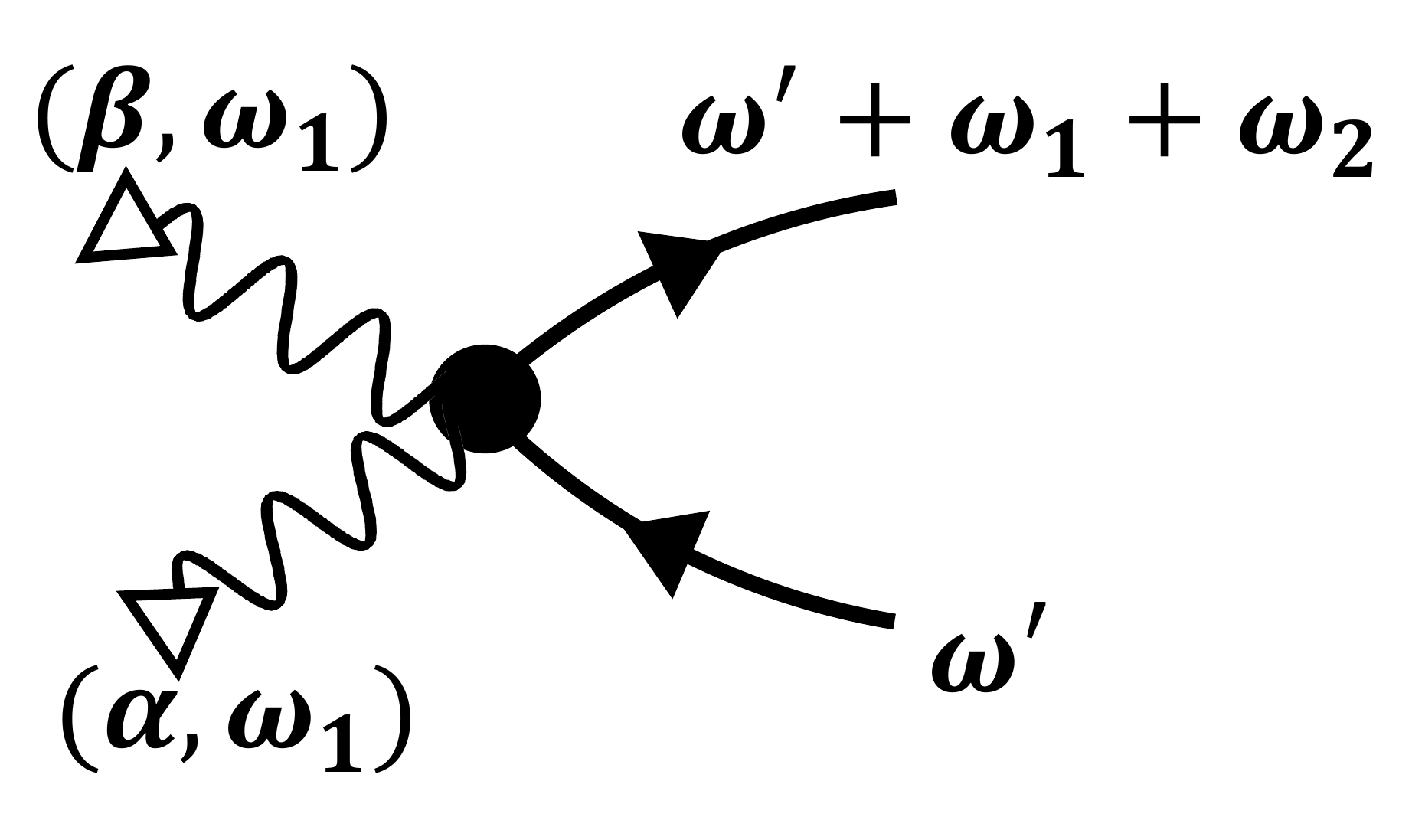}};
\end{tikzpicture}
&
$\frac{1}{i\omega_1}\frac{1}{i\omega_2}\m{J}_{\alpha\beta}$\\[2em]
\begin{minipage}[h]{2.2cm}
	One-Photon\\
	Output Vertex
\end{minipage}&
\begin{tikzpicture}[baseline=(a.center)]
	\node (a) {\includegraphics[width=0.3\linewidth]{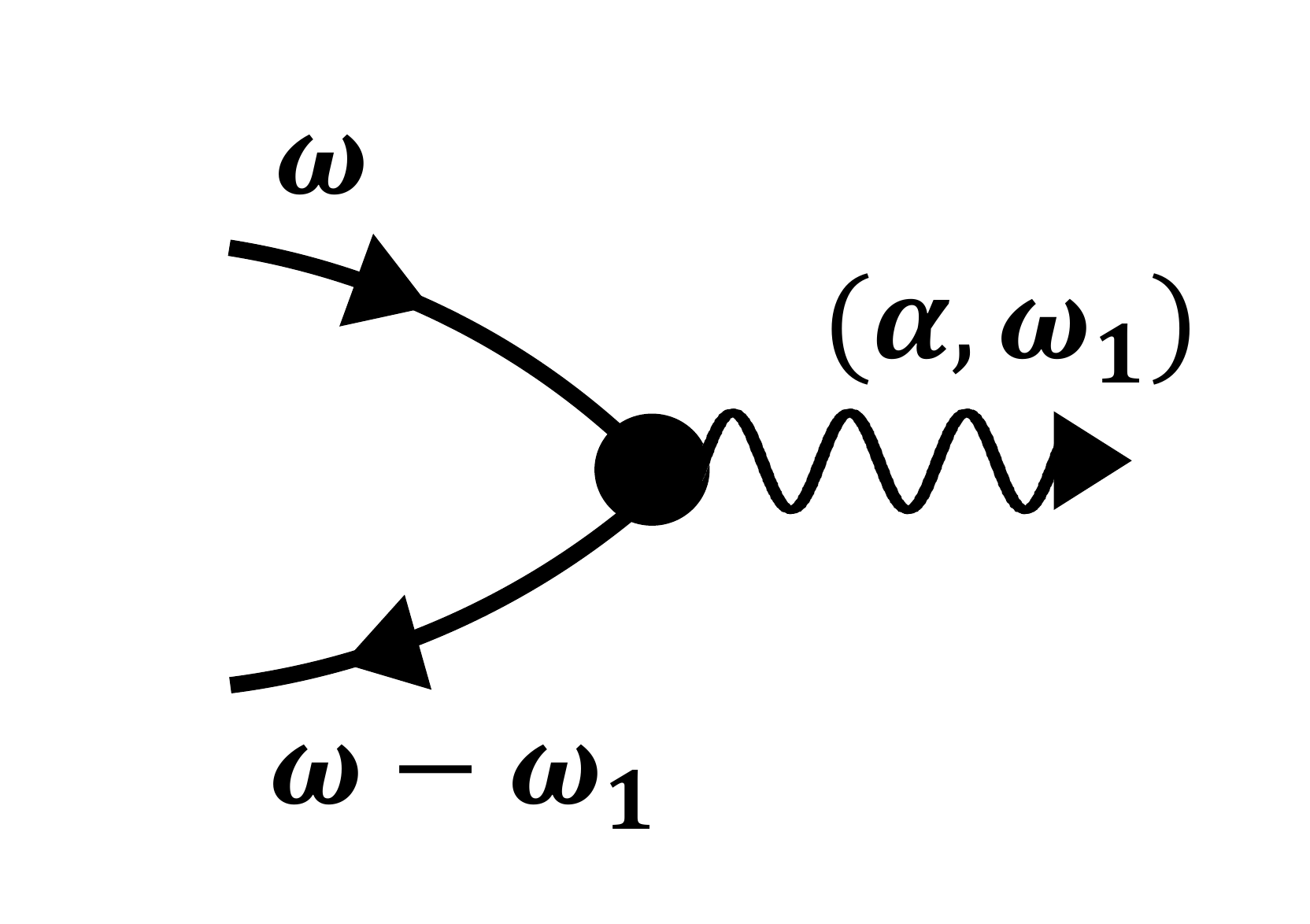}};
\end{tikzpicture}
&
$\m{J}_{\alpha}$\\
\end{tabular}
\end{ruledtabular}
\caption{Objects to construct Feynman diagrams for nonlinear electromagnetic perturbations in a crystal at finite temperature --
A new vertex with $N$ incoming photons will appear in a diagram of the $N$-th order response function. The input vertex can appear with any number of photons with a coefficient $(-i\omega_i)^{-1}$ for each 
photon. The right(left)-handed arrows represent  retarded(advanced) Green's functions.  The direction of the arrow changes at the distribution function object and the output vertex, but not at the input vertices. We note that the input can occur at the same place as the output, such as in the first term of Eqs.~(\ref{app:Diag1}) and (\ref{app:Diag2}). In that case, the value for the $n$-th order  vertex becomes $\prod_i^{n-1} \frac{1}{i\omega_i} \m{J}_{\alpha_1\dots \alpha_n}$.} 
\label{tab:feynman_rules}
\end{table}

Parker $et \ al.$ introduced a diagrammatic method for nonlinear responses in \cite{PhysRevB.99.045121}, and  Jo$\tilde{\mr{a}}$o {\it et al.}\cite{Jo_o_2019} introduced a diagrammatic method using Keldysh Green's functions. In this section, with the results from the previous section in mind, we construct an extension to this diagrammatic method for finite temperatures using real frequencies, which is summarized in Table~\ref{tab:feynman_rules}.
Each diagram for the $N$-th order response function includes $N$ incoming photons and one vertex for an outgoing photon.
For each incoming photon a coefficient $(i\omega_i)^{-1}$ is multiplied. The frequencies of the input vertices need to sum up to the output frequency.  Furthermore, each diagram includes one object corresponding to the distribution function. Finally, retarded and advanced Green's functions are used to connect all vertices in a single loop.
The difference of our results to the results by Parker $et~  al.$\cite{PhysRevB.99.045121} is the presence of the distribution function and the distinction between the retarded and advanced Green's functions. 
For calculating the $N$-th order response, we construct all distinct diagrams using these rules.
 We then can easily evaluate the diagrams tracing the objects anticlockwise starting from the output vertex.

For example, the linear optical conductivity can be described using diagrams as
\begin{align}
&\sigma^{(1)}_{\alpha\beta}(\omega;\omega_1)\nonumber\\
\nonumber
	&=\begin{tikzpicture}[baseline=(a.center)]
		\node (a) {\includegraphics[width=2.3cm]{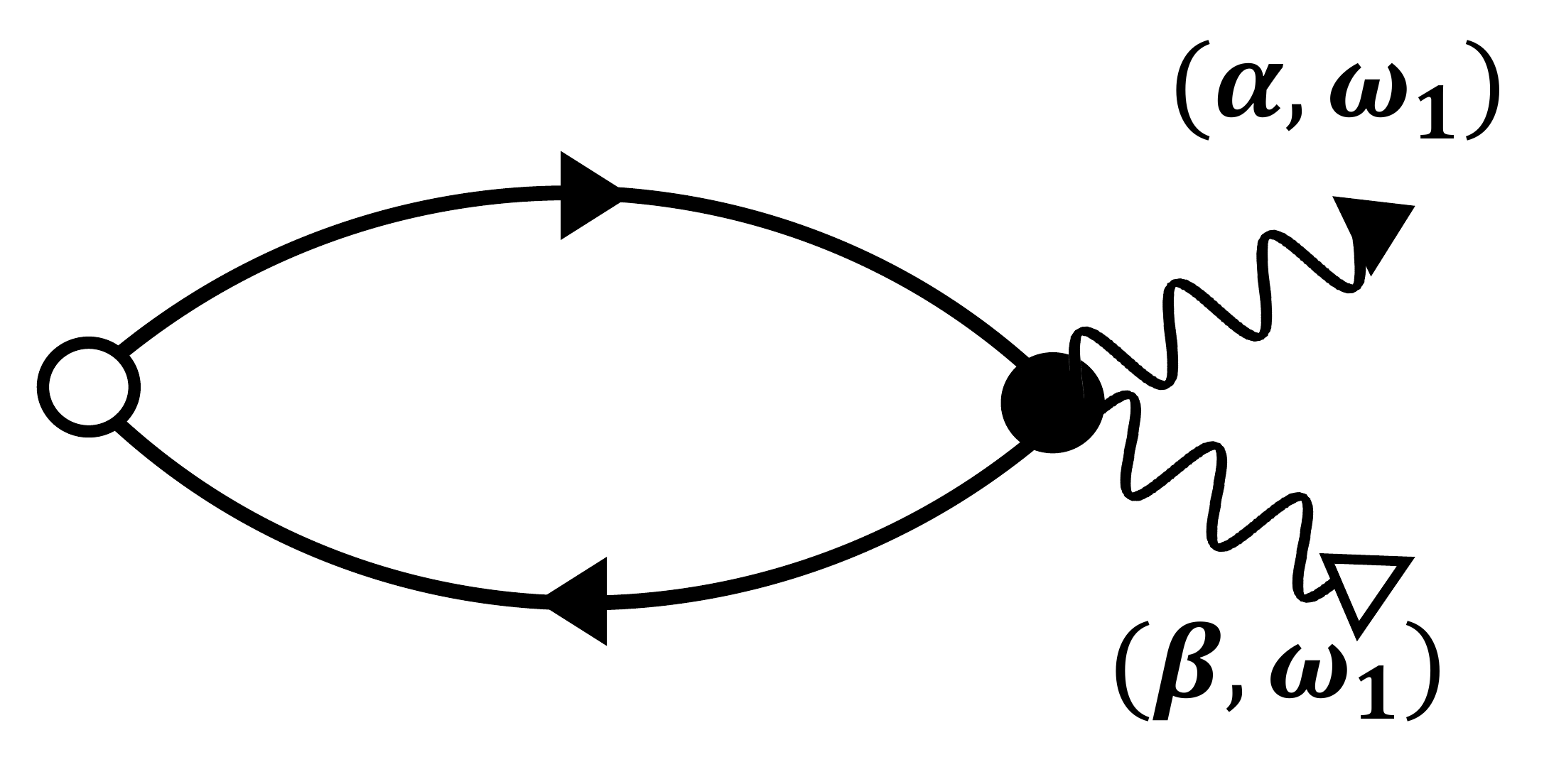}};
	\end{tikzpicture}
	+ 
	\begin{tikzpicture}[baseline=(a.center)]
		\node (a) {\includegraphics[width=2.3cm]{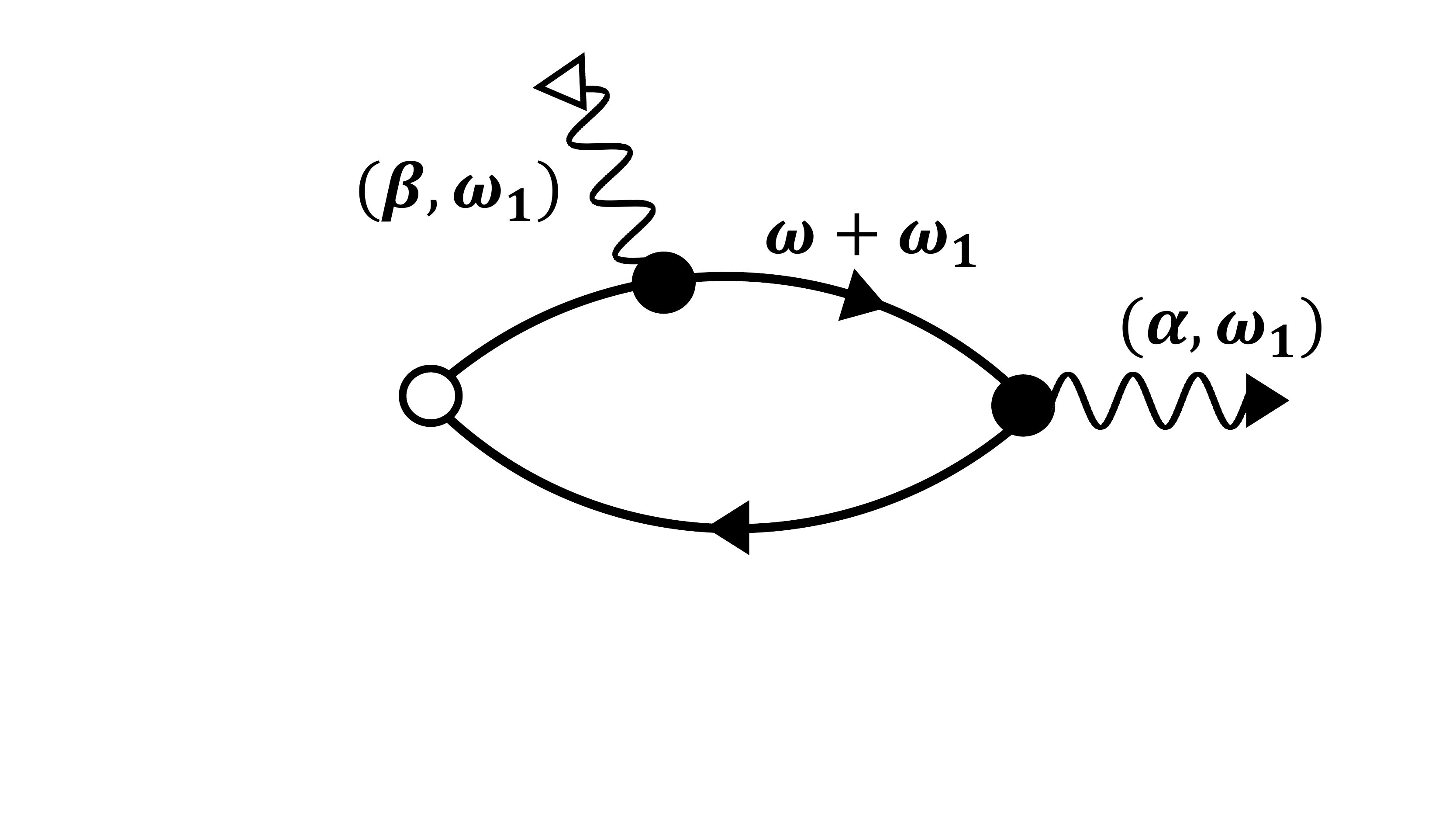}};
    \end{tikzpicture}
    + 
	\begin{tikzpicture}[baseline=(a.center)]
		\node (a) {\includegraphics[width=2.3cm]{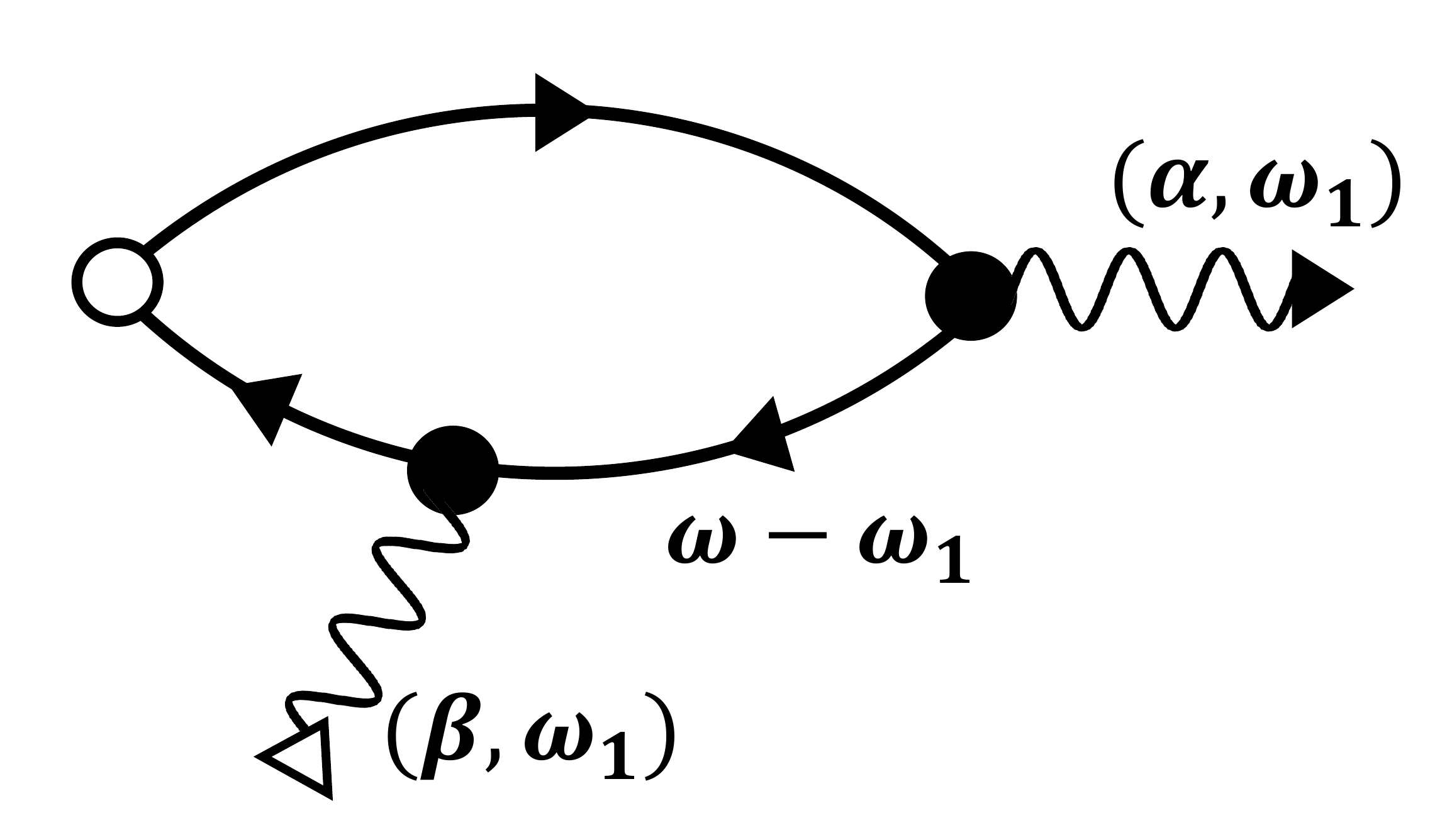}};
    \end{tikzpicture}
\end{align}
\begin{align}
&=-\frac{1}{\omega_1}\int_{-\infty}^{\infty} \frac{d\omega}{2\pi}f(\omega)\sum_{\bk}\Biggl\{\tr{ \m{J}_{\alpha\beta}(\bk) \bigl(G^R(\omega,\bk)\!-\!G^A(\omega,\bk)\bigr)}\nonumber\\
 & \ \ \ + \tr{\m{J}_{\alpha}(\bk)G^R(\omega\!+\!\omega_1,\bk)\m{J}_{\beta}(\bk) \Bigl(G^R(\omega,\bk)\!-\! G^A(\omega,\bk)\Bigr)\nonumber\\
 & \ \ \ +\m{J}_{\alpha}(\bk)\Bigl(G^R(\omega,\bk)\!-\! G^A(\omega,\bk)\Bigr)\m{J}_{\beta}(\bk) G^A(\omega\!-\!\omega_1,\bk)}\Biggr\},\label{app:Diag1}
\end{align}

\ 
The diagrams for the second-order optical conductivity are given as
\begin{align}
    &\sigma^{(2)}_{\alpha\beta\gamma}(\omega;\omega_1,\omega_2)\nonumber\\
\nonumber
	&=\begin{tikzpicture}[baseline=(a.center)]
		\node (a) {\includegraphics[width=2.3cm]{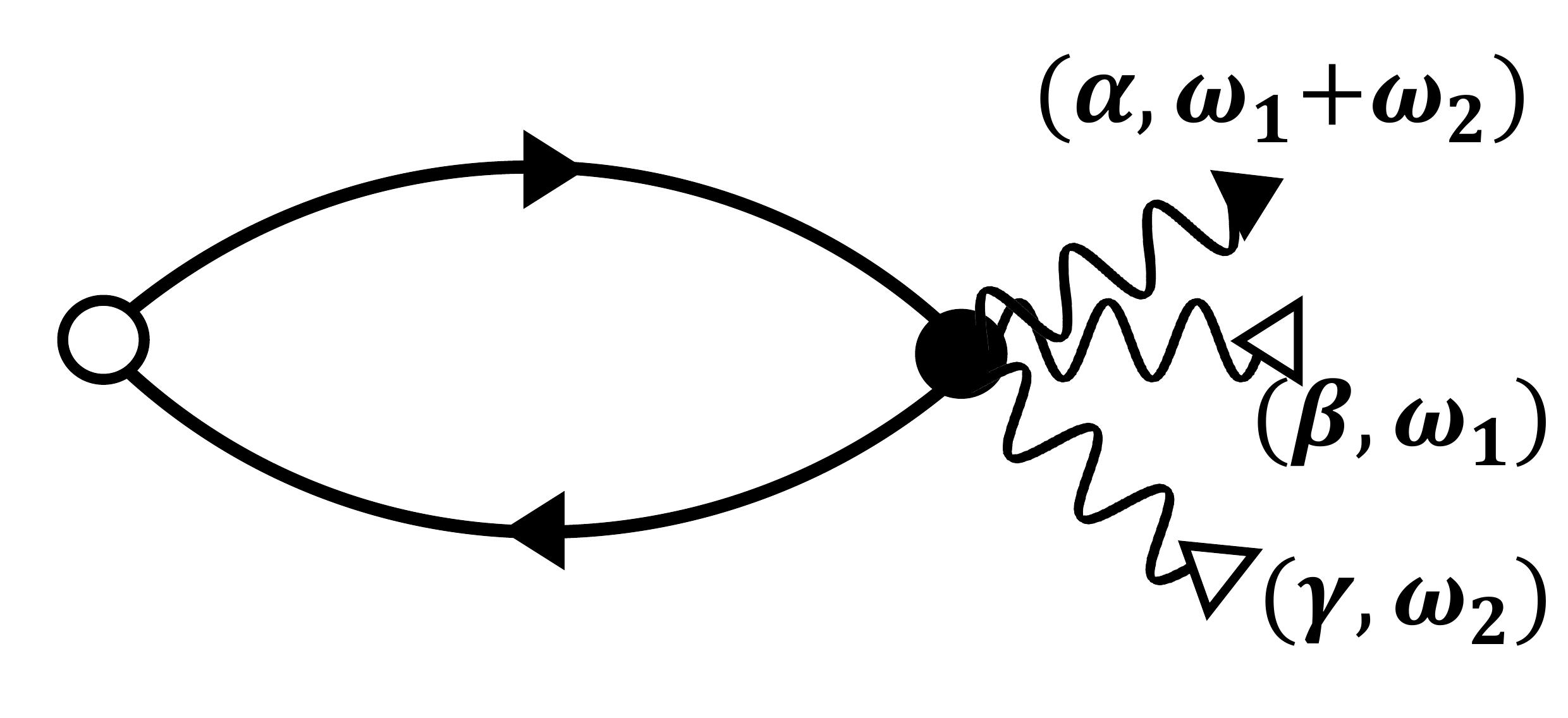}};
	\end{tikzpicture}\nonumber\\
	& \ \ \ \ + 
	\begin{tikzpicture}[baseline=(a.center)]
		\node (a) {\includegraphics[width=2.3cm]{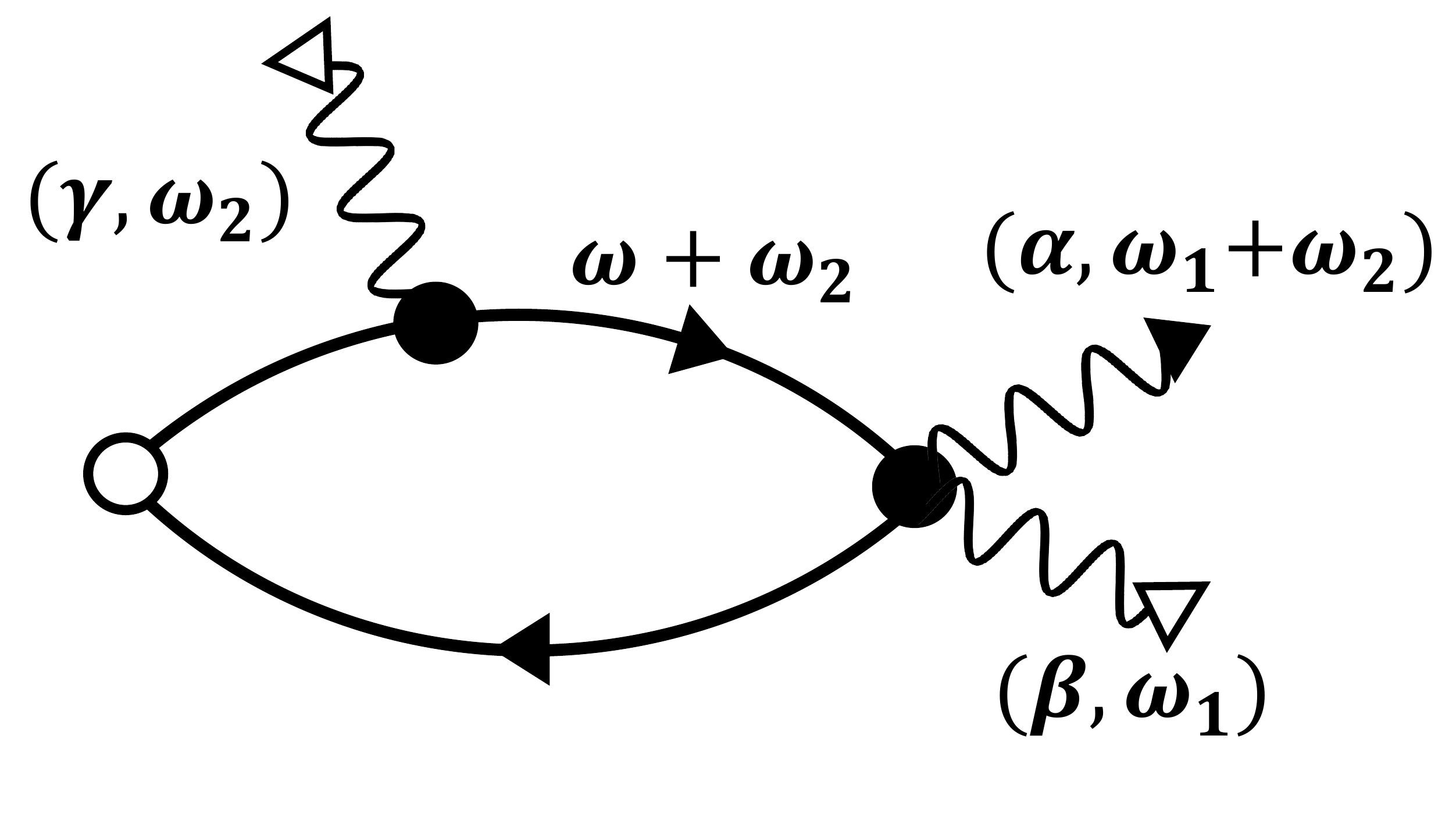}};
    \end{tikzpicture}
    + 
	\begin{tikzpicture}[baseline=(a.center)]
		\node (a) {\includegraphics[width=2.3cm]{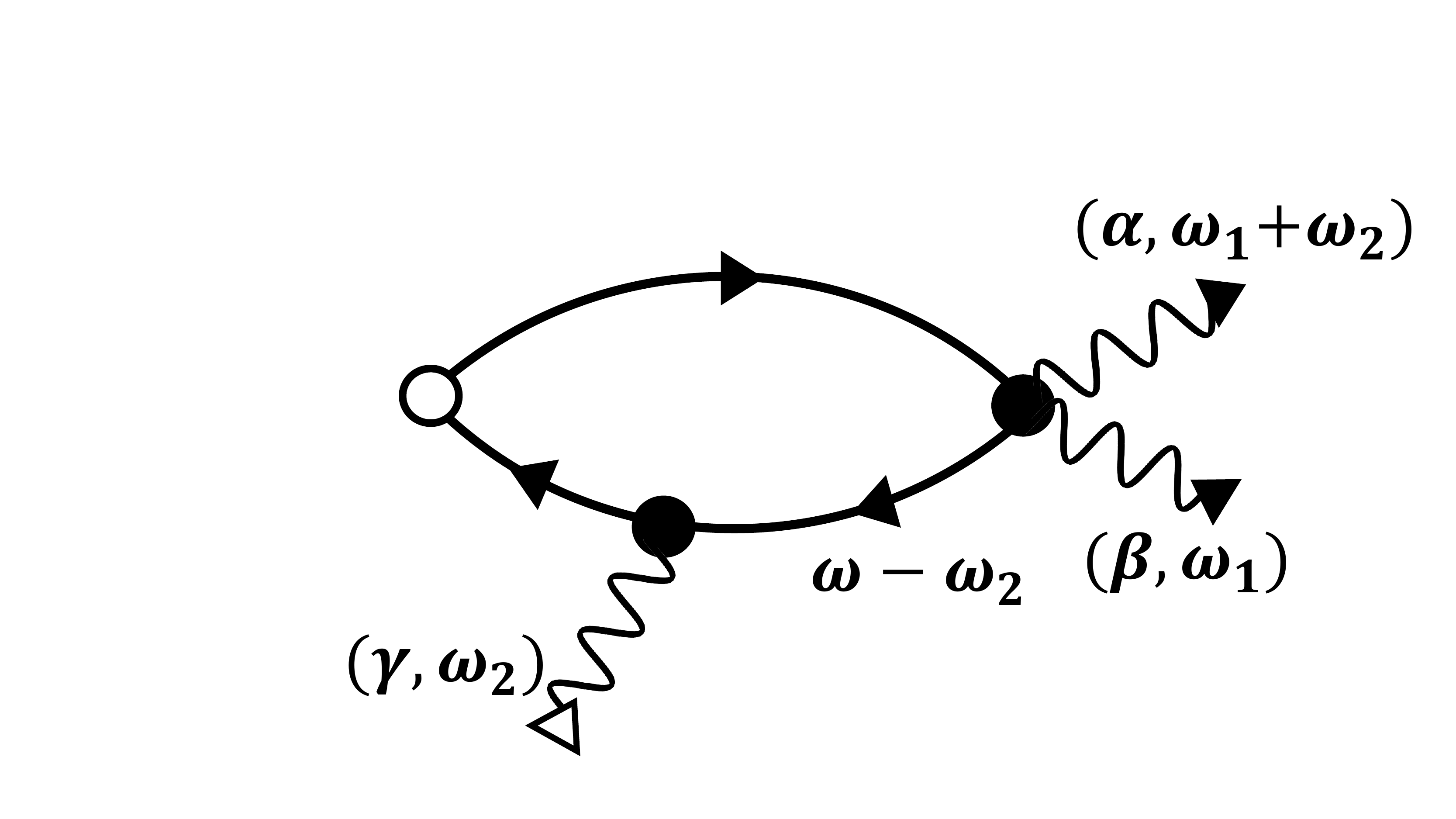}};
    \end{tikzpicture}\nonumber\\
    & \ \ \ \ + 
	\begin{tikzpicture}[baseline=(a.center)]
		\node (a) {\includegraphics[width=2.3cm]{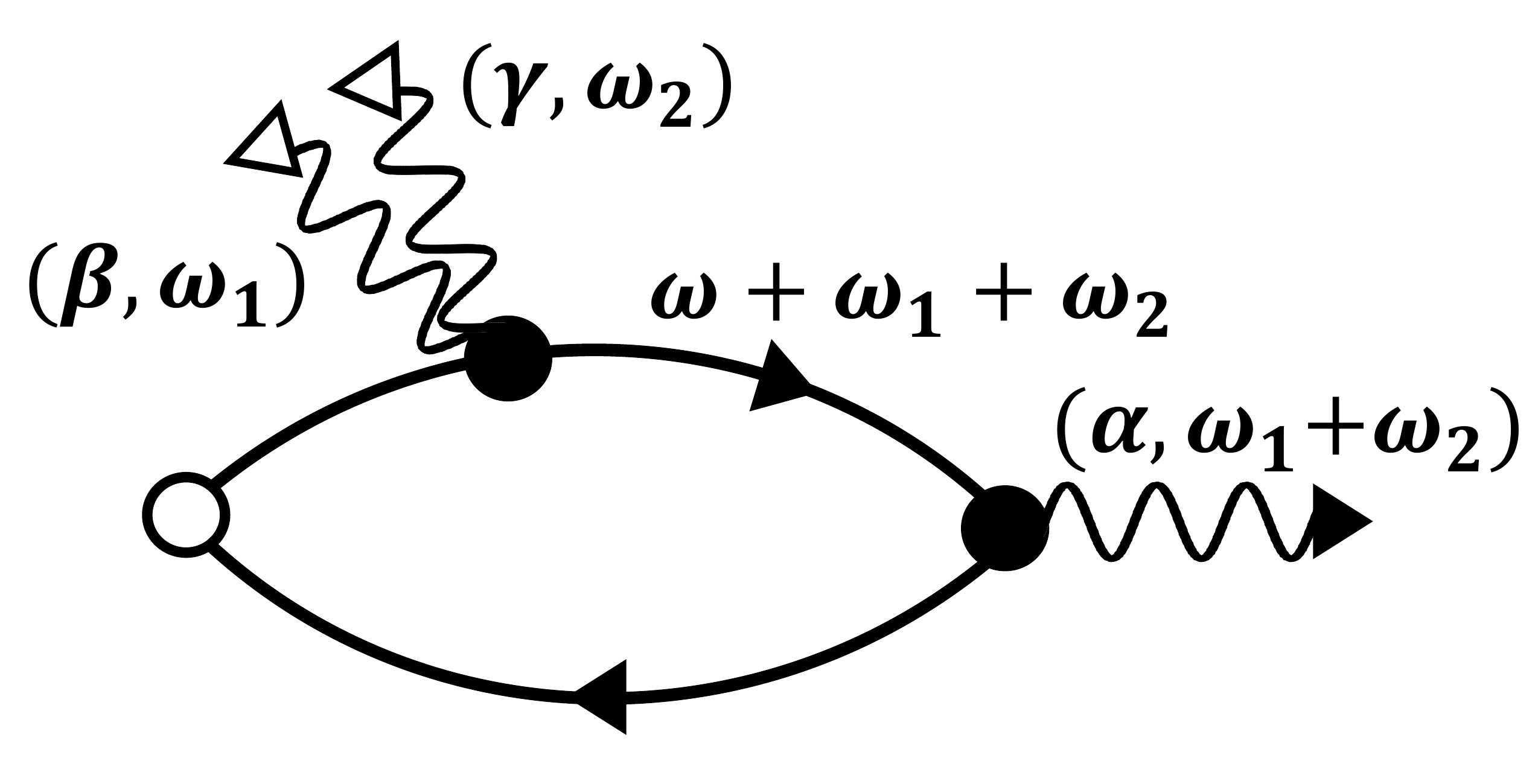}};
    \end{tikzpicture}
    + 
	\begin{tikzpicture}[baseline=(a.center)]
		\node (a) {\includegraphics[width=2.3cm]{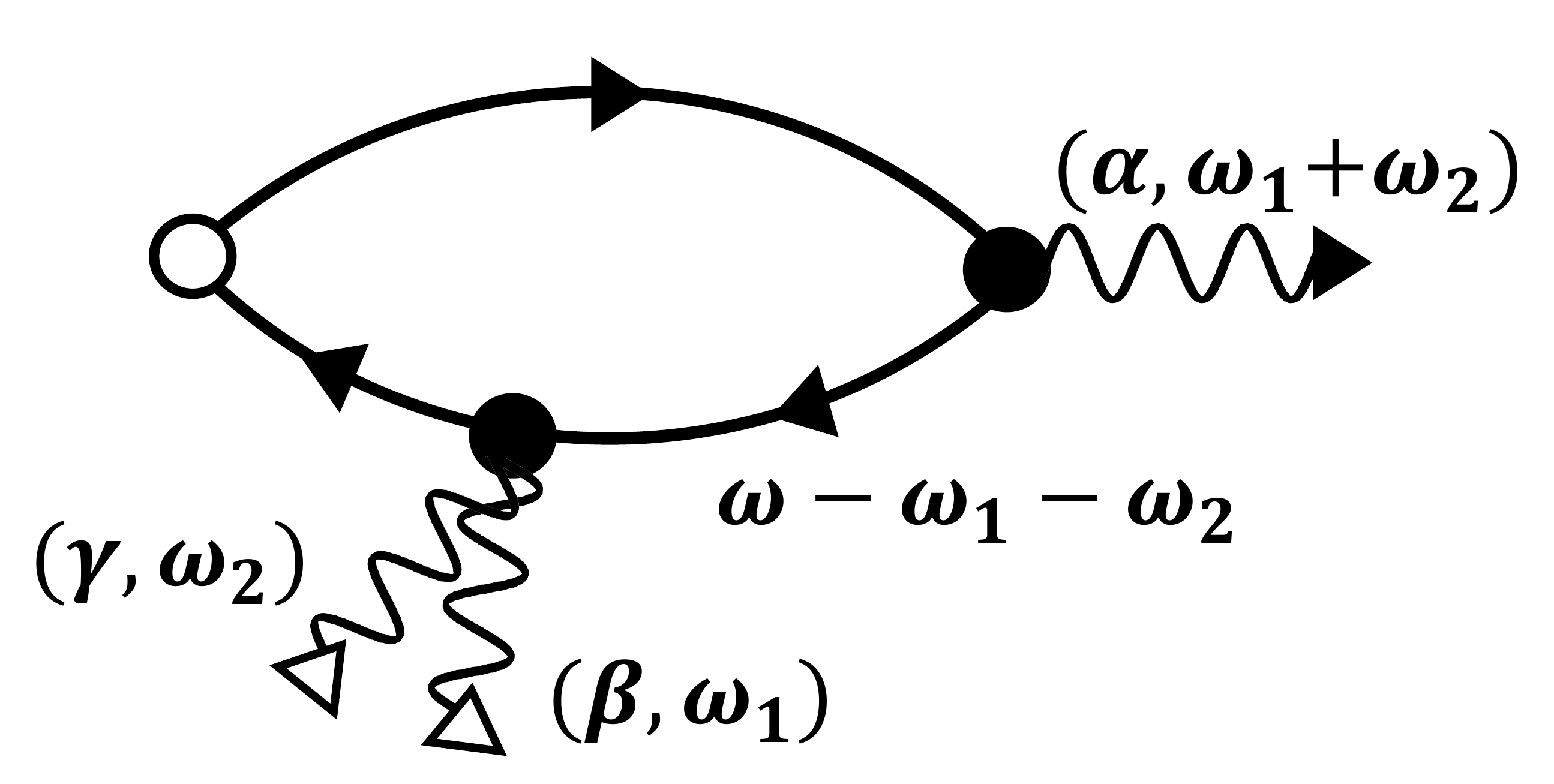}};
    \end{tikzpicture}\nonumber\\
    & \ \ \ \ + 
	\begin{tikzpicture}[baseline=(a.center)]
		\node (a) {\includegraphics[width=2.3cm]{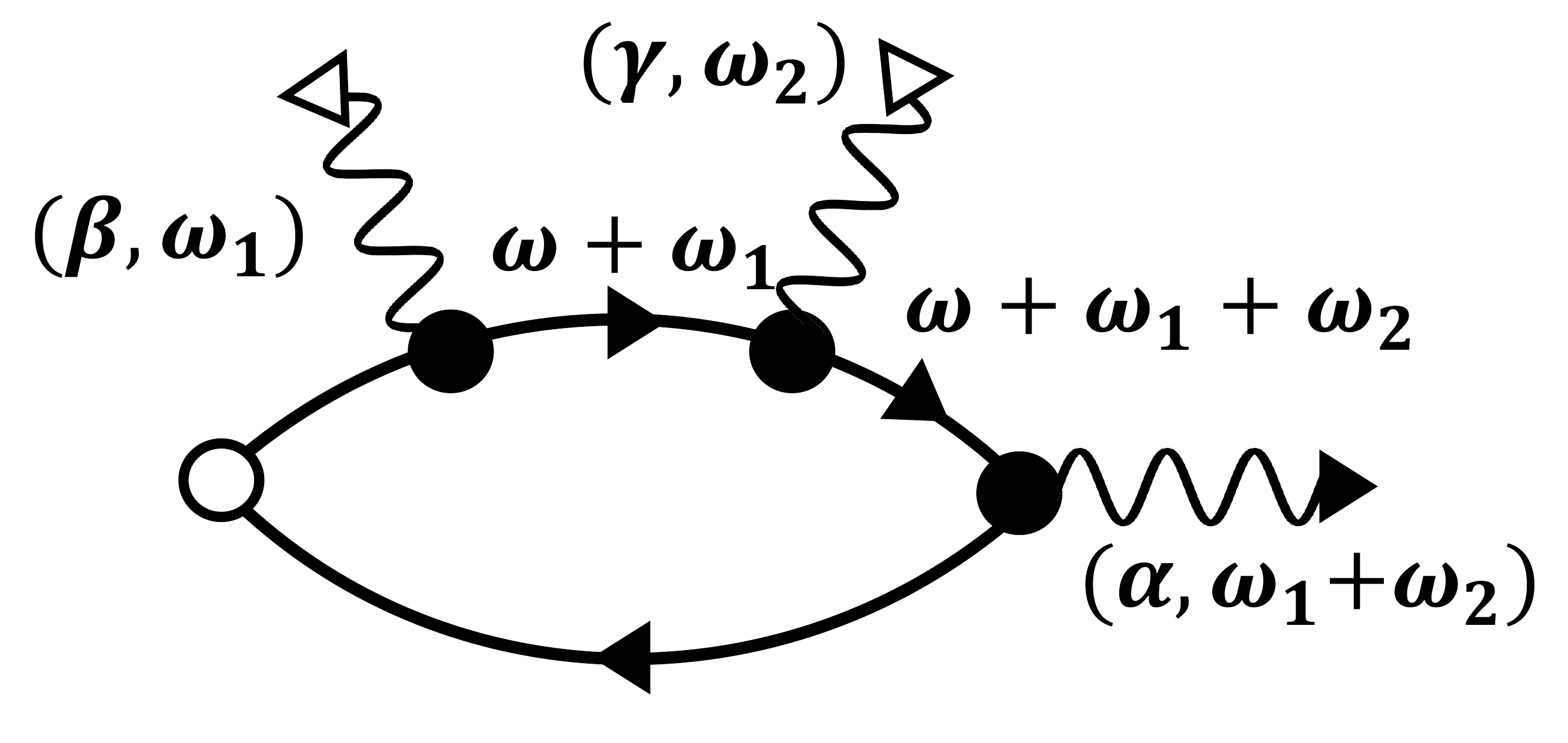}};
    \end{tikzpicture}
    + 
	\begin{tikzpicture}[baseline=(a.center)]
		\node (a) {\includegraphics[width=2.3cm]{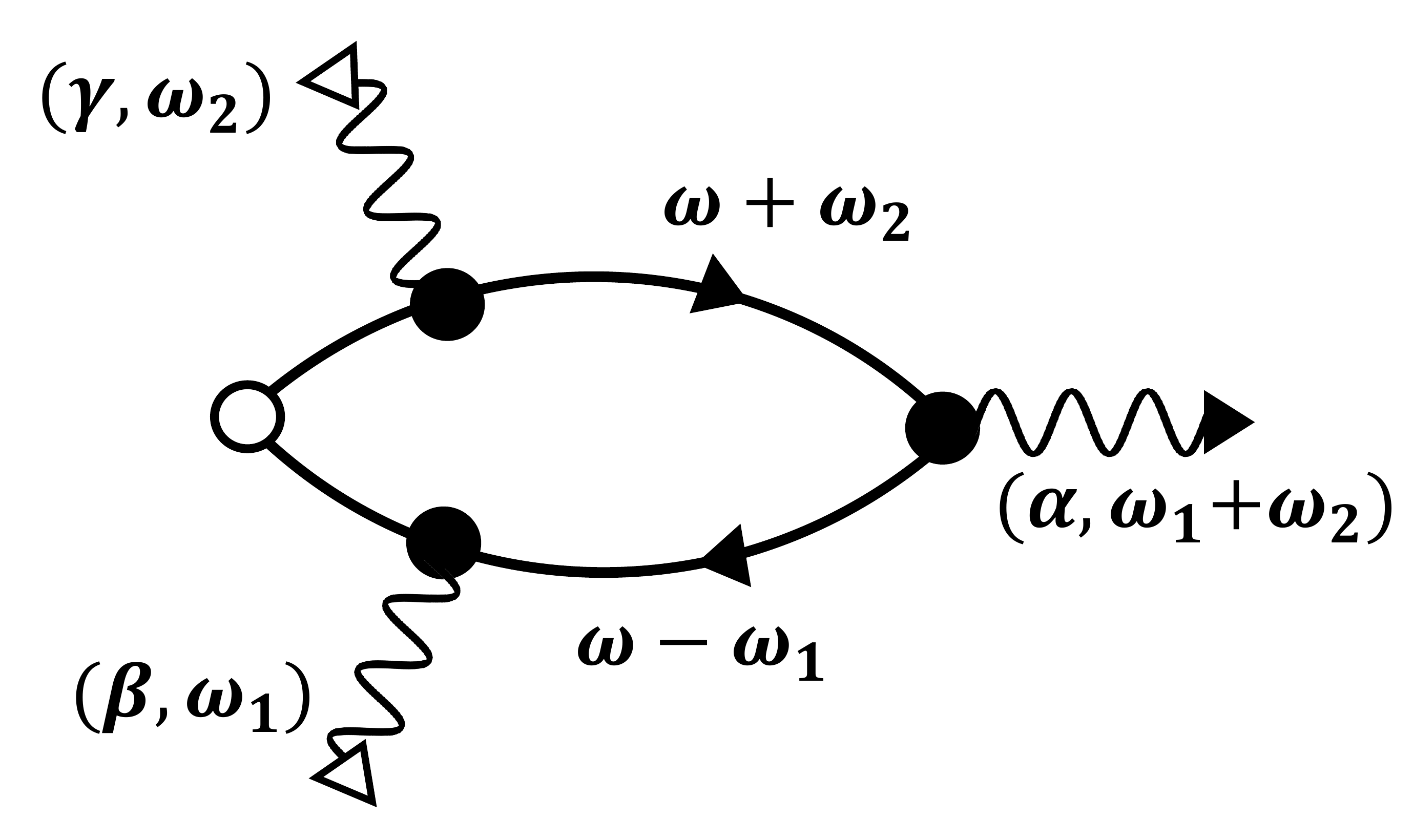}};
    \end{tikzpicture}
    + 
	\begin{tikzpicture}[baseline=(a.center)]
		\node (a) {\includegraphics[width=2.3cm]{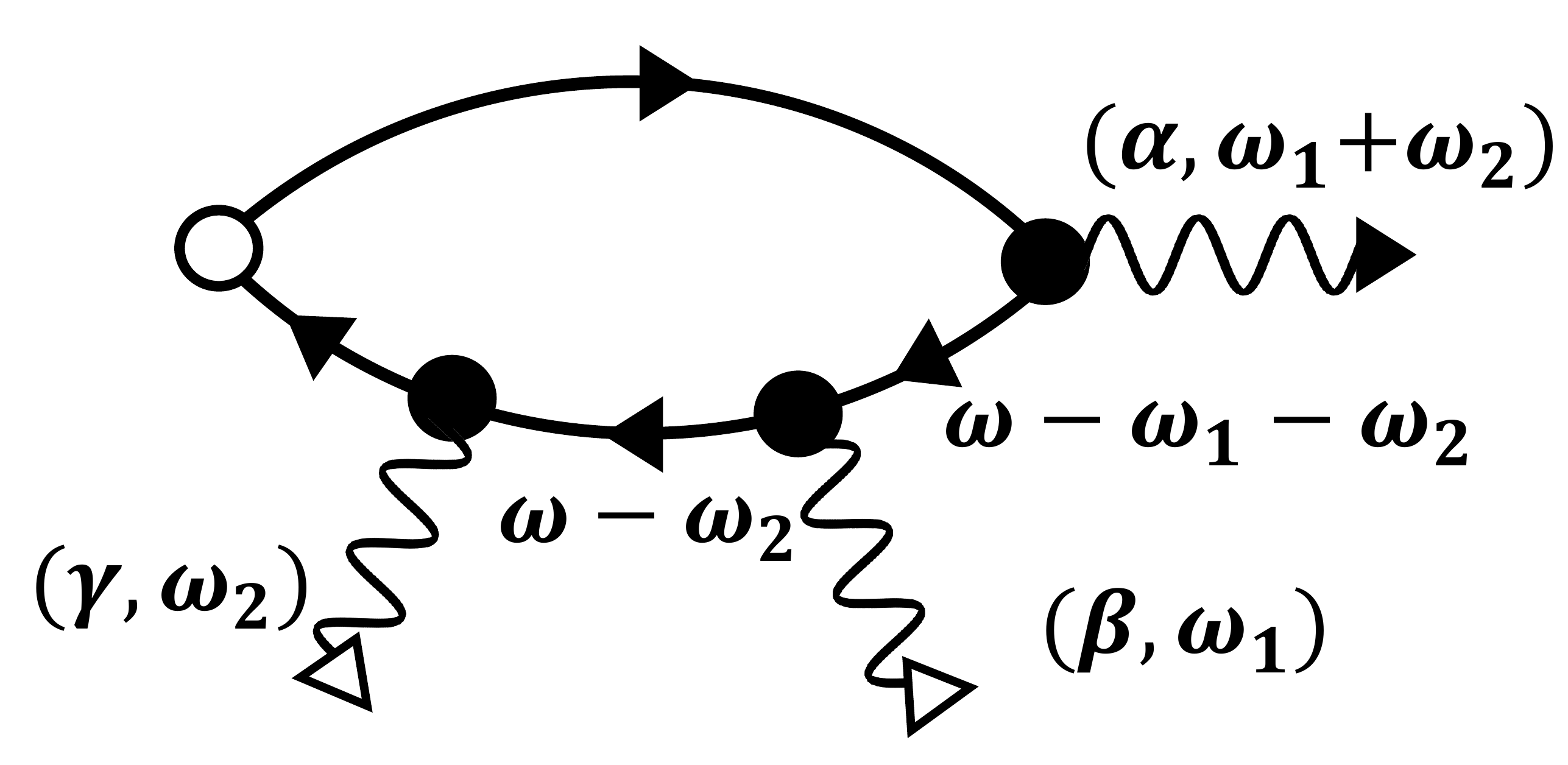}};
    \end{tikzpicture}\nonumber\\
& \ \ + \bigl((\beta,\omega_1)\leftrightarrow (\gamma,\omega_2)\bigr).\label{app:Diag2}
\end{align}

\section{Weak-scattering limit in the Green's function method\label{app:free_limit}}
When considering the weak-scattering limit where $G^R(\omega)=1/(\omega-\m{H}-i\gamma/2)$ and $\gamma\ll1/\beta,\omega_1,\epsilon_{nm}$, we can perform the frequency integration by using 
\begin{align}
&\int d\omega A(\omega,\{\omega_i\}\{\epsilon_m\})(G^R_n(\omega)-G^A_n(\omega))f(\omega)\nonumber\\
&\simeq-2\pi i A(\epsilon_n\pm i\gamma/2,\{\omega_i\}\{\epsilon_m\})f(\epsilon_n\pm i\gamma/2)\nonumber\\
&\simeq-2\pi i A(\epsilon_n\pm i\gamma/2,\{\omega_i\}\{\epsilon_m\})f(\epsilon_n),
\end{align}
where $A(\omega,\{\omega_i\}\{\epsilon_m\})$ is a product of Green's functions and velocities, and
 the sign takes $\pm$ when $A(\omega,\{\omega_i\}\{\epsilon_m\})$ is an analytical function in the
 upper/lower plane of the complex $\omega$-space. The plane is chosen such that $A(\omega,\{\omega_i\}\{\epsilon_m\})$ is analytic. Other poles than $\omega=\epsilon_n\pm i\gamma/2$ can be ignored because  $G^R_n(\epsilon)-G^A_n(\epsilon)\simeq0$ at those due to the assumption $\gamma\ll1/\beta,\omega_1,\epsilon_{nm}$.
Then we can derive the linear and nonlinear optical conductivities as
\begin{widetext}
\begin{eqnarray}
    \sigma^{(1)}_{\alpha\beta}(\omega_1;\omega_1) &\simeq&  \frac{i}{\omega_1}\sum_{\bk} \Biggl\{\m{J}^{nn}_{\alpha\beta}f(\epsilon_n)+ \frac{\m{J}^{nm}_{\alpha}\m{J}^{mn}_{\beta}}{\omega_1-\epsilon_{mn}+i\gamma}f_{nm}\Biggr\}\label{G1_free}\\
    \sigma^{(2)}_{\alpha\beta\gamma}(\omega_1+\omega_2;\omega_1,\omega_2)&\simeq& -\frac{1}{\omega_1\omega_2}\sum_{\bk}\Biggl\{\frac{1}{2}\Bigl(\m{J}^{nn}_{\alpha\beta\gamma}f_n + \frac{\m{J}^{nm}_{\alpha}\m{J}^{mn}_{\beta\gamma}}{\omega_{12}-\epsilon_{mn}+i\gamma}f_{nm}\Bigr) + \frac{\m{J}^{nm}_{\alpha\beta}\m{J}^{mn}_{\gamma}}{\omega_{2}-\epsilon_{mn}+i\gamma}f_{nm}\nonumber\\
    && \ \ \ \ \ + \frac{\m{J}^{nm}_{\alpha}\m{J}^{ml}_{\beta}\m{J}^{ln}_{\gamma}\Bigl\{(\omega_1-\epsilon_{ml}+i\gamma)f_{nl}+(\omega_2-\epsilon_{ln}+i\gamma)f_{ml}\Bigr\}}{(\omega_{12}-\epsilon_{mn}+i\gamma)(\omega_{2}-\epsilon_{ln}+i\gamma)(\omega_1-\epsilon_{ml}+i\gamma)} + \Bigl((\beta,\omega_1)\leftrightarrow(\gamma,\omega_2)\Bigr)\label{G2_free}
\end{eqnarray}
\end{widetext}
where $\m{J}^{nm}=\bra{n}\m{J}\ket{m}$ and $G^R_n(\omega)=\bra{n}G^R(\omega)\ket{n}= 1/(\omega-\epsilon_n+i/2\tau)$, $\epsilon_{nm}=\epsilon_n-\epsilon_m$, and $f_{nm}=f(\epsilon_n)-f(\epsilon_m)$. We also use the approximation $(\omega_{i}-\epsilon_{nl})/(\omega_{i}-\epsilon_{nl}+i\gamma)\simeq1$ to derive Eq.~(\ref{G2_free}). We note that these equations diverge in the DC limit, where the assumption $\gamma \ll\omega_i$ is not satisfied. These results correspond to the results by the RDM method with RTA under the velocity gauge.
Under the assumption $\gamma \ll\omega_i$, we can regard $(\omega_1+i\gamma)/\omega_1\simeq1$ and derive the same results by the RDM methods under the length gauge from Eqs.~(\ref{G1_free}) and (\ref{G2_free}). 

Finally, we analyze the DC limit by first taking the limit $\omega_i\rightarrow 0$ and assuming $\omega\ll\gamma$.
Then we can derive the DC conductivity as
\begin{widetext}
\begin{eqnarray}
    \sigma^{(1)}_{\alpha\beta;DC} &=&\sum_{\bk} \Biggl\{\tau\m{J}_{\alpha}^{nn}\m{J}_{\beta}^{nn}\Bigl(-\frac{\p f}{\p\omega}\Bigr)\Bigr\vert_{\epsilon_n} - \frac{\m{J}_{\alpha}^{nm}\m{J}_{\beta}^{mn}}{(\epsilon_{nm}+i\gamma)^2}f_{nm}\Biggr\}\label{G1_DC_free}\\
    \sigma^{(2)}_{\alpha\beta\gamma;DC}&=& -\sum_{\bk}\Biggl\{\frac{\tau^2}{2}\Bigl(\m{J}^{nn}_{\alpha}\m{J}^{nn}_{\beta}\m{J}^{nn}_{\gamma}\frac{\p^2f_n}{\p\epsilon_n^2}\!+\!\m{J}^{nn}_{\alpha}\m{J}^{nn}_{\beta\gamma}\Bigl(-\frac{\p f(\epsilon_n)}{\p\epsilon_n}\Bigr)\Bigr)+ \frac{i\tau\m{J}^{nm}_{\alpha}\m{J}^{mn}_{\beta}}{(\epsilon_{nm}+i\gamma)^2}\Bigl(\m{J}^{nn}_{\gamma}\frac{\p f(\omega)}{\p\omega}\Bigr|_{\epsilon_n}\!-\!\m{J}^{mm}_{\gamma}\frac{\p f(\omega)}{\p\omega}\Bigr|_{\epsilon_m}\Bigr)\nonumber\\
    && \ \ \ \ \ \ \ \ +\m{J}^{nm}_{\alpha}\m{J}^{ml}_{\beta}\m{J}^{ln}_{\gamma}\Bigl(\frac{f_l}{(\epsilon_{lm}+i\gamma)^2(\epsilon_{nl}+i\gamma)^2} + \frac{f_n}{(\epsilon_{nm}+i\gamma)^2(\epsilon_{nl}+i\gamma)^2} + \frac{2f_n}{(\epsilon_{nm}+i\gamma)^3(\epsilon_{nl}+i\gamma)}\Bigr)\nonumber\\
    && \ \ \ \ \ \ \ \  +\frac{1}{2}\m{J}^{nm}_{\alpha}\m{J}^{mn}_{\beta\gamma}\frac{f_{n}}{(\epsilon_{nm}+i\gamma)^2} + \Bigl(\beta\leftrightarrow\gamma\Bigr)\Biggr\}\label{G2_free_DC}
\end{eqnarray}
\end{widetext}
The first terms, which are proportional to $\tau$ for $\sigma^{(1)}$ and proportional to $\tau^2$ for $\sigma^{(2)}$, represent the Drude term. The other terms for the second-order conductivity represent the Berry curvature dipole term and the Fermi sea terms.

\section{Semi-classical Boltzmann equation \label{app:OtherMethods}\label{app:BE}}
In the semi-classical Boltzmann treatment, transport phenomena are analyzed by calculating the distribution function for particles near equilibrium\cite{RevModPhys.82.1959,PhysRevLett.115.216806,NatureComm.10.3047}. 
The effect of the vector potential is taken into account as
\begin{eqnarray}
    \m{H}(\bm{p}) \rightarrow \m{H}(\bm{k}(\bm{p},t)) = \m{H}(\bm{p}-q\bm{A}(\bm{x},t)).
\end{eqnarray}
By taking the Coulomb gauge $\bm{A}(\bm{x},t)=\bm{A}(t)$, the translational symmetry is preserved, and the following equations are satisfied:
\begin{eqnarray}
    \dot{\bm{k}} &=& -q\frac{\partial \bm{A}(t)}{\partial t} = q\bm{E}\\
    \frac{\p}{\p p^{\alpha}} &=& \frac{\p}{\p k^{\alpha}}, \ \ \ \frac{\p}{\p t} = \dot{\bm{k}}\cdot\bm{\nabla_k} = q\bm{E}\cdot\bm{\nabla_k}.
\end{eqnarray}
where $p$ is the wavenumber of the particle without the electric field, $E$ the electric field described by the vector potential $A$, and $q$ is the wavenumber under the electric field.
Considering the change of the eigenstates and the band velocity induced by the vector potential up to the first order of the vector potential, we find
\begin{align}
    &\ket{n(\bm{p})}\nonumber\\
    &\rightarrow \ket{\tilde n(\bm{k}(t))} \simeq \ket{n(\bm{p})} \!-\!i\sum_{n'\neq n} \frac{\ket{n'(\bm{p})}\braket{n'(\bm{p})|\frac{\p}{\p t}n(\bm{k}(t))}}{\epsilon_n-\epsilon_{n'}}\\
    &v_{n\alpha}(\bm{k}(t))=\bra{\tilde n}\frac{\p\m{H}}{\p k^{\alpha}}\ket{\tilde n}\nonumber\\
    &\simeq \frac{\p\epsilon_n(\bm{p})}{\p p^{\alpha}}\nonumber\\
    & \ \ -i\sum_{n'\neq n}\Bigl(\frac{\bra{n(\bm{p})}\frac{\p\m{H}}{\p q^{\alpha}}\ket{n'(\bm{p})}\braket{n'(\bm{p})|q\bm{E}\cdot\bm{\nabla_k}|n(\bm{p})}}{\epsilon_n-\epsilon_{n'}}  - \mr{c.c.}\Bigr)\nonumber\\
    &=  v^0_{n\alpha}(\bm{p}) -q\Bigl(\bm{E}\times\bm{\Omega_n}(\bm{p})\Bigr)_{\alpha},\label{result1}
\end{align}
\begin{align}
    &\bm{\Omega_n}(\bm{k})= \bm{\nabla}_{\bm{k}}\times\bm{\m{A}_n}, \ \ \ \ \ \bm{\m{A}_n} = -i\braket{u_{n\bm{k}}|\bm{\nabla}u_{n\bm{k}}},
\end{align}
where $\ket{n}$ is the eigenstate of the Hamiltonian without the vector potential, $\epsilon_n(\bm{p})$ is the eigenvalue, and  $\m{H}(\bm{p})\ket{n(\bm{p})} = \epsilon_n(\bm{p})\ket{n(\bm{p})}$ holds. 
By taking into account the correction of the band velocity, we obtain the semi-classical equation of motion, which reads, 
\begin{align}
    &\bm{\dot k_n}=q\bm{E}(t), \ \ \ \ \ {\bm{\dot r_{n}}} = \frac{\partial \epsilon_{n}(\bm{p})}{\partial\bm{p}} - q\bm{E}(t)\times \bm{\Omega_{n}}(\bm{p})
\end{align}
Finally, the distribution function in the  Boltzmann formalism with applied electric field using the relaxation time approximation(RTA) is given by the following equation
\begin{eqnarray}
    \frac{df_n(t)}{dt} = \frac{\p f_n(t)}{\p t} + \dot{\bm{k}}\cdot \bm{\nabla_k}f_n(t) = -\frac{f_n(t)-f^{(0)}_n}{\tau}
\end{eqnarray}
    which can be solved as
\begin{eqnarray}
    f(t)&=&f^{(0)}(t)+f^{(1)}(t)+f^{(2)}(t)+\ldots\label{asymm}\\
    &\Rightarrow& \tau\frac{\p f^{(m)}_n(t)}{\p t} + f^{(m)}_n(t) = -q\tau\bm{E}(t)\cdot\bm{\nabla_k}f^{(m-1)}_{n}(t),\nonumber
\end{eqnarray}
where $f^{(0)}_n = 1/(1+\exp[\beta \epsilon_n(\bm{k})])$ is the Fermi distribution function, $\beta$ is the inverse of the temperature, and $f^{(m)}$ represent the $m$-th order non-equilibrium perturbative distribution function for the electric field.

The first and second order term of the distribution function become
\begin{eqnarray}
    &&f^{(1)}_n(\omega,\alpha)=\frac{-q\tau}{1-i\omega\tau}E^{\alpha}\p_{\alpha}f^{(0)}_n\\
    &&f^{(2)}_n\bigl((\omega_1,\beta),(\omega_2,\gamma)\bigr)\nonumber\\
    &&=\frac{-q\tau E^{\beta}\p_{\beta}}{1-i(\omega_1\!+\!\omega_2)\tau}f^{(1)}_n(\omega_2,\gamma) + \Bigl((\omega_1,\beta) \leftrightarrow (\omega_2,\gamma)\Bigr)\nonumber\\
    &&= \frac{(q\tau)^2 E^{\alpha}E^{\beta}\p_{\alpha}\p_{\beta}f^{(0)}_n}{(1-i(\omega_1\!+\!\omega_2)\tau)(1-i\omega_2\tau)} +\Bigl((\omega_1,\beta)\leftrightarrow (\omega_2,\gamma)\Bigr)\nonumber\\
\end{eqnarray}
By combining the recurrence relation in Eq.~(\ref{asymm}) with the velocity corrected by the electric field in Eq.~(\ref{result1}), we can derive the second order nonlinear conductivity as
\begin{widetext}
\begin{eqnarray}
    \sigma^{B(2)}_{\alpha\beta\gamma}(\omega_1+\omega_2;\omega_1,\omega_2) &=& q^3\sum_{n,\bk}\Bigl\{\frac{\p\epsilon_n}{\p p^{\alpha}}\frac{\tau^2 \p_{\beta}\p_{\gamma}f_0}{(1-i\omega_{12}\tau)(1-i\omega_2\tau)} +\frac{\tau}{2(1-i\omega_2\tau)}\epsilon_{\alpha\beta\mu}\Omega_{n\mu} \p_{\gamma}f_0 + \Bigl((\omega_1,\beta)+ \leftrightarrow (\omega_2,\gamma)\Bigr)\Bigr\}\nonumber\\
    &&\label{Boltzmann}\\
    \sigma^{B(2)}_{DC;\alpha\beta\gamma} &=& q^3\sum_{n,\bk}\Bigl\{\frac{\p\epsilon_n}{\p p^{\alpha}}\tau^2 \p_{\beta}\p_{\gamma}f_0 +\tau\epsilon_{\alpha\beta\mu}\Omega_{n\mu} \p_{\gamma}f_0 + \Bigl(\beta \leftrightarrow \gamma\Bigr)\Bigr\}\label{app:Boltzmann_DC}
\end{eqnarray}
\end{widetext}
Then, we compare our results with the semi-classical Boltzmann treatment. For the sake of comparison, we set the self-energy in the Green's function as $G^R(\omega)=1/(\omega-\m{H}+i/2\tau)=1/(\omega-\m{H}+i\gamma/2)$. In this case, the Green's function can be diagonalized with the eigenvalue of the free Hamiltonian, and therefore, the nonlinear conductivity calculated by the semi-classical Boltzmann treatment can be written using Green's functions. First, we focus on the Green's function representation of $\sigma^{B(2)}_{DC;\alpha\beta\gamma}$ in the DC limit, which reads
\begin{widetext}
\begin{eqnarray}
    \sigma^{B(2)}_{DC;\alpha\beta\gamma} &=& -\sum_{n,m(\neq n),\bk}\int\frac{d\omega}{2\pi i}\Bigl\{\frac{1}{2}\m{J}^{nn}_{\alpha}\Bigl(\frac{\p G^R_n(\omega)}{\p\omega}\Bigr)\m{J}^{nn}_{\beta\gamma}G^A_n(\omega)\Bigl(-\frac{\p f(\omega)}{\p\omega}\Bigr) + \m{J}^{nn}_{\alpha}G^R_n(\omega)\m{J}^{nn}_{\beta}G^R_n(\omega)\m{J}^{nn}_{\gamma}G^A_n(\omega)\frac{\p^2 f(\omega)}{\p\omega^2}\nonumber\\
    && \ \ \ \ \ \ \ \ \ \ \ \ \ \ \ \ \ \ + \m{J}^{n}_{\alpha}\frac{\p G^R_n(\omega)}{\p\omega}\m{J}^{nm}_{\beta}G^R_m(\omega)\m{J}^{mn}_{\gamma}G^A_n(\omega)\Bigl(-\frac{\p f(\omega)}{\p\omega}\Bigr)\nonumber\\
    && \ \ \ \ \ \ \ \ \ \ \ \ \ \ \ \ \ \ +\m{J}^{nm}_{\alpha}\Bigl(\frac{\p G^R_m(\omega)}{\p\omega}\Bigr)\m{J}^{mn}_{\beta}G^R_n(\omega)\m{J}^{nn}_{\gamma}G^A_n(\omega)\Bigl(-\frac{\p f(\omega)}{\p\omega}\Bigr) + \Bigl(\beta \leftrightarrow \gamma\Bigr)\Bigr\},\label{app:Boltzmann_G_DC}
\end{eqnarray}
\end{widetext}
where $\m{J}^{nm}=\bra{n}\m{J}\ket{m}$ and $G^R_n(\omega)=\bra{n}G^R(\omega)\ket{n}= 1/(\omega-\epsilon_n+i/2\tau)$. 
We use $q\p_{\alpha}\m{J}^{nn}_{\beta} = \m{J}^{nn}_{\alpha\beta}+ (\m{J}^{nm}_{\alpha}\m{J}^{mn}_{\beta}+\m{J}^{nm}_{\beta}\m{J}^{mn}_{\alpha})/(\epsilon_{nm})$ to derive Eq.~(\ref{app:Boltzmann_G_DC}).
Here, we suppose that $\beta\gamma$ is small and $G^R_n(\omega)G^A_n(\omega)=1/[(\omega-\epsilon_n)^2 + \gamma^2/4]\simeq 2\pi\delta(\omega-\epsilon_n)/\gamma$ is justified.
Then, doing the frequency integration in Eq. (\ref{app:Boltzmann_G_DC}), we can obtain the original result Eq.~(\ref{app:Boltzmann_DC}).

This Green's function representation of the Boltzmann equation  Eq.~(\ref{app:Boltzmann_G_DC}) can be directly derived from the original Green's function method  shown in the main text, Eq.~(\ref{cond_second}), by ignoring the Fermi sea terms and the interband transitions $\m{J}_{mn}(m\neq n)$ except for the second and third term in Eq.~(\ref{app:Boltzmann_G_DC}), which is justified when $\epsilon_{nm}\tau \gg 1$.

Next, we consider the AC case. We can recover a finite frequency $\omega_i$ from the DC limit in Eq.~(\ref{app:Boltzmann_G_DC}), which can be derived from Eq.~(\ref{cond_second}) under the following assumptions:
\begin{itemize}
\item{approximate $\omega_i\simeq \omega_i+i\gamma$  which is justified in the limit $\omega_i\tau\gg1$}
\item{approximate $f(\omega+\omega_i)-f(\omega)\simeq [\p f(\omega)/\p\omega] \omega_i$ and $\bigl[\p f(\omega)/\p\omega\bigr]-\bigl[\p f(\omega-\omega_i)/\p\omega\bigr]\simeq [\p^2 f(\omega)/\p\omega^2]\omega_i$ which is justified when $\beta\omega_i\ll 1$.}
\item{approximate $G^R(\omega+\omega_{12})-G^R(\omega+\omega_2)\simeq [\p G^R(\omega+\omega_2)/\p\omega]\omega_1$ and $1/(\omega_2-\epsilon_{nm})\sim1/(-\epsilon_{nm})$ which is justified when $\omega_i \ll \epsilon_{nm} \ \ (\epsilon_{nm}=\epsilon_n-\epsilon_m)$.}
\end{itemize}
Therefore, in the case of AC electric fields, there are severe approximations. Thus, the semi-classical Boltzmann equation is applicable at high temperatures or when the frequency $\omega_i$ is very small so that the above conditions are satisfied. We note that we can also derive Eq.~(\ref{Boltzmann}) from Eq.~(\ref{cond_second}) by supposing $\gamma\rightarrow0$, which corresponds to the condition $\omega_i\gg\gamma$ for the RDM method.
We note that taking the DC limit in this situation leads to a diverging conductivity.
Moreover, the relaxation time in most materials is usually about $1\sim100 [\mr{ps}]$\cite{NatureComm.10.3047}. Thus, when analyzing a Terahertz laser as input force, $\omega_i \tau \sim1$, the conditions are not fulfilled.
On the other hand, for a DC electric field in which $\omega_i=0$, the only condition for the semi-classical Boltzmann treatment are $\epsilon_{nm}\tau\gg1$ and $\beta\gamma\ll 1$.

We note that, by considering higher-order corrections of the eigenstates by the electric field in Eq.~(\ref{result1}), we can derive a more precise semi-classical Boltzmann equation. In this way, it is possible to get rid of the approximation $\epsilon_{nm}\tau \gg 1$ and to include the Fermi sea terms in the Boltzmann equation. The other approximations listed above, however, remain necessary due to the relaxation time approximation. 

\section{Gauge invariance with the dissipation in quantum master equation formalism\label{app:Gauge}}
In this section, we analyze the correspondence between the length gauge and the velocity gauge in the quantum master equation in Eq.~(\ref{QME3}) in the main text.
By using  Eq.~(\ref{LvsV}), we can describe Eq.~(\ref{QME2}) under the velocity gauge as
\begin{align}
    &\frac{d}{dt}\rho_{A\bk}(t) +i[q\bm{E}\cdot\bm{r},\rho_{A\bk}(t)]\nonumber\\
    &= -i[\m{H}_0(\bk-q\bm{A})-q\bm{E}\cdot\bm{r},\rho_{A\bk}(t)] -\lambda^2\int_{t_0}^{t}ds\nonumber\\
    & \ \ \ \ \times\Bigl(\mr{Re}\Bigl[\Bigl\{iG^{l}_B(t\!-\!s)\psi^{\dagger }_{\bk}\tilde{U}(t,s)\psi_{\bk},\rho_{A\bk}(s),\tilde{U}^{\dagger}(t,s)\Bigr\}\Bigr)^{(n)},\label{app:XX}\\
    &\tilde{U}(t,s) = T_{\rightarrow}U^{-1}(t)\exp[-i\int_{s}^{t}dt'(\m{H}_0-q\bm{E}(t')\cdot\bm{r})]U(s),
\end{align}
where $\{\m{O}_1,\rho,\m{O}_2^{\dagger}\}=\m{O}_1\rho\m{O}_2^{\dagger}+\m{O}_2\rho\m{O}^{\dagger}_1$. If we can show that $T_{\rightarrow}\tilde{U}(t,s)=\exp[-i\int_{s}^{t}\m{H}_0(\bm{k}-q\bm{A}(t))]$, the second term on the right side in Eq.~(\ref{app:XX}) can be written in the interaction representation in the velocity gauge Hamiltonian and the gauge invariance holds true in the open system. This can be verified by calculating the $s$ and $t$ derivatives of $\tilde{U}(t,s)$ as
\begin{align}
    &\frac{\p}{\p t}\tilde{U}(t,s)\nonumber\\
    &= U^{-1}(t) \Bigl\{ -iq\bm{E}(t)\cdot\bm{r} -i \Bigl(\m{H}_0(\bk)-q\bm{E}(t)\cdot\bm{r}\Bigr)\Bigr\}\nonumber\\
    & \ \ \ \ \times\exp[-i\int_{s}^{t}dt'(\m{H}_0-q\bm{E}(t')\cdot\bm{r})]U(s)\nonumber\\
    &= U^{-1}(t) \Bigl(-i\m{H}_0(\bk)\Bigr)\exp[-i\int_{s}^{t}dt'(\m{H}_0-q\bm{E}(t')\cdot\bm{r})]U(s)\nonumber\\
    &=-i\m{H}_0(\bk-q\bm{A}(t))\tilde{U}(t,s)\label{app:Eq1}\\
    &\frac{\p}{\p s}\tilde{U}(t,s) = \tilde{U}(t,s)\Bigl(i\m{H}_0(\bk-q\bm{A}(s))\Bigr)\label{app:Eq2}\\
    &\tilde{U}(t,t) = 1.\label{app:Eq3}
\end{align}
We use the relation $U^{-1}(t)\m{H}_{0}(\bk)U(t)=\m{H}(\bk-q\bm{A}(t))$ to derive Eqs.~(\ref{app:Eq1}) and (\ref{app:Eq2}). From the equality in Eqs.~(\ref{app:Eq1}), (\ref{app:Eq2}), and (\ref{app:Eq3}), we can identify $\tilde{U}(t,s)=T_{\rightarrow}\exp[-i\int_{s}^{t}dt'\m{H}_0(\bm{k}-q\bm{A}(t'))]$. Therefore, the correspondence between the length gauge and the velocity gauge holds true in the quantum master equation, while it is broken when introducing the RTA at finite frequency.

\section{Models used in the main text\label{app:Model}}
In the main text, we use the following two models to numerically confirm our general results. In this section, we introduce the effective Hamiltonian $\m{H}_{\mr{eff}}=\m{H}_0+\Sigma^R$, which includes the dissipation effect.
\subsection{One-dimensional Rice-Mele model with  sublattice-dependent dissipation}
We start from the Hermitian 1D Rice-Mele model, but assume that the dissipation depends on the sublattice. Such an effective non-Hermitian Hamiltonian can also be derived from the non-Hermitian matrix describing the single-particle Green's function.
The effective non-Hermitian Hamiltonian reads\cite{Morimoto2018}
\begin{eqnarray}
    &&\m{H}_{\mr{eff}}(\bk)\nonumber\\
    &&=\sum_{a,b}\psi^{\dagger}_a \Bigl(\tau^0i\eta\!+\!\tau^xt\cos{k}\!+\!\tau^y\delta t\sin{k}\!+\! \tau^z(\Delta+i\Gamma) \Bigr)_{ab} \psi_{b},\nonumber\\
    \label{app:RM_NH}
\end{eqnarray}
where $\psi^{(\dagger)}_{A/B}$ describes the annihilation (creation) operator in sublattice $A/B$, $\tau$ represents the Pauli matrices, $\eta$ is the average of the dissipation strength at each sublattice, $t$ is an intra-lattice hopping, $\delta t$ is an inter-lattice hopping, $\Delta$ is the difference of the chemical potential between the sublattices, and $\Gamma$ is the difference of the dissipation strength at each sublattice.

\

\subsection{Monolayer TMD materials with a spin-dependent dissipation}
This model is commonly used to describe 
transition metal dichalcogenide(TMD) monolayers.
The effective non-Hermitian Hamiltonian, which can again be understood as the non-Hermitian matrix describing the single-particle Green's function, can be written as\cite{PhysRevB.102.094507,PhysRevApplied.13.024053}
\begin{align}
    &\mathcal{H}_{eff} = \sum_{\bm{k},s} \left(\epsilon(\bm{k})-\mu-i\eta -i\Gamma_s\right) c_{\bm{k},s}^{\dag} c_{\bm{k},s} \nonumber\\
& \ \ \ \ \ + \sum_{\bm{k},s,s'}
\bm{g}(\bm{k}) \cdot \bm{\sigma}_{ss'} c_{\bm{k},s}^{\dag} c_{\bm{k},s'}\label{app:TMD_NH}\\
& \epsilon(\bk) = 2t\Bigl(p\cos(\bk\cdot\bm{a_1})\!+\!\cos(\bk\cdot\bm{a_2})\!+\!\cos(\bk\cdot(\bm{a_1}\!+\!\bm{a_2}))\Bigr)\\
& g^{x}(\bk) = \frac{\alpha_1}{2}\Bigl[\sin(\bk\cdot(\bm{a_1}+\bm{a_2})) + \sin(\bk\cdot\bm{a_2})\Bigr]\\
& g^{y}(\bk) = -\frac{\alpha_1}{\sqrt{3}}\Bigl[ \sin(\bk\cdot\bm{a_1})\!+\!\frac{\sin(\bk\cdot(\bm{a_1}\!+\!\bm{a_2}))\!-\!\sin(\bk\cdot\bm{a_2})}{2}\Bigr]\\
& g^z(\bk) = \frac{2\alpha_2}{3\sqrt{3}}\Bigl[\sin(\bk\cdot\bm{a_1})\!+\!\sin(\bk\cdot\bm{a_2})\!-\!\sin(\bk\cdot\bm{a_1}\!+\!\bm{a_2})\Bigr]
\end{align}
where $c^{(\dagger)}_{\bk,s}$ is the annihilation(creation) operator for a conduction electron whose momentum is $\bk$ and spin is $s$. $\mu$ is the chemical potential, $\Gamma_{\uparrow/\downarrow}=\pm\Gamma$ is the spin-dependent dissipation, $p$ is the effect of the strain\cite{PhysRevApplied.13.024053} and $\bm{\sigma}$ are the Pauli matrices and $\bm{g}$ represents the spin-orbit coupling. The lattice vectors are $\bm{a_1}=(1,0)$ and $\bm{a_2}=(-1/2,\sqrt{3}/2)$.

\section{Details of the numerical calculations\label{app:Numerical}}

In this section, we write in detail how to numerically calculate the results shown in the figures of the main text. The codes used for the numerical calculations in this paper are published in \footnote{https://github.com/YoshihiroMichishita/Test\_Codes/}. 

\subsection{Green's function method\label{app:Num_Green}}
Here, we describe the procedure of how to perform the numerical calculation using the Green's function method.
\begin{itemize}
    \item A tight-binding Hamiltonian, $\m{H}(\bm{k})$, describing the single-electron part of the model, such as Eq.~(\ref{app:RM_NH}) or Eq.~(\ref{app:TMD_NH}), must be obtained.
    \item Starting from this tight-binding Hamiltonian, current operators $\m{J}_{\alpha\beta\dots}$ can be calculated by Eq. (\ref{CO}).
    \item For accounting for correlation effects, self-energies must be calculated. In this paper, we have used the dynamical mean-field theory\cite{RevModPhys.68.13}. 
    \item Using $\m{H}(\bm{k})$ and the self-energies, retarded and advanced Green's functions can be calculated.
    \item Having these Green's functions and current operators, one can use the Green's function formalism to calculate nonlinear response in strongly correlated systems. 
\end{itemize}

To calculate the effect of renormalization of the band structure, we set $\Sigma^R(\omega) = -(1/Z-1)\omega - \Sigma^R_0$, where $\Sigma^R_0$ is the real-part of the self-energy at $\omega=0$. Then, one can analyze the renormalization effects on the linear response and the nonlinear response. We note that, when calculating the optical conductivity for a small input frequency ($\omega_i$), one should do the momentum integration before the frequency ($\omega$) integration.
Furthermore,  one should use Eqs. (\ref{app:D6}), (\ref{app:A00}), (\ref{app:A11}), and (\ref{app:A11p}).

\subsection{RDM methods using the RTA \label{app:Num_RDM}}
When using the RDM for calculating the (non)linear conductivity, one first needs to diagonalize the free Hamiltonian $\m{H}(\bm{k})$. Using the eigenvectors, one calculates the velocity operators for different bands and calculates the (non)linear conductivity by Eqs. (82) in Ref. \cite{PhysRevB.96.035431}.

\section{ Proof that $\gamma_{NH}\geq 1$\label{app:NH_proof}}
The left and right eigenvectors $\bra{n_L}$, $\ket{n_R}$ can be described as $\ket{n_R}=(a_1,\dots,a_l)^{\mr{T}}$ and $\bra{n_L} = (b_1,\dots,b_l)$. Then, the following quantity must be larger than zero. Therefore, the non-Hermitian factor $\gamma_{NH}$ must be  larger than 1:
\begin{eqnarray}
    &&\braket{n_R|n_R}\braket{n_L|n_L} - \braket{n_L|n_R}\braket{n_R|n_L}\nonumber\\
    && = \Bigl(\sum_s |a_s|^2\Bigr)\Bigl(\sum_s |b_s|^2\Bigr) - |\sum_s(a_sb_s)|^2\nonumber\\
    && \geq \Bigl(\sum_s |a_s|^2\Bigr)\Bigl(\sum_s |b_s|^2\Bigr) - \bigl(\sum_s|a_s||b_s|\bigr)^2\nonumber\\
    && = \sum_{s,t}\bigl(|a_s||b_t|-|a_t||b_s|\bigr)^2/2 \geq 0\\
    && \ \ \Leftrightarrow \gamma_{NH;n} = \braket{n_R|n_R}\braket{n_L|n_L}/\braket{n_L|n_R}\braket{n_R|n_L}\geq1\nonumber\\
\end{eqnarray}

\bibliography{nonlinear.bib}

\end{document}